\documentclass[twocolumn,preprintnumbers,prd,superscriptaddress,nofootinbib,floatfix,showpacs,showkeys]{revtex4-1}

\usepackage{float}
\usepackage{graphicx}

\usepackage{multirow}
\usepackage{natbib}
\usepackage{pifont}
\usepackage{amssymb}
\usepackage{bm}
\usepackage{amsfonts}
\usepackage{amsmath}
\usepackage{amssymb}
\usepackage[titletoc]{appendix}
\usepackage{color}
\usepackage{hyperref}
\usepackage{cleveref}
\usepackage[english]{babel}

\usepackage{booktabs}
\usepackage{mathtools}
\usepackage{subfigure}
\usepackage{enumerate}

\Crefname{equation}{Eq.}{Eqs.}
\Crefname{figure}{Fig.}{Figs.}
\Crefname{section}{Sec.}{Secs.}

\definecolor{vividviolet}{rgb}{0.62, 0.0, 1.0}
\definecolor{amaranth}{rgb}{0.9, 0.17, 0.31}
\definecolor{palatinateblue}{rgb}{0.15, 0.23, 0.89}
\definecolor{brightpink}{rgb}{1.0, 0.0, 0.5}
\definecolor{cornflowerblue}{rgb}{0.39, 0.58, 0.93}
\definecolor{deepcarminepink}{rgb}{0.94, 0.19, 0.22}
\definecolor{radicalred}{rgb}{1.0, 0.21, 0.37}

\hypersetup{ linktoc=all,
    colorlinks, linkcolor={palatinateblue},
    citecolor={brightpink}, urlcolor={amaranth}
}

\begin{document}

\title{Quasi-quintessence inflation with non-minimal coupling to curvature in the Jordan and Einstein frames}

\author{Orlando Luongo}
\email{orlando.luongo@unicam.it}
\affiliation{University of Camerino, Via Madonna delle Carceri, Camerino, 62032, Italy.}
\affiliation{Department of Mathematics and Physics, SUNY Polytechnic Institute, Utica, NY 13502, USA.}
\affiliation{INAF - Osservatorio Astronomico di Brera, Milano, Italy.}
\affiliation{Istituto Nazionale di Fisica Nucleare (INFN), Sezione di Perugia, Perugia, 06123, Italy.}
\affiliation{Al-Farabi Kazakh National University, Almaty, 050040, Kazakhstan.}

\author{Tommaso Mengoni}
\email{tommaso.mengoni@studenti.unicam.it}
\affiliation{University of Camerino, Via Madonna delle Carceri, Camerino, 62032, Italy.}

\begin{abstract}
We here explore a specific class of scalar field, dubbed quasi-quintessence which exhibits characteristics akin to ordinary matter. Specifically, we investigate under which conditions this fluid can mitigate the classical cosmological constant problem. We remark that, assuming a phase transition, it is possible to predict inflationary dynamics within the metastable phase triggered by the symmetry breaking mechanism. During this phase, we study inflationary models incorporating this cancellation mechanism for vacuum energy within the context of quasi-quintessence. There, we introduce four novel potentials, categorized into two main groups, \emph{i.e.}, the Starobinsky-like and symmetry breaking paradigms. Afterwards, we consider two distinct cases, the first without coupling with the curvature, while the second exhibiting a Yukawa-like interacting term. Hence, we compute the inflationary dynamics within both the Jordan and Einstein frames and discuss the objective to unify  old with chaotic inflation into a single scheme. We therefore find the tensor-to-scalar ratio and the spectral terms and conclude that the most suited approach involves the Starobinsky-like class of solution. Indeed, our findings show that small field inflationary scenarios appear disfavored and propose \emph{de facto} a novel technique to reobtain the Starobinsky potential without passing through generalizations of Einstein's gravity. Last but not least, we conjecture that vacuum energy may be converted into particles by virtue of the geometric interacting term and speculate about the physics associated with the Jordan and Einstein frames.
\end{abstract}

\maketitle
\tableofcontents

\section{Introduction}\label{intro}

Inflation is a theoretical scenario of great significance in addressing the main issues related to the standard Big Bang paradigm \cite{giap,baumann2012tasi,riotto,Bassett_2006,Vazquez_Gonzalez_2020}. Despite its importance, there is presently no consensus toward the potential that definitively describes the inflationary epoch \cite{starob,higgs,hilltop,natural,universality,Linde_1994}.

Originally, inflation was described by incorporating a ``de Sitter" phase, which relied on a scalar field  undergoing a first-order phase transition\footnote{In a first-order phase transition, there is a discontinuity in the first derivatives of the ``Gibbs enthalpy" during the transition, while the latter remains continuous in the first derivatives but discontinuous in the second derivatives. For a thermodynamic perspective of inflation see e.g. \cite{Andrei,gtd}.}. Within this scenario, the scalar field is initially trapped in a local minimum of the potential and then eventually reaches the true minimum of the potential.

This initial description, dubbed \emph{old inflation}, was soon abandoned due to the challenges it faced in explaining the end of inflation and the resulting highly chaotic nature of the universe. As an alternative, \emph{new inflation} models were proposed\footnote{Here, the scalar field begins in a state of thermal equilibrium in the false vacuum and gradually rolls down into degenerate minima through a second-order transition to the true vacuum.} \cite{guth,linde,newinfl}. Subsequently, an improved version, called \emph{chaotic inflation}, was introduced, which successfully addresses both the initial conditions and the exit problems \cite{LINDE1983177}. Here, the combination of chaotic initial conditions and evolution of the scalar field naturally leads to the end of inflation without requiring any external mechanism, referred to as the \emph{graceful exit}.

After these initial attempts, a wide variety of inflationary models have been proposed \cite{Odintsov:2023weg}. Generally, single-field theories can be categorized into classes based on their properties, say a) \emph{small field models}: in these models, inflation is driven by a scalar field that evolves from small values to larger ones, moving towards the minimum of the potential \cite{hilltop,natural,natural1}; b) \emph{large field models}: in these models, inflation is driven by a scalar field that evolves from large values to smaller ones, moving towards the minimum of the potential  \cite{starob,1983SvAL....9..302S,ferrara}.

Nevertheless, according to the Planck satellite results \cite{planck}, the Starobinsky potential seems to be the most promising inflationary framework \cite{starob}. Notably, one of the significant findings from the Planck satellite's Bayesian analysis on various potential models is that the quartic potential $V(\phi)=\frac{1}{4}\lambda\phi^4$, and more broadly power-law potentials, are strongly disfavored. However, including a \emph{Yukawa-like} non-minimal coupling to curvature, $R$, leads to statistically significant improvements. In this regard, the concept of non-minimally coupled inflation has been extensively explored and clearly cannot be excluded \emph{a priori} \cite{improvement,hertzberg,density,complete,obscons,preheating,futamase,futamase1,LUCCHIN1986163}.

Thus, based on our current understanding, the following points can be made regarding the best models of inflation:  $1)$ successful models  transport  vacuum energy, behaving as quasi-de Sitter phase and allowing a sufficient release of energy; $2)$ the Starobinsky potential can be obtained through a $\phi^4$-potential non-minimally coupled with curvature, passing from the Jordan to Einstein frame\footnote{The original formulation of the Starobinsky potential involves a quadratic extended gravity Lagrangian, $\mathcal L\sim R+\alpha R^2$. So far, no clear evidences have been found in favor of extended theories of gravity, but only stringent cosmological or gravitational limits have been put, see e.g. \cite{Capozziello:2019cav,Aviles:2012ay,Capozziello:2014zda,Calza:2019egu,Aviles:2013nga,Capozziello:2020ctn,Capozziello:2017ddd,Capozziello:2018aba,Aviles:2016wel}.}; $3)$ polynomial frameworks that involve non-minimal coupling to curvature are more effective in describing the stages of inflation, suggesting that such coupling may indeed exist; $4)$ the Higgs inflation model is mathematically equivalent to change the frame, specifically transitioning from the Jordan frame to Einstein frame, using the quartic term model \cite{higgs}.

Motivated by the above points, we here explore specific  inflationary models that integrate a mechanism to counteract vacuum energy, triggered by a phase transition. Specifically, we introduce a particular category of scalar field, dubbed \emph{quasi-quintessence}, that exhibits characteristics resembling those of ordinary matter. This scalar field differs from pure quintessence since its pressure arises solely from the potential term. In this context, our objective is to scrutinize the circumstances under which quasi-quintessence can provide a solution to the classical cosmological constant problem. To do so, we undertake a symmetry breaking fourth-order potential through which the cosmological constant can be erased after the transition, by invoking the shift symmetry and standard thermodynamics. We show that during the phase transition, it is possible to predict a strong cosmic speed up reinterpreted in terms of inflation. In this context, we study the corresponding inflationary dynamics that incorporates the above-quoted cancellation mechanism. Since during the transition the corresponding potential may change its form, we assume that it might be continuous at the end of inflation matching the fourth-order potential evaluated in its minimum. In this respect, we present four distinct models, classified into two main categories. The first pertains to potentials that resemble the Starobinsky one, while the second concerns more general symmetry breaking potentials. Regarding the first category, we introduce a Starobinsky-like potential that varies from the pure Starobinsky potential due to certain constants. Following this, we examine a $\pi$ model, which incorporates a discrete symmetry applied to the inflaton field. In the second category, we first present a $W$ model, characterized by finite potential walls, and then a $\Omega$ model, showing infinite potential walls. In all these scenarios, we emphasize how to provide a graceful exit, unifying \emph{de facto} the old with chaotic inflation under the same standards through our proposed potentials. We work out minimal and non-minimal couplings, where the latter is performed by virtue of an effective \emph{Yukawa-like} term. Following this recipe, we discuss how the geometric contribution  ensures that vacuum energy may be converted into particles, likely different from baryons. As a result, we thoroughly investigate the characteristics of these potentials in both the Jordan and Einstein frames. For each case, we calculate the slow roll parameters, the tensor-to-scalar ratio and the spectral index, in order to determine the compatibility of these models with observational data. We conclude that the small field class of potentials appears to be disfavored based on our analysis and also confirm that coupling the potential with curvature improves the quality of our overall results in analogy with the quartic potential. We demonstrate that the most prominent potential appears the Starobinsky-like, constructed in a different way that does not involve any generalization of Einstein's gravity. In this regard, we propose an alternative perspective to generate the potential itself involving directly the quasi-quintessence picture. Moreover, based on our findings, we conjecture that the Jordan and Einstein frames may be interchangeable and that the description of our inflationary stages simply requires formulation in the most suitable frame. As final conclusion, since  both frames yield consistent results in terms of inflationary dynamics, we speculate on how the abundance of generated particles could align with present observations, producing a bare cosmological constant as byproduct of our treatment.

The paper is structured as follows. In Sect. \ref{sec 2}, we start introducing the main features of  quasi-quintessence. In Sect. \ref{sec 3}, we provide the cancellation mechanism through which we construct our quasi-quintessence fluid. In Sect. \ref{sec 4}, our novel four potentials are exploited, categorizing them into classes. The dynamics of inflation within the quasi-quintessence paradigm is emphasized in Sect. \ref{sec 5}, whereas our results obtained with minimal and non-minimal couplings are presented in Sect. \ref{sec 6}. In Sect. \ref{sec 7}, the dynamics of the models is established whereas in Sect. \ref{Sec 8}, we emphasize the main findings derived from our paradigms in relation with current observations. Conclusions and perspectives are reported in Sect. \ref{Sec 9}.


\section{From quintessence to quasi-quintessence}\label{sec 2}

The aim of a quasi-quintessence fluid is to establish an alternative quintessence model characterized by a vanishing sound speed. As zero sound speed is typically associated with components resembling dust, one can speculate that the physical properties of quasi-quintessence yield some sort of \emph{matter with pressure} \cite{Luongo_2018,Lim:2010yk}. In this context, particles expanding the standard model of particle physics, or even those connected to the Higgs boson itself, emerge as plausible candidates for describing the quasi-quintessence field with dust-like characteristics, albeit exhibiting non-zero pressure.

In this regard, labelling with $Q$ and $QQ$ the subscripts indicating quintessence and quasi-quintessence respectively, the effective Lagrangian descriptions for both the approaches read \cite{D_Agostino_2022}

\begin{align}
\label{eq:1}
\mathcal{L}^Q &= X -V\left(\phi\right)\,,\\
\mathcal{L}^{QQ} &= K\left(X,\phi\right) +\lambda\, Y\left[X,\nu\left(\phi\right)\right]-V\left(\phi\right)\,,
\end{align}
where $K(X,\phi)$ is a generalized kinetic term written in terms of  $X \equiv\frac12 g^{\alpha\beta}\partial_\alpha \phi \partial_\beta\phi$,  $V(\phi)$ is the potential that drives the dynamics and  $\lambda$ is a  Lagrange multiplier which enforces the total energy constraint of the universe.

Those functions are not predetermined and do not have, \emph{a priori},  specified forms, but rather the entire Lagrangian relies on a scalar field $\phi$, whereas the function $\nu(\phi)$ governs the specific inertial mass of the scalar field itself. Apparently, quasi-quintessence seems more general than quintessence, although it can be recovered as a particular case where the kinetic  term is constant. Specifically, quintessence is fully-recovered assuming the Lagrange multiplier to vanish and $K\rightarrow X$. Hence, a crucial role to single out quasi-quintessence is played by the Lagrange multiplier. Without including it, there is not chance to obtain a quasi-quintessence field from a fundamental Lagrangian, since the pressure would exhibit a kinetic contribution.

To better focus on this point, let us now introduce the corresponding effective $4$-velocity $v_\alpha \equiv \partial_\alpha\phi/\sqrt{2X}$, obtaining the energy-momentum tensor
\begin{equation}
\label{eq:no10}
T_{\alpha\beta} = 2X \mathcal{L}_{,X} v_\alpha v_\beta - \left(K -  V \right) g_{\alpha\beta}\,,
\end{equation}
where the density and the pressure terms yield
\begin{subequations}\label{eq:no11}
\begin{align}
\rho =\, &2X \mathcal{L}_{,X} - \left(K -  V \right)\,,\\
\label{eq:no12}
P =\, &K - V\,.
\end{align}
\end{subequations}
At this stage, Eqs. \eqref{eq:no11} appear analogous to those obtained using a quintessence field and turn out to be identical if the generalized kinetic energy is not a constant.

Particularly, if $K=K_0=const$ and $\lambda=0$, then $\mathcal L_X=0$ and $P/\rho=-1$ \emph{always}, \emph{i.e.}, the quasi-quintessence recipe is not applicable to a Lagrangian made by kinetic plus potential terms only. However, if $K=const$ and $\lambda\neq0$, then $\mathcal L_X=\lambda Y_X$ and, in fact, we require Eqs. \eqref{eq:no11} to read
\begin{align}
\label{eq:no11bis}
\rho =\,&2X \mathcal{L}_{,X} + \mathcal V(\phi)\,,\\
\label{eq:no12bis}
P =\, &- \mathcal V(\phi)\,,
\end{align}
where $\mathcal{V}(\phi) \equiv V - K_0$, with $\mathcal L_X\equiv\lambda Y_X$.

Conversely, attempts to get quasi-quintessence modifying \emph{directly} the energy-momentum tensor have been developed in Ref. \cite{2010PhRvD..81d3520G}, trying not to pass through the use of a Lagrange multiplier and by assuming

\begin{equation}
    T_{\mu\nu}\rightarrow T_{\mu\nu}+V(\phi)g_{\mu\nu}-X\,.
\end{equation}

While appealing, this prescription suffers  from the thorny caveat of not having a Lagrangian description, appearing  unphysical from a fundamental description.

The main differences between the quintessence and quasi-quintessence fields consist in the consequences that the shift in the energy-momentum tensor implies on the  sound speed, $c_s\equiv\sqrt{\frac{\partial P}{\partial \rho}}$. In particular, indicating with $c_s^Q$ and $c_s^{QQ}$ the sound speeds for quintessence and quasi-quintessence respectively, we have
\begin{subequations}
    \begin{align}
        c_s^{QQ}&=0,\\
        c_s^Q&=1.
    \end{align}
\end{subequations}

The quintessence sound speed exhibits similarities to stiff matter, whereas quasi-quintessence behaves as matter with zero perturbations but with a non-vanishing equation of state, resembling dust. Hereafter, we will refer to this property as a matter-like fluid.

This model bears resemblance to the dark fluid, characterized by a constant pressure, and varying density and equation of state, see e.g. \cite{Luongo:2014nld}.


\section{The cosmological constant problem in quasi-quintessence phase transition}\label{sec 3}

We can here wonder under which physical frameworks  quasi-quintessence emerges. It is worth noting that this framework seems applicable when the generalized kinetic term remains constant, independently of $X$. We here present a significant scenario in which quasi-quintessence emerges, playing a central role in mitigating the cosmological constant  determined by quantum fluctuations. This results in a cancellation mechanism that fine-tunes the constant's magnitude, leading to the observed value we measure today.

If we consider  quasi-quintessence  as the responsible for carrying vacuum energy and invoke a \emph{specific phase transition}, it becomes possible to \emph{eliminate the cosmological constant}. Specifically, during the phase transition the universe speeds up by means of the vacuum energy itself and, accordingly, to ensure that inflation comes to an end, the corresponding potential can effectively counteract and cancel out the quantum fluctuations of vacuum energy, being responsible for the aforementioned \emph{cancellation mechanism}.

To show that, we first start with a standard fourth-order  potential prototype, embedded in a thermal bath  \cite{Martin:2012bt}
\begin{equation}
\label{eq:pottemperature}
V(\phi)=V_0+\frac{\chi}{4}\phi_0^4+\frac{m_{eff}^2}{2}\phi_0^2\phi^2+\frac{\chi}{4}\phi^4,
\end{equation}
where the symmetry breaking occurs as the effective mass changes sign. Indeed, since it depends upon the critical temperature, $T_c$, we write
\begin{equation}
m_{eff}^2=\chi\left[\frac{T^2}{T^2_c}-1\right],
\end{equation}
being positive or negative whether $T>T_c$ or $T<T_c$, respectively, having $\chi>0$. Specifically, $\chi$ is a dimensionless coupling constant and $\phi_0$ represents the value of $\phi$ at the minimum.

Consequently,  the critical temperature induces the phase transition, acting on the mass sign and, particularly,

\begin{itemize}
    \item[-] before the transition, when  $T>T_{\rm c}$, the minimum associated with the potential lies on $\phi=0$, having that the potential reads $V_0 + \frac{\chi \phi_0^4}{4}$;
    \item[-] during the transition, inflation takes place since the universe is pushed up by the presence of vacuum energy. Hence, vacuum energy behaves as a source for a quasi-de Sitter phase;
    \item[-] after the transition, when the temperature $T$ drops below $T_{\rm c}$, the minimum of the potential is then shifted to $\phi=\phi_0$. Here, the potential is given by $V_0$ at $\phi=\phi_0$.
\end{itemize}

The \emph{classical cosmological constant problem} is intimately related to the value of the potential offset, namely $V_0$ \cite{weinberg, peebles,Sahni_2002}. The issue arises when considering whether vacuum energy should become zero before or after the transition. Indeed, if we set $V_0 = -\frac{\chi\phi_0^4}{4}$, then the vacuum energy density, denoted as $\rho_{\text{vac}}$, becomes zero before the transition, implying that $\rho_{\text{vac}}$ is non-zero after the transition. Conversely, if we choose $V_0 = 0$,  vacuum energy is set to zero after the transition, but before the transition, $\rho_{\text{vac}}$ is non-zero. In the two cases, it is important to emphasize that vacuum energy cannot be zero at the same time before and after the transition as the offset $V_0$ cannot disappear during those periods.

However, to revising the classical cosmological constant problem with quasi-quintessence, as demonstrated in Ref. \cite{Luongo_2018}, it is possible to employ a constant generalized kinetic term by

\begin{enumerate}
    \item ensuring the shift symmetry invariance, namely $\phi\rightarrow\phi+c_0$, with $c_0$ a generic constant,
    \item fixing the conserved currents from the Noether theorem,
    \item requiring structures to form at all scales, invoking standard thermodynamics to hold.
\end{enumerate}

Considering the previously-mentioned requirements, it becomes possible to set the associated kinetic energy to a constant value, $K_0$. Further, incorporating the sign of $K_0$ to be the opposite of the potential reaching its minima may provide a resolution to the cosmological constant problem \cite{Belfiglio:2023eqi}, showing an effective dark fluid representation under the form of quasi-quintessence. This effective dark fluid is also well-suited for describing small perturbations, where the Jeans length remains identically zero across all scales, see e.g.   \cite{Aviles:2011ak}.

Bearing this in mind, we thus re-explore below the two possibilities to fix the offset and  delete the cosmological constant.
\begin{itemize}
\item[1)] If we select $V_0= -\chi\varphi_0^4/4$, then before the transition we obtain $V=0$ and therefore
\begin{align}
\label{s8bbb}
P_1&= \left\{
\begin{array}{ll}
K_0\quad &\quad{\rm (BT)}\\
K_0+\chi \varphi_0^4/4\quad&\quad{\rm (AT)}
\end{array}
\right.,\\
\label{s8bisbbb}
\rho_1&= \left\{
\begin{array}{ll}
2X\lambda Y_X-K_0\quad &\quad{\rm (BT)}\\
2X\lambda Y_X-K_0-\chi \varphi_0^4/4\quad&\quad{\rm (AT)}
\end{array}
\right.,
\end{align}
where we labeled the pressure and density with the subscript ``1" to indicate that we are employing the first possible case. Here, we have in addition that $K_0$ may turn into $K_0<-\chi \varphi_0^4/4$.
\item[2)] If we select $V_0=0$, after the transition we find  $V=0$, so that
\begin{align}
\label{s8ccc}
P_2&= \left\{
\begin{array}{ll}
K_0-\chi \varphi_0^4/4\quad &\quad{\rm (BT)}\\
K_0\quad&\quad{\rm (AT)}
\end{array}
\right.,\\
\label{s8bisccc}
\rho_2&= \left\{
\begin{array}{ll}
2X\lambda Y_X-K_0+\chi \varphi_0^4/4\quad &\quad{\rm (BT)}\\
2X\lambda Y_X-K_0\quad&\quad{\rm (AT)}
\end{array}
\right.,
\end{align}
where again we labeled the pressure and density with the subscript ``2" to indicate the second occurrence, having this time $K_0<0$.
\end{itemize}
Consequently, in both cases $K_0<0$, albeit their magnitude upper values
 change accordingly.

Thus, both  these possibilities are, in principle, plausible and the common mechanism circumvents the contribution of the vacuum energy to the cosmological constant by employing a fluid resembling matter, which possesses non-zero pressure. Notably, examples of such fluids can be found in the literature, often delving into the realm of \emph{unified dark energy-dark matter models}, see e.g., \cite{buchdahl1970non,becca2007dynamics,2012IJMPD..2130002Y,avelino2008linear,bento1999compactification,hu1999structure,2006tmgm.meet..840G,2011PhLB..694..289F,2011PhRvD..84h9905A,2015EPJP..130..130C,2021PhRvD.104b3520B,2014IJMPD..2350012L,Dunsby:2023qpb}. Nevertheless, it appears evident that after the transition the universe accelerates \emph{because of the presence of a negative matter pressure}, that acts as \emph{emergent cosmological constant}.

However, if $V_0=0$ our matter-like fluid does not provide a cancellation  since, after the transition, the term $\chi\phi_0^4/4$ disappears without the action of $K_0$. This fact produces a discontinuity in $\rho_2$ and $P_2$. So, again, the coincidence problem persists as we require a \textit{ad hoc} value for $K_0$ to justify the observed acceleration of the universe. As a consequence, this case resembles the actual  $\Lambda$CDM paradigm and, then, we deduce that only the first case remains the most suitable candidate to cancel vacuum energy out.

The mechanism provides insights into how to erase the vacuum energy excess but does not elucidate some additional points, that are our ongoing investigations, as outlined below.

\begin{itemize}
    \item[-] Which kind of potential is expected \emph{during the phase transition}?
    \item[-] Is that possible to reconcile the cancellation mechanism to the Planck measurements on inflationary potential \cite{planck}? And, particularly, can we recover the most suitable potential, namely the Starobinsky model?
    \item[-] Are the two choices for the offset both compatible with small and large field domains?
    \item[-] What are the main differences, expected once passing from Jordan to Einstein frame, while coupling the above Lagrangian to scalar curvature?
\end{itemize}

The purpose of what follows is to answer the above questions and to explore each point in order to provide a unified description of quasi-quintessence and phase transition.

\section{Quasi-quintessence potential during the transition} \label{sec 4}

Remarkably, during the transition, the shift symmetry breaks down, rendering the fourth-order potential, Eq. \eqref{eq:pottemperature}, no longer valid. Essentially, this potential remains applicable only before and after the transition, while during it Eq. \eqref{eq:pottemperature} appears as a limiting case of some more complicated potentials, required to account for the properties of the phase transition itself.

In our search for a novel, and likely more complicated form of the potential, we require that it reduces to the standard $\phi^4$ potential after the transition. More precisely,
\begin{itemize}
    \item[-] the potential under consideration should ensure continuity in the energy budget of the universe through a first-order phase transition, implying that its behavior before and after the transition  should be proportional to $\phi^4$;
    \item[-] once the metastable phase ends, the universe successfully escapes the transition and settles into the symmetry minimum, ensuring a \emph{graceful exit} from the transition, \emph{i.e.}, from inflation;
        \item[-] the potential should be compatible with current observations and the coupling constant $\chi$ should be associated with   vacuum energy.

\end{itemize}

For the sake of completeness, the cancellation mechanism bears resemblance to the concept of \emph{old inflation}, characterized by a phase transition. However, in our case the potential can escape the metastable phase with the graceful exit. Particles are thus produced as due to the coupling with curvature  \cite{gravbarion,Dolgov_1997}. Accordingly, our  picture enables particles to form even during inflation, being reinterpreted in terms of non-baryonic constituents, see e.g. \cite{Belfiglio:2022cnd,Belfiglio:2023eqi,Belfiglio_2023,belfiglio2023particle}.

\begin{figure*}[p]
  \centering
  \textbf{a)}\hspace{0.mm}
    \includegraphics[width=.47\linewidth]{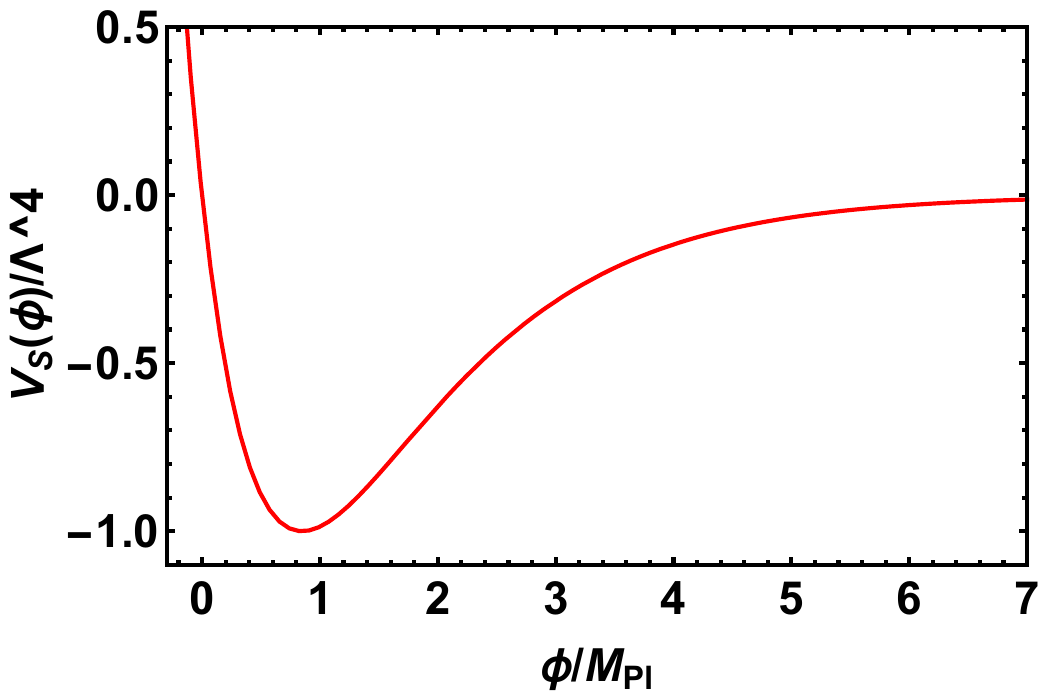}
  \hspace{1.8mm}
  \textbf{b)}\hspace{0mm}
    \includegraphics[width=.46\linewidth]{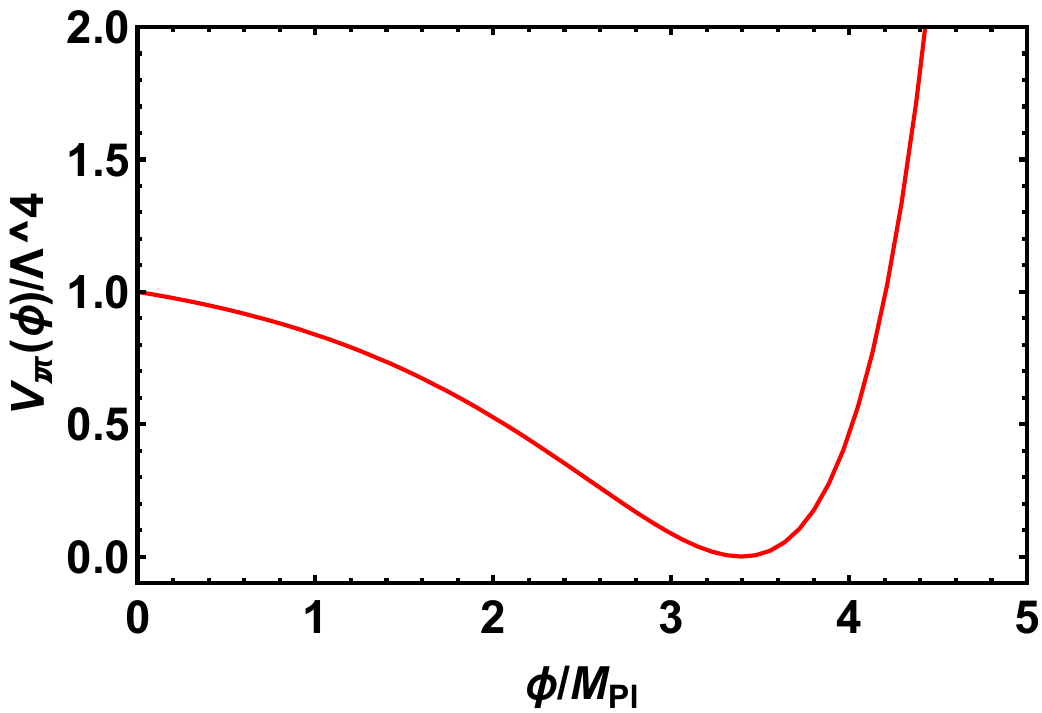}\\
    \vspace{7mm}
  \textbf{c)}
    \includegraphics[width=.46\linewidth]{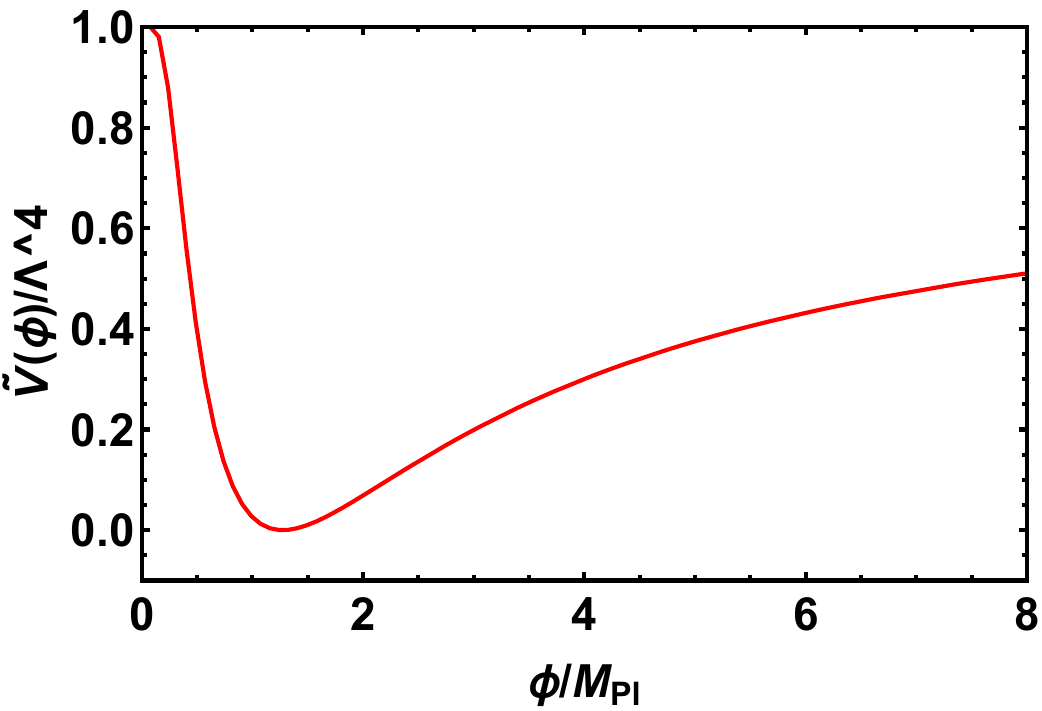}
  \hspace{4mm}
  \textbf{d)}
    \includegraphics[width=.45\linewidth]{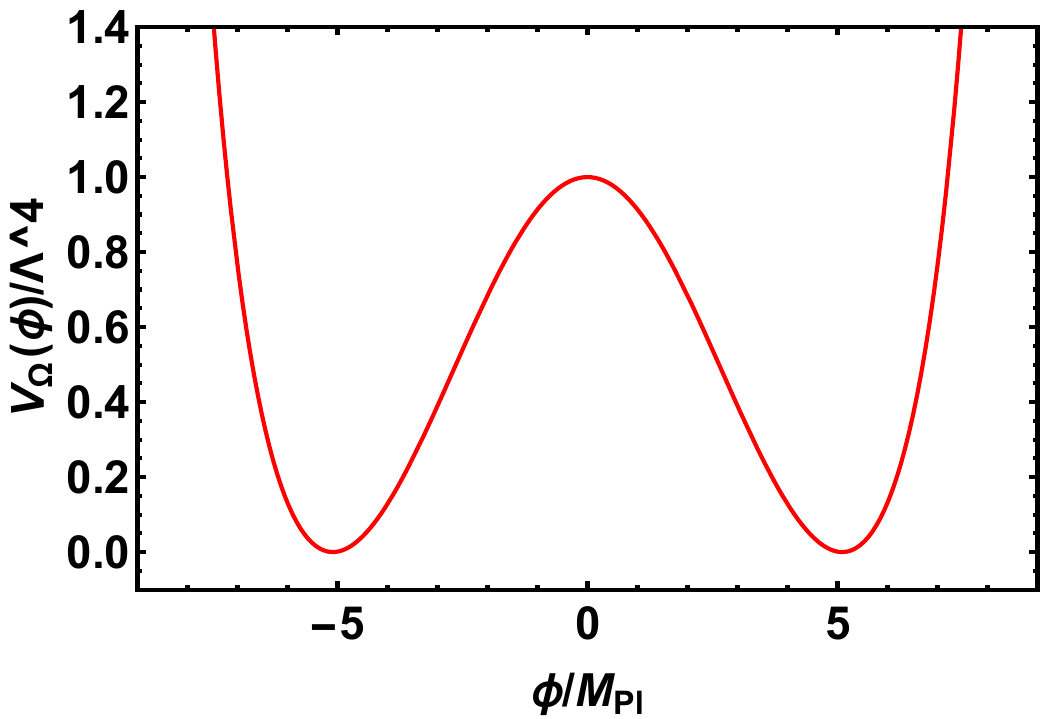}\\
    \vspace{7mm}
  \textbf{e)}
    \includegraphics[width=.46\linewidth]{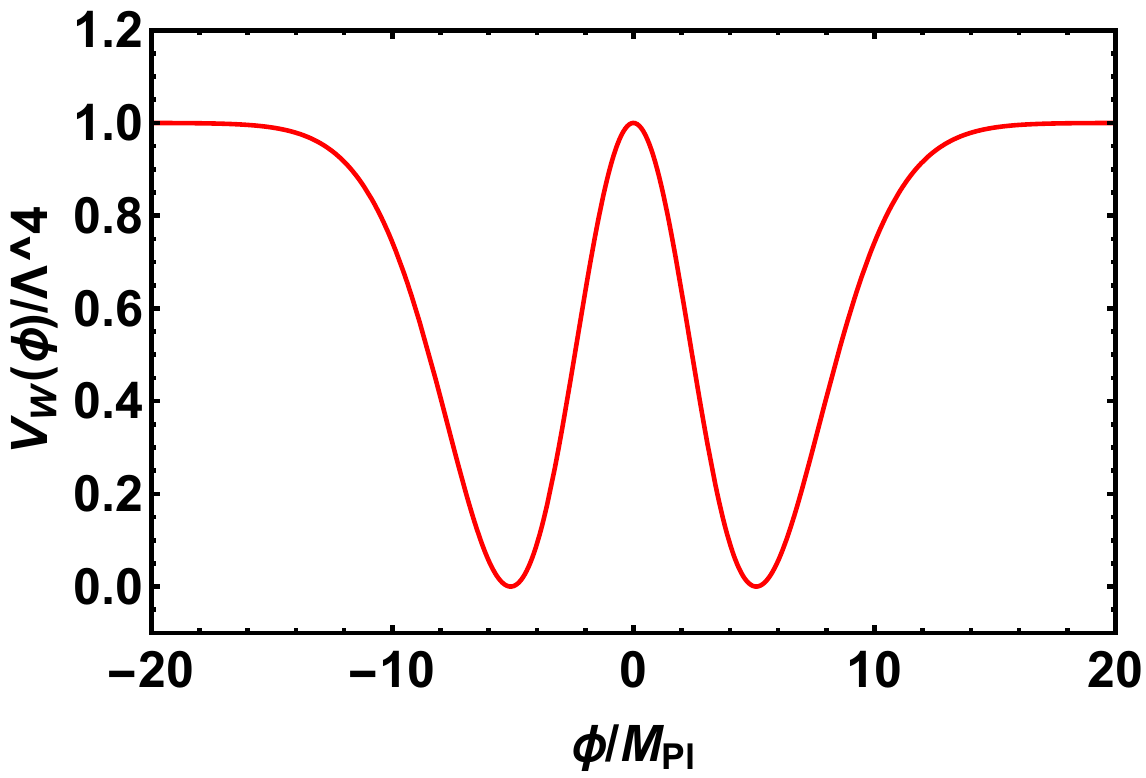}
  \hspace{3mm}
  \textbf{f)}
    \includegraphics[width=.46\linewidth]{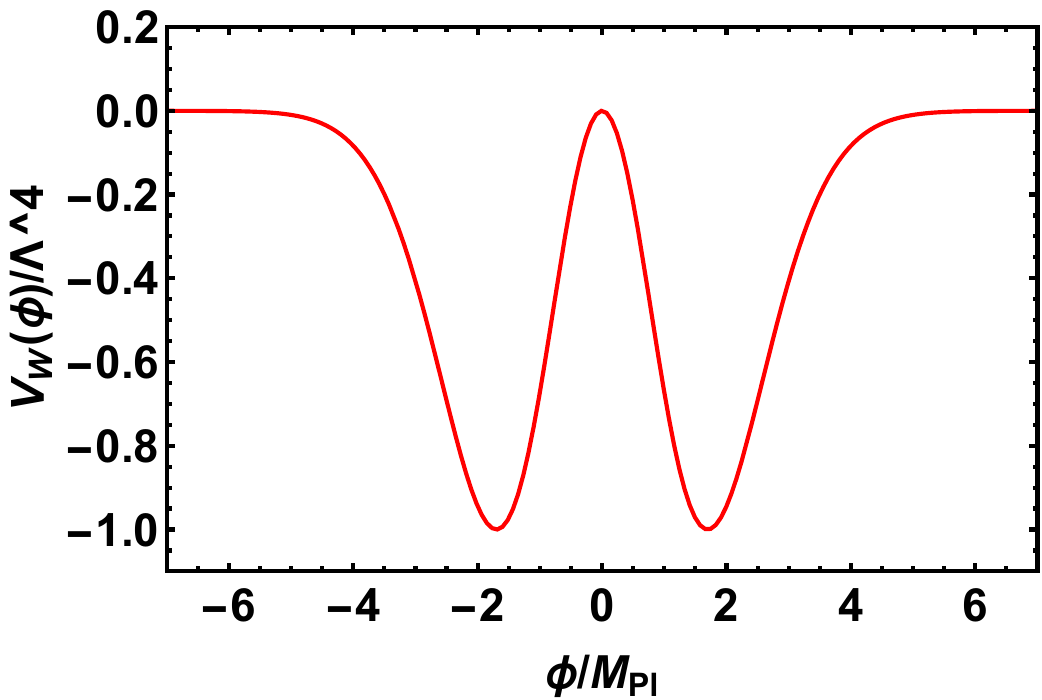}
  \caption{Inflationary potentials introduced in Sect. \ref{sec 4}: \textbf{a)} $V_S(\phi)$, Eq. \eqref{star_like}, with $\phi_0=\sqrt{3/2}M_{Pl}\ln2$, $\chi=1.66$, $\alpha=\sqrt{2/3}/{M_{Pl}}$  and $V_0=-{\chi\phi_0^4}/{4}$; \textbf{b)} $V_\pi(\phi)$, Eq. \eqref{pi model}, with $\phi_0=4\sqrt{3/2}M_{Pl}\ln2$, $\chi=6.48\cdot10^{-3}$, $\alpha=\sqrt{2/3}/{M_{Pl}}$ and $V_0=0$; \textbf{c)} $\Tilde{V}(\phi)$, Eq. \eqref{morse model}, with $\phi_0=4\sqrt{3/2}M_{Pl}\ln2$, $\chi=1.66$, $\alpha=\phi_0\ln2$ and $V_0=-{\chi\phi_0^4}/{4}$;  \textbf{d)} $V_\Omega(\phi)$, Eq. \eqref{v3}, with $\phi_0=6\sqrt{3/2}M_{Pl}\ln2$, $\chi=1.28\cdot10^{-3}$, $\alpha=\ln2/\phi_0^2$ and $V_0=0$; \textbf{e),f)} $V_W(\phi)$, Eq. \eqref{v2}, with respectively $\phi_0=6\sqrt{3/2}M_{Pl}\ln2,\>2\sqrt{3/2}M_{Pl}\ln2$, $\chi=1.28\cdot10^{-3},\>0.10$, $\alpha=\ln2/\phi_0^2$ and $V_0=0,\>-{\chi\phi_0^4}/{4}$. In these plots we consider $\Lambda^4=\frac{\chi\phi_0^4}{4}$. Further, the plots refer to as the potentials $V$ without considering the kinetic term, namely $\mathcal V=V-K_0$. }
  \label{fig pot}
\end{figure*}

Thus, we start from the most general double exponential form \cite{D_Agostino_2022,Dolgov_1997}
\begin{equation}\label{proposed}
    V(\phi)=V_0+\frac{\chi\phi_0^4}{4}\left[\frac{e^{\beta\phi^d}-e^{\beta\phi_0^d}e^{\alpha\left(\phi^c-\phi_0^d\right)}}{1-e^{\beta\phi_0^d-\alpha\phi_0^d}}\right]^2.
\end{equation}
This proposal is general and contains all the information to construct a wide classes of possible potentials, encoded into the constants $V_0,\chi,\phi_0,\alpha,\beta,c,d$. It has several applications that span from theoretical up to solid state physics. For examples, it is possible to reobtain the Morse potential by invoking a benchmark scenario as in Eq. \eqref{proposed} \cite{delSolMesa:1997ddp}. In addition, we leave all these constants \emph{a priori} free, fixing by physical motivations each of them, in order to realize the best choice to drive up the universe to accelerate.

For example, if  we handle  $\beta=0,\>\alpha<0$ and $c=1$, we obtain
\begin{equation}
    V_S(\phi)=V_0+\frac{\chi\phi_0^4}{4}\left[\frac{1-e^{-|\alpha|\left(\phi-\phi_0\right)}}{1-e^{|\alpha|\phi_0}}\right]^2,\label{staroblike}
\end{equation}
reducing, through a simple shift over the field in terms of  $\phi_0$, and by virtue of the following further positions:
\begin{equation}
    \Lambda^4\simeq\frac{\chi\phi_0^4}{4}\left({1-e^{|\alpha|\phi_0}}\right)^{-2},\qquad \alpha\simeq\sqrt\frac{2}{3}\frac{1}{M_{Pl}}.
\end{equation}
to the Starobinsky model, written under the form,
\begin{equation}
V_\mathcal{S}=\Lambda^4\left(1-e^{-\sqrt{2/3}\phi/M_{Pl}}\right)^2,\label{star_like}
\end{equation}
where  $M_{Pl}$ is the Planck mass. Among the various inflationary models, the Starobinsky potential has been found to be the best-suited scenario according to current observational data. Our picture, however, is slightly different than a genuine Starobinsky potential, due to the coupling between $\alpha$ and $\Lambda$ and then deserves further investigation.

Further, the Starobinsky-like picture, above obtained, appears a large field potential. Interestingly, if we incorporate symmetry considerations on the inflaton field, new models with distinct properties can be easily computed.

Thus, from Eq. \eqref{staroblike}, considering $V_0=0$ and $\Lambda^4=\frac{\chi\phi_0^4}{4}\left({1-e^{|\alpha|\phi_0}}\right)^{-2}$ we obtain
\begin{equation}
    V_S(\phi)=\Lambda^4\left(1-e^{-|\alpha|\left(\phi-\phi_0\right)}\right)^2,
\end{equation}
and considering a discrete and a continuous transformation, under the form
\begin{equation}
    \phi^\prime=-\phi,\quad\phi^\prime=\pm1/\phi, \label{tranformations on v}
\end{equation}
we find two interesting scenarios. Within $\phi\rightarrow1/\phi$, we recover a potential given by
\begin{equation}
    \Tilde{V}(\phi)=\Lambda^4\left(1-e^{-|\alpha|\left(\frac{1}{\phi}-\frac{1}{\phi_0}\right)}\right)^2.\label{morse model}
\end{equation}
Here, despite the encouraging dynamical analyses and the fact that the potential is small field, differently from the Starobinsky-like potential that, as claimed, is large field, the disagreement with experimental data emphasizes that the functional form under exam cannot be used. The coefficients $c$ and $d$ in Eq. \eqref{proposed} are thus favored to be positive definite.

Conversely, with the discrete symmetry of the field, $\phi\rightarrow-\phi$, we physically invoke a rotation by a phase of $\pi$, we get a small field potential such as
\begin{equation}
    V_\pi(\phi)=\Lambda^4\left(1-e^{|\alpha|\left(\phi-\phi_0\right)}\right)^2,\label{pi model}
\end{equation}
where $\Lambda^4=\frac{\chi\phi_0^4}{4}\left({1-e^{-|\alpha|\phi_0}}\right)^{-2}$. For the sake of convenience we  baptize this potential \emph{$\pi$ potential}. The discrete symmetry is imposed with the aim of passing from large to small field. Clearly, this can be mathematically performed once the field is complex. However, we infer the functional form of the potential imposing the shift $\phi\rightarrow-\phi$, but keeping $\phi$ real. Remarkably, for the previous models we emphasize that the choice of $V_0=0$ is due to the fact that the transformations lead to a small field inflationary scenario.

Two more choices of parameters leading to symmetry breaking potentials are $\beta=0,\>\alpha<0$ and $c=2$. This class of models provides two symmetric minima, resulting into a \emph{$W$ model} \cite{D_Agostino_2022}
\begin{equation}
    V_W(\phi)=V_0+\frac{\chi\phi_0^4}{4}\left[\frac{1-e^{-|\alpha|\left(\phi^2-\phi^2_0\right)}}{1-e^{|\alpha|\phi^2_0}}\right]^2.\label{v2}
\end{equation}
This model presents finite walls in analogy to the more popular class of T-models \cite{universality}. Here, in principle, an apparent problem seems to occur, as the energy required to exit from the minimum is finite. However, the inflaton, during the accelerated phase, loses the initial energy because of the Hubble friction term, Eq. \eqref{infl dyn}, that, at the same time, dampens the oscillations around the minimum at $\phi_0$. In other words, to have a well-defined graceful exit, the friction should be enough to decrease the energy of the inflaton not to exceed the wall. Moreover, in order to reduce the complexity of the potential, the height of the walls is fixed as $V_W(0) = V_W(\pm\infty)$, \emph{i.e.}  $|\alpha| = \frac{\ln2}{\phi_0^2}$.

Instead, by considering $\beta\neq0$, and, for $\alpha<0,\>|\alpha| = \beta$ and $c = d = 2$ we get the $\Omega$ potential
\begin{equation}\label{v3}
    V_\Omega(\phi)=V_0+\frac{\chi\phi_0^4}{4}e^{2|\alpha|\phi^2}\left[\frac{1-e^{-2|\alpha|\left(\phi^2-\phi_0^2\right)}}{1-e^{2|\alpha|\phi_0^2}}\right]^2.
\end{equation}
It also provides a symmetry breaking with two symmetric minima, but this potential has infinite walls. Let us highlight that the requirements $|\alpha| = \beta$ and $c = d$ are fundamental to avoid nonphysical or complex potentials.

The last two proposals involve chaotic potentials with a symmetry breaking functional form. The $W$ model exhibits both small and large field behaviors due to its finite potential walls, while the latter exhibits small field behavior. In this context, inflation occurs during the metastable phase of a first-order phase transition.  Consequently, the offset is $V_0=0$ for the $\Omega$ potential while the $W$ potential can be studied with $V_0=0$ and $V_0\neq0$.

The conclusion of this phase transition is characterized by a chaotic graceful exit, where the inflaton field loses energy while it is oscillating around the minimum of the potential.

Hence, this picture provides a unification of old and new inflation through a chaotic scheme.

Summing up, we thus introduced two main classes of potentials:
\begin{itemize}
    \item[-] Starobinsky-like potentials, starting from a general double exponential potential, Eq. \eqref{proposed}, within the quasi-quintessence hypothesis, we derived a large field potential denoted as $V_S$, Eq. \eqref{star_like}, that mimes the Starobinsky features \cite{D_Agostino_2022}. Additionally, by considering a discrete field symmetry, we introduced a new small field potential named $\pi$ potential, Eq. \eqref{pi model}.
    \item[-] Symmetry breaking potentials, namely $V_W$ and $V_\Omega$, respectively Eqs. \eqref{v2} and \eqref{v3}. The latter displays infinite potential walls and a small field nature, while the former, due to its finite potential walls, can exhibit both small and large field behaviors.
\end{itemize}

Consequently, we examined the inflationary dynamics induced by these potentials in both minimally and non-minimally coupled scenarios.


\section{Inflationary dynamics within the quasi-quintessence picture}\label{sec 5}

The inflationary dynamics is slightly modified by the presence of a quasi-quintessence fluid. Particularly, it can be described starting from the continuity equation
\begin{equation}
    \ddot\phi+\frac{3}{2}H\dot\phi+\frac{\mathcal{V}^\prime(\phi)}{2\mathcal{L}_{,X}}=0,\label{infl dyn}
\end{equation}
where $\mathcal{V}^\prime(\phi)\equiv\frac{d\mathcal{V}(\phi)}{d\phi}$.

The so-obtained equation is a forced harmonic oscillator, where the force  is provided by  $\mathcal V^\prime(\phi)$, whereas the viscous term, $\sim H\dot\phi$, is related to the universe expansion.

Noticeably, Eq. \eqref{infl dyn} shows up minimal modifications with respect to the quintessence equation of motion, that  instead reads

\begin{equation}
\ddot\phi+3H\dot\phi+V^\prime(\phi)=0.
\end{equation}
The numerical factors of difference are not relevant in the slow roll regime, once $\mathcal L_X$ is conveniently fixed to unity.

In addition, the first Friedmann equation of quasi-quintessence is
\begin{equation}\label{1f}
    H^2=\frac{8\pi}{3M^2_{Pl}}\rho=\frac{8\pi}{3M_{Pl}^2}\left(\mathcal{L}_{,X}\dot\phi^2+\mathcal{V}(\phi)\right).
\end{equation}

\subsection{Slow roll in quasi-quintessence scenarios}

We have now all the ingredients to better understand the conditions under which a quasi-quintessence field can drive an inflationary period. Thus, we consider a \emph{slow roll} stage, where the potential energy of the inflaton dominates over its kinetic energy. Hence, the slow roll conditions in our case become
\begin{equation}
    \mathcal{L}_{,X}\dot\phi^2\ll\mathcal V(\phi)\>\Rightarrow\>\ddot\phi\ll\frac{3}{2}H\dot\phi,\label{2 slowroll}
\end{equation}
implying that $H^2\gg|\dot H|\simeq0$, \emph{i.e.}, giving rise to a de quasi-Sitter phase  \cite{Riotto:2010jd}, where the corresponding Friedmann and continuity equations yield respectively
\begin{align}
    H^2&\simeq\frac{8\pi}{3M_{Pl}^2}\mathcal{V}(\phi),\label{srfr}\\
    3H&\dot\phi\simeq-\mathcal{V}^\prime(\phi),\label{srdin}
\end{align}
encoded in the Hubble \emph{slow roll parameters} defined as
\begin{subequations}
\begin{align}
    \epsilon_H(\phi)&\equiv\frac{M_{Pl}^2}{4\pi\mathcal{L}_{,X}}\left(\frac{H^\prime(\phi)}{H(\phi)}\right)^2,\label{heps}\\
    \eta_H(\phi)&\equiv\frac{M_{Pl}^2}{4\pi\mathcal{L}_{,X}}\frac{H^{\prime\prime}(\phi)}{H(\phi)}.\label{heta}
\end{align}
\end{subequations}
The subscript ``H" refers to Hubble slow roll. Here,  we made a more suitable choice of slow roll parameters than the widely-used \emph{potential} slow roll parameters:
\begin{subequations}
\begin{align}
    \epsilon_V&\equiv-\frac{\dot H}{H^2}\simeq\frac{M_{Pl}^2}{16\pi\mathcal{L}_{,X}}\left(\frac{\mathcal V^\prime}{\mathcal V}\right)^2,\label{eps}\\
    \eta_V&\equiv-\frac{\ddot \phi}{H\dot\phi} -\frac{\dot H}{H^2} \simeq\frac{M_{Pl}}{8\pi\mathcal{L}_{,X}}\left(\frac{\mathcal V^{\prime\prime}}{\mathcal V}\right).\label{eta}
\end{align}
\end{subequations}
Commonly the latter relations appear overused in the literature, as due to their simplicity. They represent  first-order approximated terms, being easier to handle from a computational viewpoint \cite{Ellis_2015}. However, as the approximation starts to break down, say if $H^2\gtrsim \dot H$, as due to the complexity of potentials, it is not convenient to work Eqs. \eqref{eps}-\eqref{eta} out, but rather to employ Eqs. \eqref{heps}-\eqref{heta}.

We notice that the above definitions are intertwined, and, in fact, $\epsilon_V$ and $\eta_V$ can be written in terms of the Hubble slow roll  parameters by
\begin{subequations}
\begin{align}
    \epsilon_V&=\epsilon_H\left(\frac{3-\eta_H}{3-\epsilon_H}\right)^2,\label{eps1}\\
    \eta_V&=\sqrt{\frac{M_{Pl}^2}{4\pi\mathcal{L}_{,X}}\epsilon_H}\frac{\eta_H^\prime}{3-\epsilon_H}+\left(\frac{3-\eta_H}{3-\epsilon_H}\right)\left(\epsilon_H+\eta_H\right),\label{eta1}
\end{align}
\end{subequations}
where the first term in Eq. \eqref{eta1} is higher order with respect to the second one, and so hereafter it will be neglected.

During the slow roll, the conditions for sustaining inflation occur when
\begin{equation}
    \ddot{a}>0\>\iff\>\epsilon_H\ll1,
\end{equation}
thus, from Eqs. \eqref{eps1} and \eqref{eta1}, we infer that inflation ends as
\begin{equation}\label{infl end}
    \epsilon_V\simeq\left(1+\sqrt{1-\frac{\eta_V}{2}}\right)^2.
\end{equation}

Considering only the potential slow roll parameters, we drop out the subscript \emph{V} and straightforwardly we compute the number of e-foldings between the beginning and end of inflation within the slow roll approximation.

If we denote the values of the inflaton field at the beginning and end of inflation as $\phi_i$ and $\phi_f$ respectively, the total number of e-foldings can be calculated as follows
\begin{equation}
    N\equiv\int_{\tau_i}^{\tau_f}{Hd\tau}\simeq-\frac{8\pi}{M_{Pl}^2}\int_{\phi_i}^{\phi_f}{\frac{\mathcal V}{\mathcal V^\prime}d\phi}.\label{efold slowroll1}
\end{equation}
To have enough amount of inflation we require that $N\gtrsim60$, as in the standard picture \cite{linde2005particle}.

For the sake of completeness, it might be remarked that  the aforementioned expression is just the first-order approximation of a more general expression, $\Tilde{N}$, that by definition satisfies $\Tilde N\leq N$ and requires a e-folding number $\Tilde N\gtrsim70$ to have sufficient inflation, see e.g. \cite{Liddle_1994}. Following this recipe, we here consider $\Tilde{N}$ up to the second order
\begin{equation}\label{efold slowroll}
    \Tilde{N}\simeq-\sqrt{\frac{4\pi}{M_{Pl}^2}}\int_{\phi_i}^{\phi_f}{\frac{1}{\sqrt{\epsilon(\phi)}}\left(1-\frac{1}{3}\epsilon(\phi)-\frac{1}{3}\eta(\phi)\right)d\phi}.
\end{equation}
Hereafter, we may confuse $N$ with $\tilde N$, without losing generality. Afterwards, as the slow roll ends, we expect that the inflaton field starts oscillating around the minimum of the potential, $\phi_0$. This phenomenon, dubbed reheating \cite{Allahverdi_2010}, does not influence the \emph{graceful exit} that depends instead on the chaotic form of the potential involved into Eq. \eqref{infl dyn}.


\subsection{Coupling inflaton with spacetime curvature}\label{subsec4}

At primordial times, as stated in Sect. \ref{intro}, there is no reason to exclude \emph{a priori} a non-minimal coupling between the inflaton and scalar curvature, $R$. An interacting Yukawa-like term in Eq. \eqref{eq:1}, representing the simplest interacting Lagrangian, can therefore be written as
\begin{align}
    \mathcal{L}_{int}&=\frac{1}{2}\xi R\phi^2,\label{L int}
\end{align}
giving rise to a new effective potential, $V_{eff}(\phi)=V(\phi)+\frac{1}{2}\xi R\phi^2$, where the coupling constant, $\xi$, is considered as a free parameter and the $1/2$ factor is arbitrary.

First, we focus on the range $-4<\log_{10}\xi<4$, namely a plausible bound provided by the Planck mission \cite{planck}, whereas later we further extend the range of $\xi$, invoking $|\xi|\ll1$, by virtue of the  analogy with particle physics, where a very small $\xi$ is essential to guarantee that gravity does not  significantly change at local scales.

Indeed, from Eq. \eqref{L int}, the effective Newtonian gravitational constant, $G_{eff}$, couples to $\phi$ by
\begin{equation}
G_{eff}=\frac{G}{1-\xi\chi\phi^2}.
\end{equation}
The effective constant might be consistent with our present constraints on $G$. However, we remark that the condition $G_{eff}\simeq G$ is  restored after the transition, while during the transition is not strictly necessary as due to the metastable phase itself, see e.g. \cite{preheating,futamase}. Regarding the sign of $\xi$, when $\xi>0$, we ask for $G_{eff}>0$, yielding
\begin{equation}\label{il}
    \phi^2<\frac{1}{\chi\xi},
\end{equation}
leading to a multiplicative degeneracy between $\chi$ and $\xi$.

It is important to note that the specific implications of positive or negative non-minimal coupling can have significant consequences for the large scale dynamics of our  resulting inflationary dynamics and depend on the structure of the involved inflationary potentials.

\section{Frame selection in quasi-quintessence inflationary models}\label{sec 6}

The coupling term imposed by the interacting Lagrangian, Eq. \eqref{L int},  written as a Yukawa-like contribution gives rise to conceptual complications. In particular, the main concern is to single out the ``right frame", either Jordan or Einstein one, to work in \cite{faraoni,Catena_2007,einjor}.

Phrasing it differently, the corresponding dynamics can be studied in the above quoted two different frames \cite{postma,Capozziello_1997} having that
\begin{itemize}
    \item[-] in the Jordan frame,  the action is left unaltered. The interacting term, $\mathcal L_{int}$, provides the type of interaction;
    \item[-] in the Einstein frame, a conformal transformation is applied on the metric, getting rid of the interaction.
\end{itemize}

In principle, transitioning from one frame to another could potentially carry implications for the associated physics, even though this aspect remains not entirely comprehended at present, see e.g. \cite{faraoni2007}.

Below, let us focus separately on each of the two frames, adopting our quasi-quintessence fluid.

\subsection{The Einstein frame}\label{subsec41}

In the Einstein frame, after performing a conformal transformation to the metric, we recover a minimally coupled action. Following Appendices \ref{appendix} and \ref{appendix1}, a new field, $h$, implies a transformed potential, $\mathcal V_E(\phi)$, defined as
\begin{align}
\begin{split}
     h(\phi)=&\sqrt{\frac{6}{{\chi}}}\tanh^{-1}{\left(\frac{\sqrt{6\chi}\xi{\phi}}{\sqrt{1+\phi^2\chi\xi(-1+6\xi)}}\right)}\\
     &-\sqrt{\frac{{-1+6\xi}}{{\chi\xi}}}\sinh^{-1}{\left(\sqrt{\chi\xi(-1+6\xi)}\phi\right)},\label{h}
\end{split}\\
    \mathcal V_{E}(h)=&\frac{\mathcal V(\phi(h))}{\left(1-\xi\chi\phi^2(h)\right)^2},
\end{align}
where the subscript ``E" recalls that the quantities are calculated in the Einstein frame.

Passing from $\phi$ to $h$ is not analytic, in general. We can therefore consider some limiting cases. For example, one can consider  the usual conformal coupling, \emph{i.e.}, $\xi=1/6$ \cite{futamase1} providing
\begin{equation}
    \phi=\frac{1}{\sqrt{\chi\xi}}\tanh(\sqrt{\chi\xi}h),\label{conformal coupling}
\end{equation}
that represents an upper limit for $\xi$, in view of the constraints that we imposed on $\xi$. Indeed, exceeding $\xi\simeq 1/6$ with larger values yields slow roll parameters that seem not to agree with the most recent  bounds. In other words, we can undertake $\xi\simeq 1/6$ as an upper cut-off scale.

Thus, dismissing strong interactions, again the condition $|\xi|\gg1$ holds, ending up with the following two main cases:
\begin{subequations}
    \begin{align}
    \phi&\simeq\frac{1}{\sqrt{\chi\xi}}\sin(\sqrt{\chi\xi}h),&0<\xi\ll{1},\label{small xi}\\
    \phi&\simeq-\frac{1}{\sqrt{\chi\xi(6\xi-1)}}\sinh\left(\frac{\sqrt{\chi\xi}h}{\sqrt{6\xi-1}}\right),&-1\lesssim\xi<0.\label{small neg xi}
    \end{align}
\end{subequations}

We can now study the inflationary dynamics, in both positive and negative weakly-interacting cases, with the same tools introduced for the minimally coupled scenario, dealing  with  $h$ and the transformed potentials, $\mathcal V_E(h)$.

\subsection{The Jordan frame}\label{subsec42}

The first Friedmann equation, obtained taking into account the non-minimal coupled interacting term, Eq. \eqref{L int}, leads to
\begin{equation}
    H^2=\frac{\chi}{3(1-\chi\xi\phi^2)}\left(\dot\phi^2\mathcal{L}_{,X}+\mathcal{V}(\phi)\right),
\end{equation}
having
\begin{equation}
    \ddot\phi+\frac{3}{2}H\dot\phi+\frac{\mathcal{V}^\prime(\phi)+\xi R\phi}{2\mathcal{L}_{,X}}\simeq0,\label{Jordan dyn}
\end{equation}
where we approximate $R^\prime=\frac{\dot R}{\dot\phi}\simeq0$, since we delve into slow roll. Indeed, since we assume inflation occurring in a quasi-de Sitter phase, the scalar curvature, $R$,
\begin{equation}
    R=\frac{\chi}{(1-\chi\xi\phi^2)}\left[(6\xi-2)\dot\phi^2+4\mathcal{V}(\phi)+6\xi\phi\ddot\phi\right],\label{scalar curvature}
\end{equation}
can be considered roughly constant. Manifestly, the third term of Eq. \eqref{scalar curvature} can be regarded as a second-order term compared to the rest. Consequently, as the field $\phi$ increases, the curvature $R$ also increases. This situation can present a significant challenge when aiming to smoothly transition out of inflation and reach the minimum of the effective potential.

On the one hand, in fact, in the case of small field inflation with a positive coupling strength, the interacting Lagrangian has an opposite sign than the potential one. Conversely, in the case of negative $\xi$ both contribution decrease as the field decreases. On the other hand, for large field inflation the situation is perfectly symmetrical to small fields.

Hence, to address this concern, we study both the negative and positive coupling for each potential. Further, we discuss whether, and for which choice of parameters, the effective potentials in our hands lead to a well-defined inflationary stages, while ensuring the consistency of the model with observational data.


\section{Consequences on dynamics}\label{sec 7}

\begin{figure*}[p]
  \centering
  \textbf{a)}\hspace{0.mm}
    \includegraphics[width=.47\linewidth]{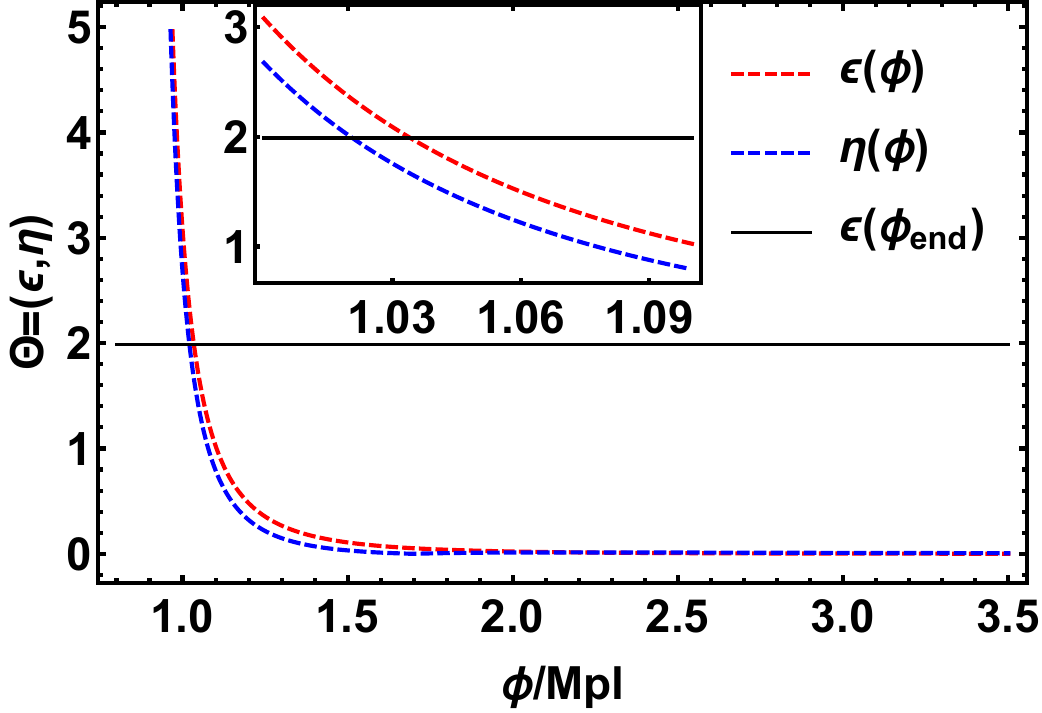}
  \hspace{0mm}
  \textbf{b)}\hspace{0mm}
    \includegraphics[width=.47\linewidth]{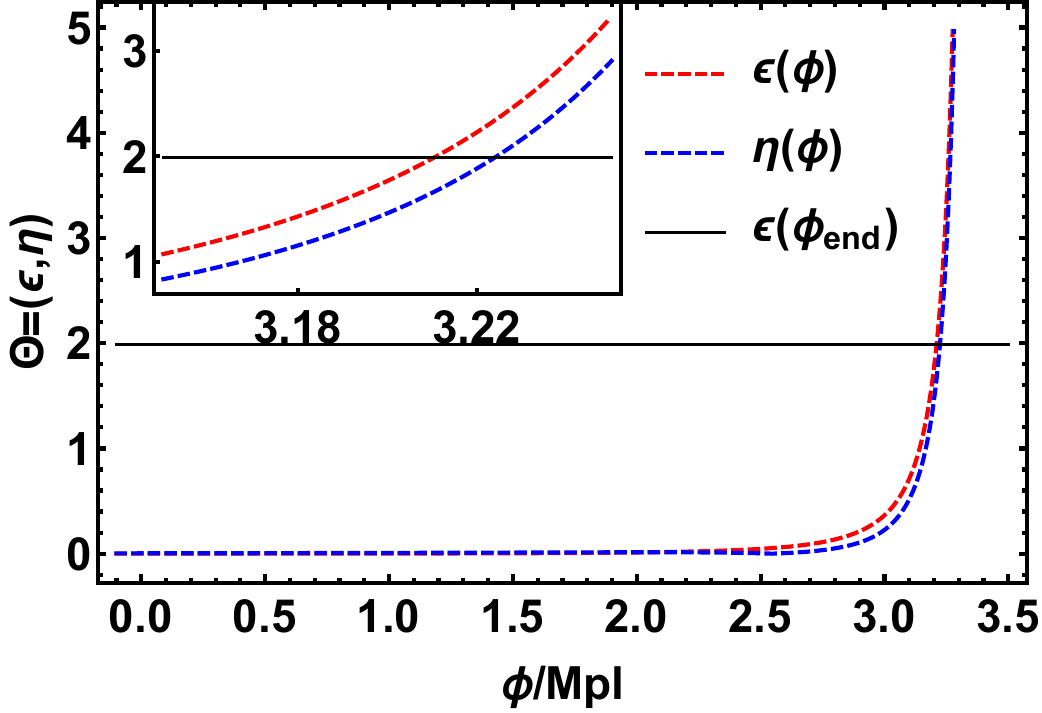}\\
    \vspace{7mm}
  \textbf{c)}
    \includegraphics[width=.47\linewidth]{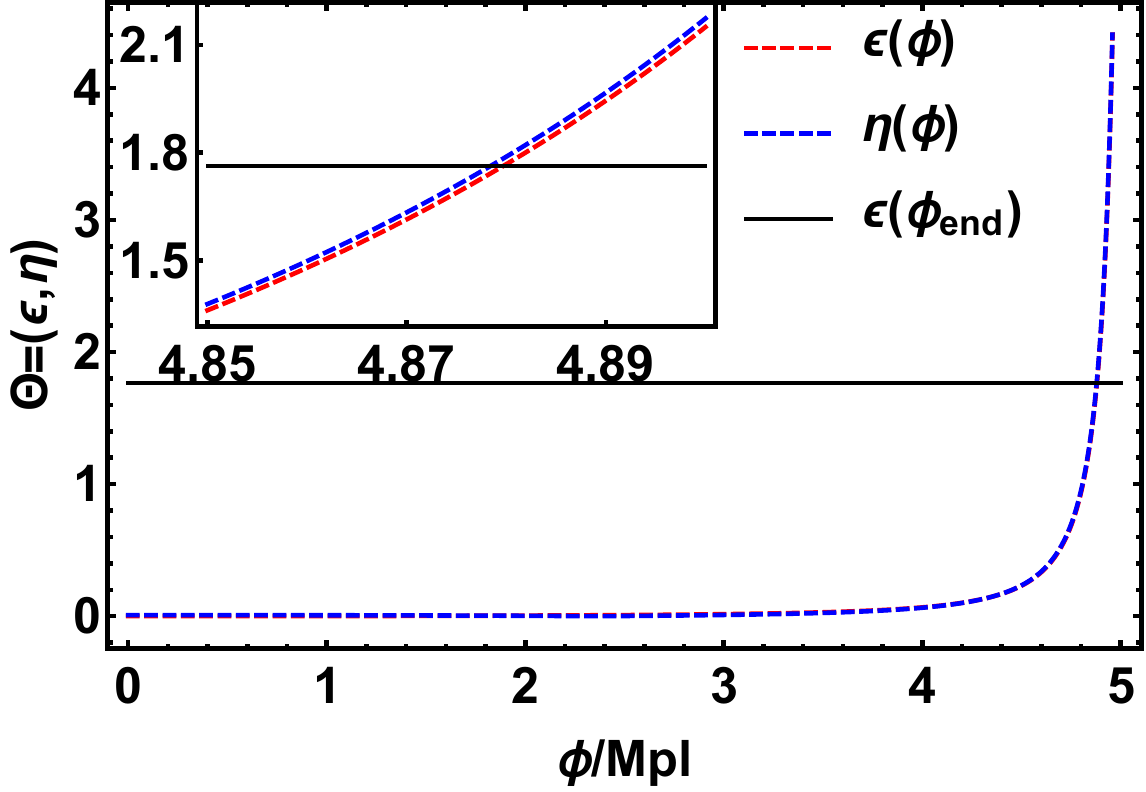}
  \hspace{0mm}
  \textbf{d)}
    \includegraphics[width=.47\linewidth]{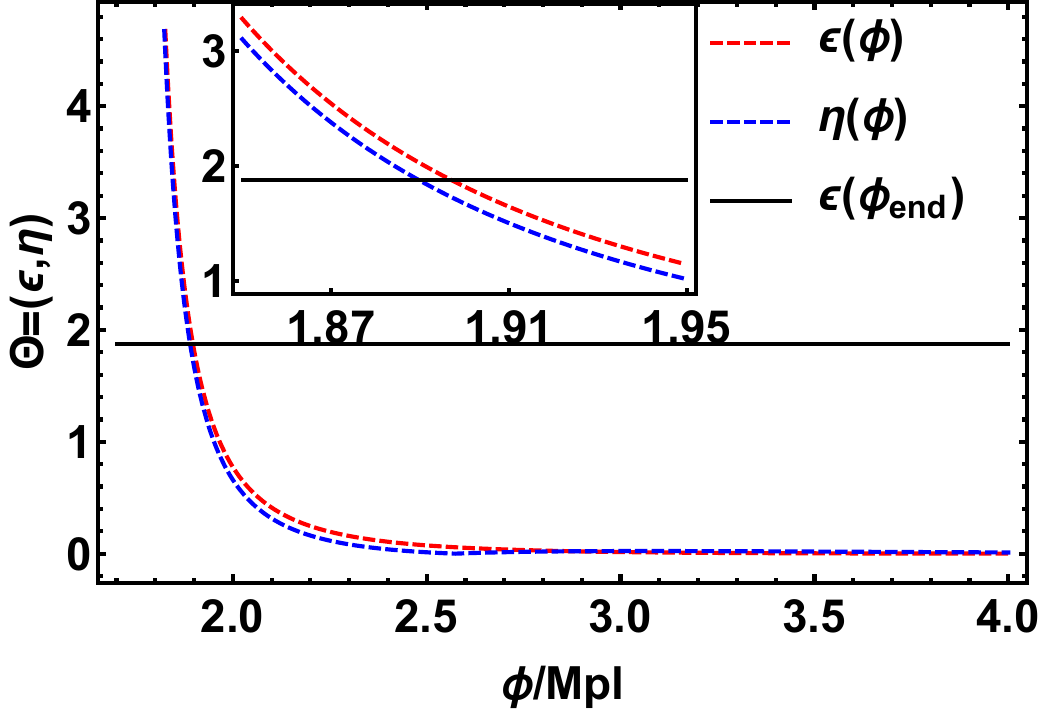}\\
    \vspace{7mm}
  \textbf{e)}
    \includegraphics[width=.47\linewidth]{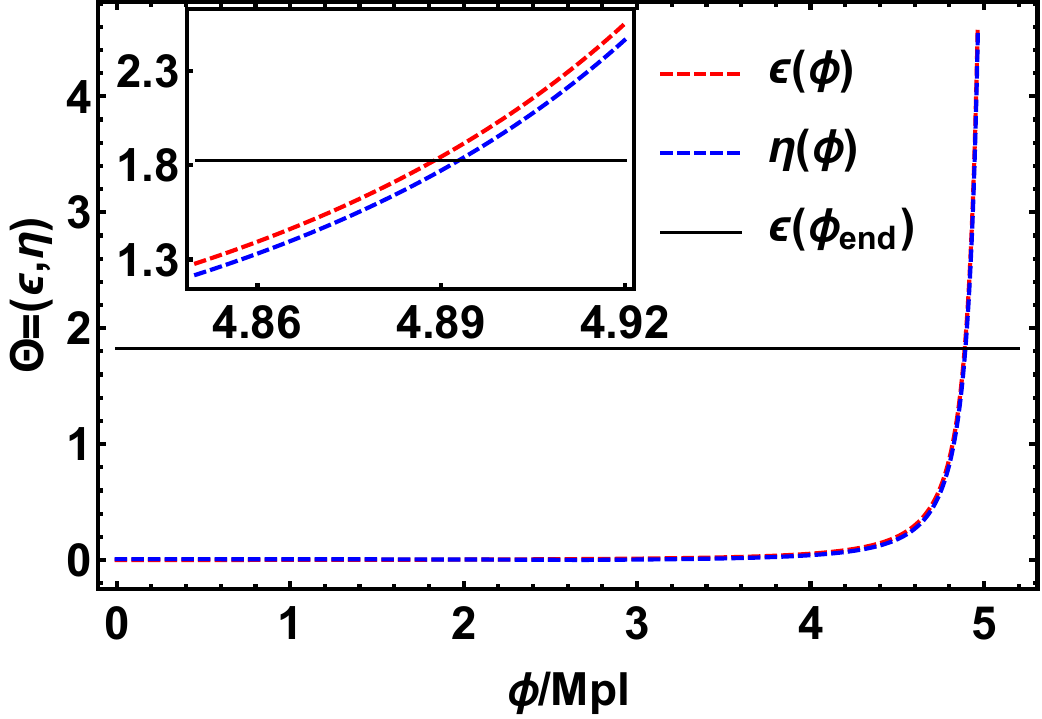}
  \hspace{0mm}
  \caption{Slow roll parameters for the potentials introduced in Sect. \ref{sec 4} in a minimally coupled scenario: \textbf{a)} $V_S(\phi)$, Eq. \eqref{star_like}, with $\phi_0=\sqrt{3/2}M_{Pl}\ln2$, $\chi=1.66$, $\alpha=\sqrt{2/3}/{M_{Pl}}$  and $V_0=-{\chi\phi_0^4}/{4}$; \textbf{b)} $V_\pi(\phi)$, Eq. \eqref{pi model}, with $\phi_0=4\sqrt{3/2}M_{Pl}\ln2$, $\chi=6.48\cdot10^{-3}$, $\alpha=\sqrt{2/3}/{M_{Pl}}$ and $V_0=0$; \textbf{c) ,d)} $V_W(\phi)$, Eq. \eqref{v2}, with respectively $\phi_0=6\sqrt{3/2}M_{Pl}\ln2,\>2\sqrt{3/2}M_{Pl}\ln2$, $\chi=1.28\cdot10^{-3},\>0.10$, $\alpha=\ln2/\phi_0^2$ and $V_0=0,\>-{\chi\phi_0^4}/{4}$; \textbf{e)} $V_\Omega(\phi)$, Eq. \eqref{v3}, with $\phi_0=6\sqrt{3/2}M_{Pl}\ln2$, $\chi=1.28\cdot10^{-3}$, $\alpha=\ln2/\phi_0^2$ and $V_0=0$.}
  \label{fig minimal sr}
\end{figure*}

Here, we examine the dynamic implications of our aforementioned four potentials. We explore both the cases of minimal and non-minimal coupling.

It is crucial to introduce a non-minimal coupling and analyze the resulting dynamics to determine whether these non-minimally coupled potentials can give rise to a suitable inflationary phase, particularly in the presence of strong gravitational fields. Particularly, recent studies have demonstrated particle creation when considering potentials arising from the interaction between the scalar field and the Ricci scalar \cite{friemanpart,cespedes}. Analogously, the interacting terms have also been utilized   to investigate the expected deviations in entanglement and particle production \cite{belfiglio2022,Belfiglio_2023}. Quite relevantly, such effective potentials have been employed to describe geometric quasi-particles \cite{belfiglio2023particle}, unifying the phenomenon of gravitational particle creation for both dark matter and baryons \cite{Ford:2021syk}.

\subsection{Conceptual choice of quasi-quintessence}

Following the idea that inflation occurs during the transition driven by the symmetry breaking mechanism, it is important to emphasize that the generalized kinetic term remains approximately constant both before and after the phase transition. As mentioned earlier, this choice has been made to address the classical cosmological constant problem.

While our model maintains self-consistency, it is crucial to distinguish between a constant term before and after the transition and the kinetic term that comes into play during inflation. During the transition, particularly when inflation takes place, the nature of the kinetic term is unspecified. Since the kinetic term is quite generic in this context, this would resemble the conventional quintessence approach rather than the notion of quasi-quintessence that applies during the transition. We  recall, in fact, that quintessence arises when the kinetic term is a generic varying function and includes quasi-quintessence as a specific case.

Hence, it becomes essential to highlight why we support a quasi-quintessence model even during the transition itself. To elucidate this decision, we can single out at least three key reasons.

Firstly, it is crucial to ensure consistency between the type of model both before and after the transition with the metastable phase.

Secondly, the field responsible for driving inflation behaves akin to a matter-like fluid with pressure. As a result, we assume that the sound speed remains zero throughout the entire evolution of the universe. Therefore, there is no necessity to complicate the scenario by introducing a varying kinetic term during the phase transition.

Last but not least,  the candidate field carries vacuum energy, which results in a large overall energy for the field constituent responsible for cancelling the cosmological constant. Previous research indicates that the corresponding mass is sufficiently large to keep the kinetic term approximately frozen, \emph{i.e.}, constant \cite{Luongo_2018}. Therefore, there is no reason to expect the kinetic term to cease being constant as third working hypothesis in favor of quasi-quintessence during the transition. So, it is likely to remain constant and ensure that the quasi-quintessence hypothesis holds even during the transition.

\subsection{Particle production and end of inflation}

The above argument is closely tied to the hypothesis of slow roll. In the conventional inflationary framework, a slow roll phase is postulated to explain the required strong acceleration. In this scenario, since vacuum energy is cancelled after the transition, then it is strictly necessary that its magnitude is converted into particles during the metastable phase, \emph{i.e.}, during the inflationary slow roll phase.

This prerogative is intimately different from the standard picture of inflation, where particles are mainly created in the reheating phase.

Hence, this property appears as a crucial difference that can \emph{falsify or support} our model from the standard puzzle. Nevertheless, as the potential diminishes due to subsequent particle production and the overall energy decreases, there will eventually be a point in time where the kinetic energy is no longer subdominant compared with the potential, tending in the end to dominate.

The equivalence point marks the end of inflation and leads to the recovery of the standard picture described by the conventional inflationary paradigm. Indeed, as the kinetic energy overcomes the remaining fraction of the potential, reheating can occur, proceeding into a phase dominated by the matter-like fluid and radiation. There, baryons can be produced, through the oscillations around the minimum \cite{Allahverdi:2010xz}, \emph{i.e.}, in a different manner than particles produced during inflation.

Importantly, the main byproduct of our approach addresses the fine-tuning problem that arises when attempting to eliminate the cosmological constant. Regarding coincidence, the final fluid entering the reheating phase is constituted by the matter-like quasi-quintessence component that acts like matter. Then, its magnitude is comparable to baryons, while the produced particles during inflation also contribute as matter field. Consequently, the coincidence problem is fulfilled, since there is no reason to invoke two very different magnitudes for the quasi-quintessence final fluid and the remaining potential at the end of inflation.

Within this picture, it should be noted that we do not claim that the kinetic term exactly cancels out the potential, but rather that there exists a time at which the two quantities become comparable. At that time, the particle production becomes subdominant and the remaining contribution induced by the fraction of potential, that did not have enough time to convert into particles, contributes to the \emph{bare cosmological constant}, emerging as dominant at current time, where matter dilutes because of the cosmic expansion. Additional details of the overall mechanism are reported in Ref. \cite{belfiglio2022}. The whole treatment considers that at late times the universe is therefore dominated by a bare cosmological constant.

The subsequent analyses delve into these subjects into minimal and non-minimal couplings and describes the settings and properties of this matter-lik fluid during the transition.


\subsection{Minimally coupled quasi-quintessence inflation}\label{subsec 51}

\begin{figure*}[p]
  \centering
  \textbf{a)}\hspace{0mm}
    \includegraphics[width=.465\linewidth]{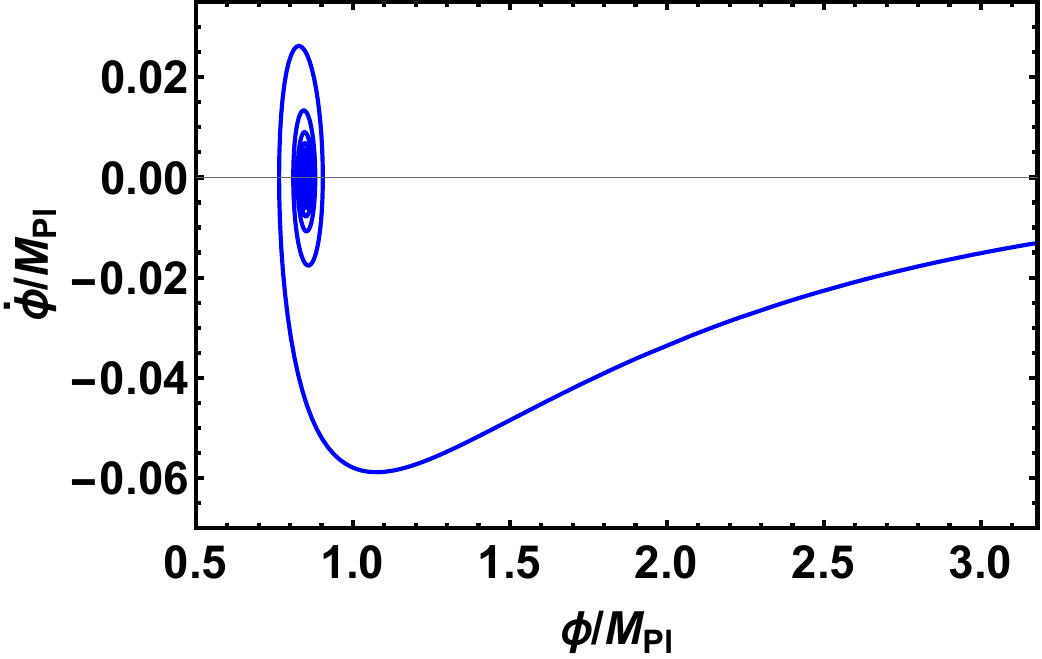}
  \hspace{1mm}
  \textbf{b)}\hspace{0mm}
    \includegraphics[width=.465\linewidth]{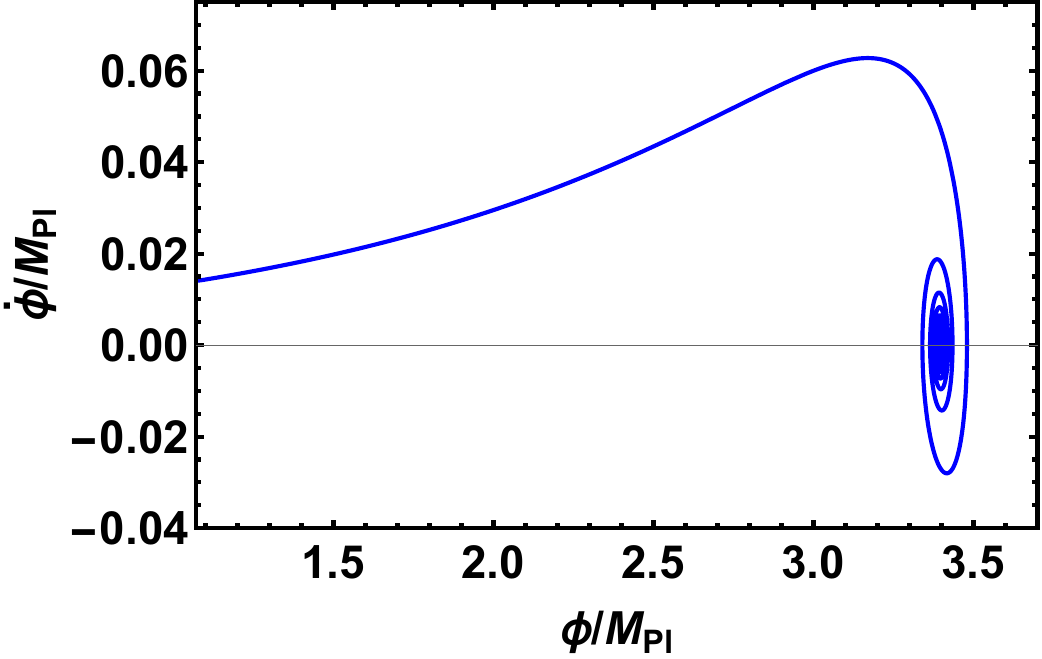}\\
    \vspace{7mm}
  \textbf{c)}
    \includegraphics[width=.475\linewidth]{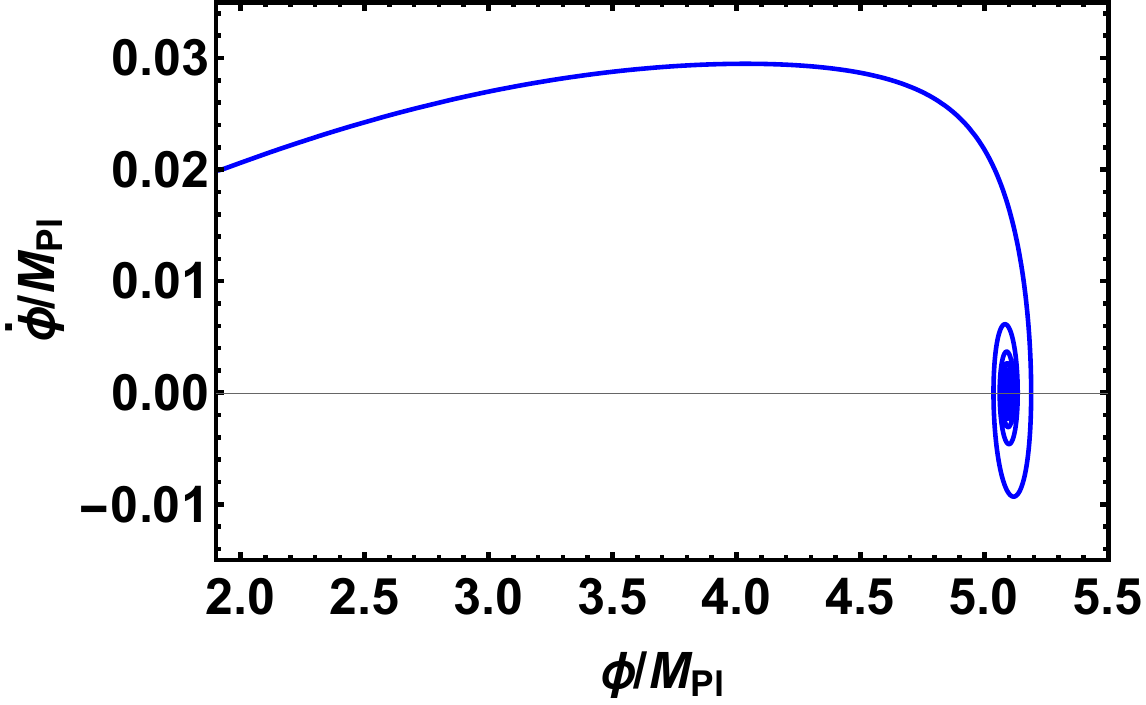}
  \hspace{0mm}
  \textbf{d)}
    \includegraphics[width=.465\linewidth]{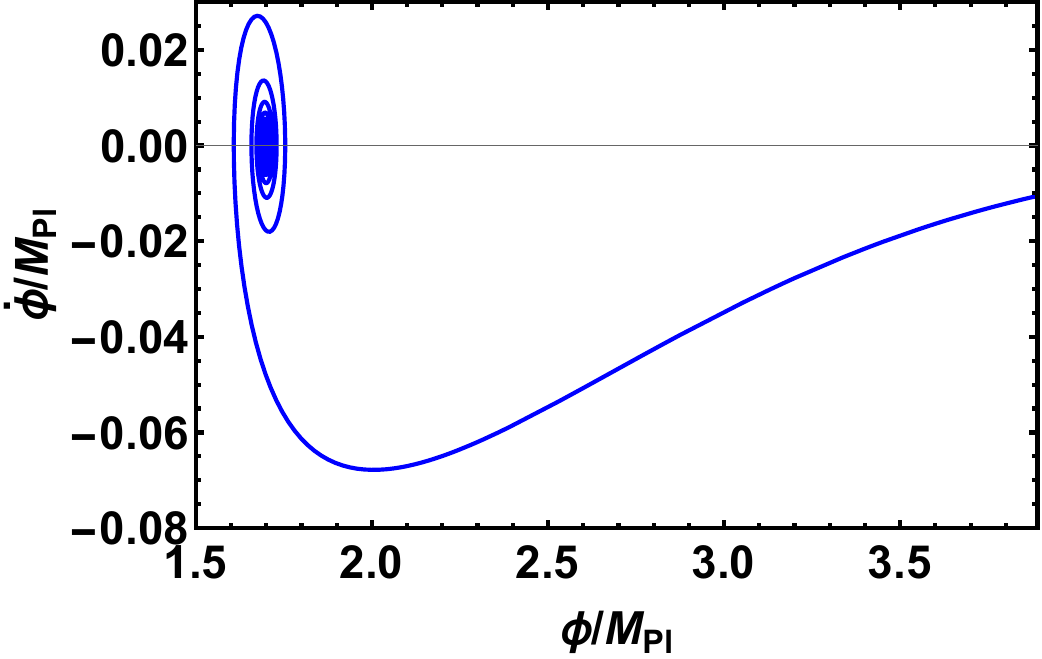}\\
    \vspace{7mm}
  \textbf{e)}
    \includegraphics[width=.47\linewidth]{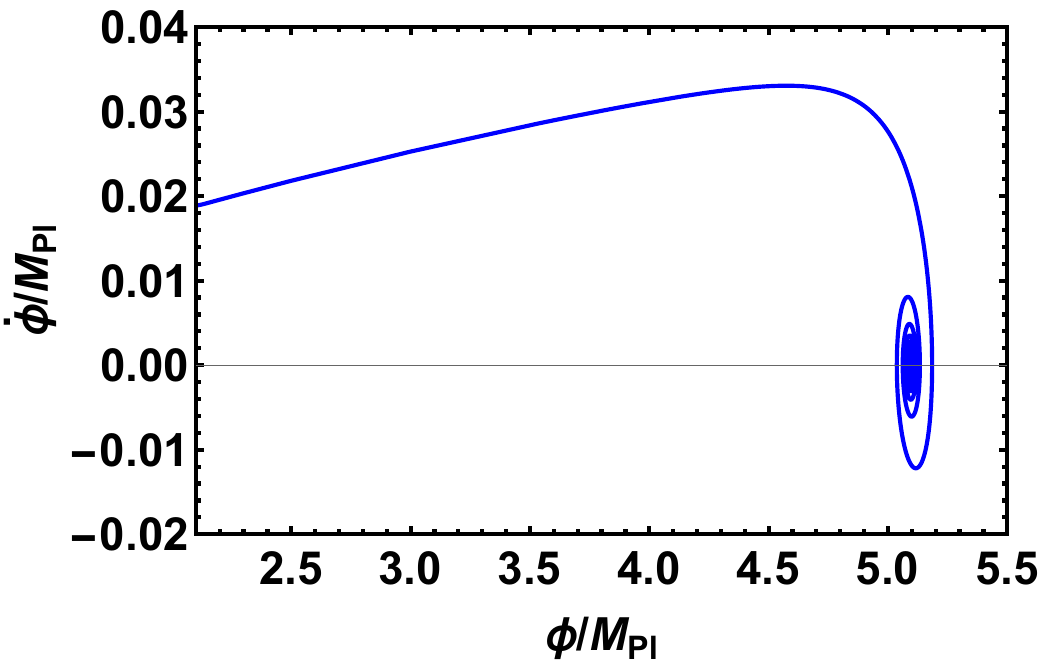}
  \hspace{0mm}
  \caption{Inflaton phase space for the potentials introduced in Sect. \ref{sec 4} in a minimally coupled scenario: \textbf{a)} $V_S(\phi)$, Eq. \eqref{star_like}, with $\phi_0=\sqrt{3/2}M_{Pl}\ln2$, $\chi=1.66$, $\alpha=\sqrt{2/3}/{M_{Pl}}$  and $V_0=-{\chi\phi_0^4}/{4}$; \textbf{b)} $V_\pi(\phi)$, Eq. \eqref{pi model}, with $\phi_0=4\sqrt{3/2}M_{Pl}\ln2$, $\chi=6.48\cdot10^{-3}$, $\alpha=\sqrt{2/3}/{M_{Pl}}$ and $V_0=0$; \textbf{c), d)} $V_W(\phi)$, Eq. \eqref{v2}, with respectively $\phi_0=6\sqrt{3/2}M_{Pl}\ln2,\>2\sqrt{3/2}M_{Pl}\ln2$, $\chi=1.28\cdot10^{-3},\>0.10$, $\alpha=\ln2/\phi_0^2$ and $V_0=0,\>-{\chi\phi_0^4}/{4}$; \textbf{e)} $V_\Omega(\phi)$, Eq. \eqref{v3}, with $\phi_0=6\sqrt{3/2}M_{Pl}\ln2$, $\chi=1.28\cdot10^{-3}$, $\alpha=\ln2/\phi_0^2$ and $V_0=0$.}
  \label{fig minimal dyn}
\end{figure*}

Let us analyze the inflationary dynamics in the minimally coupled scenario, \emph{i.e.} $\xi=0$. In this respect we individuate the parameters that describe the inflationary dynamics, \emph{i.e.} the slow roll parameters, \eqref{eps}, \eqref{eta}, the e-folding number, Eq. \eqref{efold slowroll}, initial and final conditions, and finally, the evolution in the phase space of the inflaton field. We emphasize that, while generally a potential is either large or small field, our $W$ potential, Eq. \eqref{v2}, shows both the small and large behavior due to the finite potential walls. Further, the choice of free parameters is the same as reported in Sect. \ref{sec 4}.

First, we compute the slow roll parameters, Eqs. \eqref{eps}-\eqref{eta}, as shown in Figs. \ref{fig minimal sr}.
In this way we identify the slow roll phase and we get the final value of the inflaton field, \emph{i.e.}, where Eq. \eqref{infl end} holds.
Then imposing that the e-folding number, Eq. \eqref{efold slowroll}, satisfies the constraint $N\gtrsim70$, we find the initial condition for the field $\phi$.

So, from  Eq. \eqref{srdin}, we can compute the initial condition on the time variation of the field, say
\begin{equation}
    \dot \phi_{in}\simeq-\frac{\mathcal V^\prime(\phi_{in})}{3H_{in}}.
\end{equation}
The corresponding computed initial and final conditions are summarized in Tab. \ref{tab inend}.

\begin{table}[h]
  \centering
  \begin{tabular}{c|c|c|c}
    \hline\hline
    Potentials & $\phi_{in}$ & $\phi_{end}$ & $\dot\phi_{in}$ \\
    \hline
    \hline
    $V_S$ & $3.18M_{Pl}$ & $1.03M_{Pl}$ & $-0.01M_{Pl}$ \\
    $V_\pi$ & $1.07M_{Pl}$ & $3.21M_{Pl}$ & $0.01M_{Pl}$ \\
    $V_W^{small}$ & $1.91M_{Pl}$ & $4.88M_{Pl}$ & $0.02M_{Pl}$ \\
    $V_W^{large}$ & $3.89M_{Pl}$ & $1.90M_{Pl}$ & $-0.01M_{Pl}$ \\
    $V_\Omega$ & $2.10M_{Pl}$ & $4.89M_{Pl}$ & $0.02M_{Pl}$ \\
    \hline\hline
  \end{tabular}
  \caption{Initial and final conditions of the minimally coupled inflationary stage.}
  \label{tab inend}
\end{table}

Finally, solving the equation of motion of the inflationary universe, Eq. \eqref{infl dyn}, we obtain the phase space evolution as shown in Figs. \ref{fig minimal dyn}.
The field naturally goes towards the minimum of the potential, and it presents a chaotic evolution.
\begin{figure*}[p]
  \centering
  \textbf{a)}\hspace{0mm}
    \includegraphics[width=.4\linewidth]{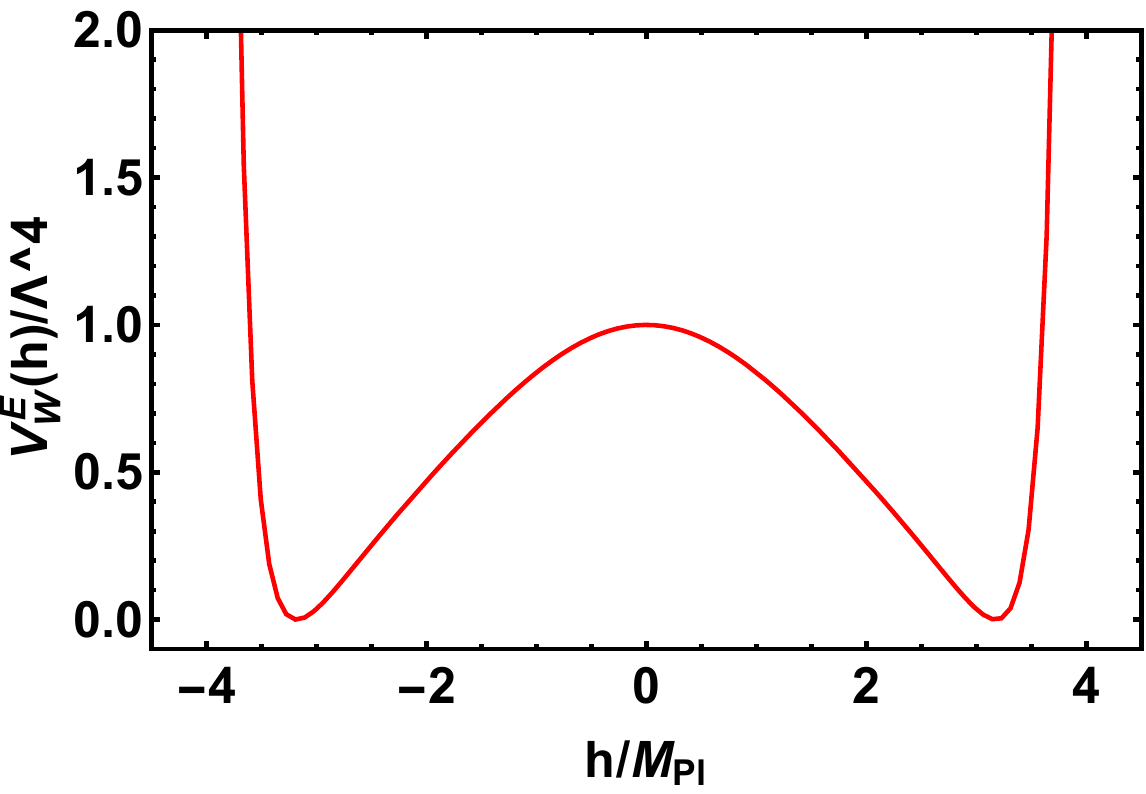}
  \hspace{0mm}
  \textbf{b)}\hspace{2mm}
    \includegraphics[width=.39\linewidth]{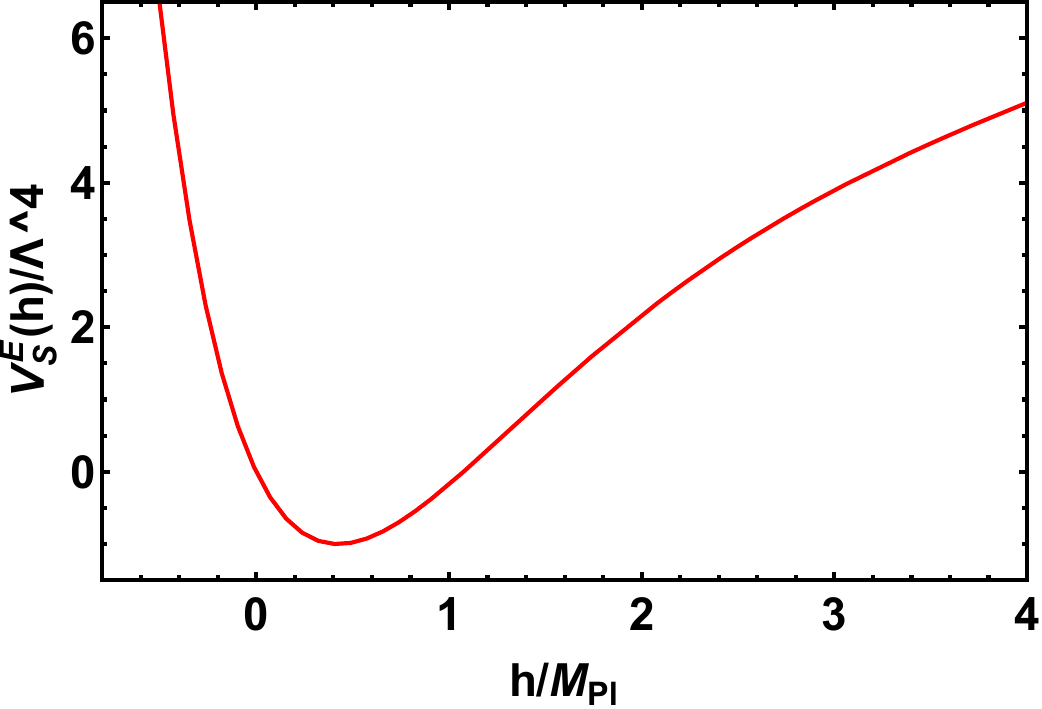}\\
    \vspace{2mm}
  \textbf{c)}
    \includegraphics[width=.4\linewidth]{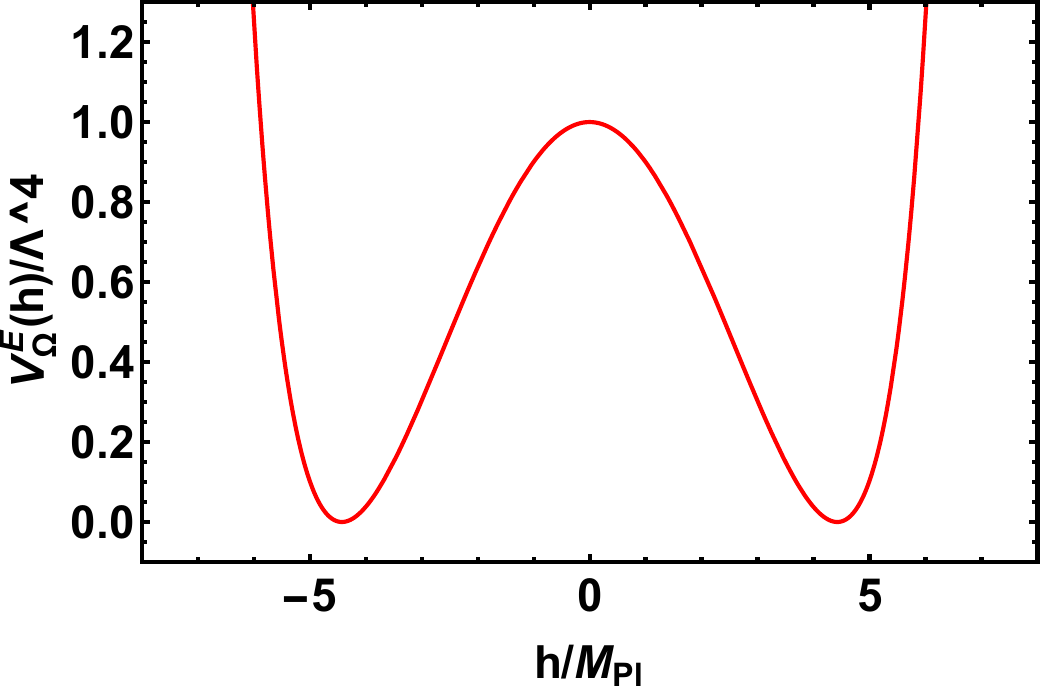}
  \hspace{0mm}
  \textbf{d)}
    \includegraphics[width=.41\linewidth]{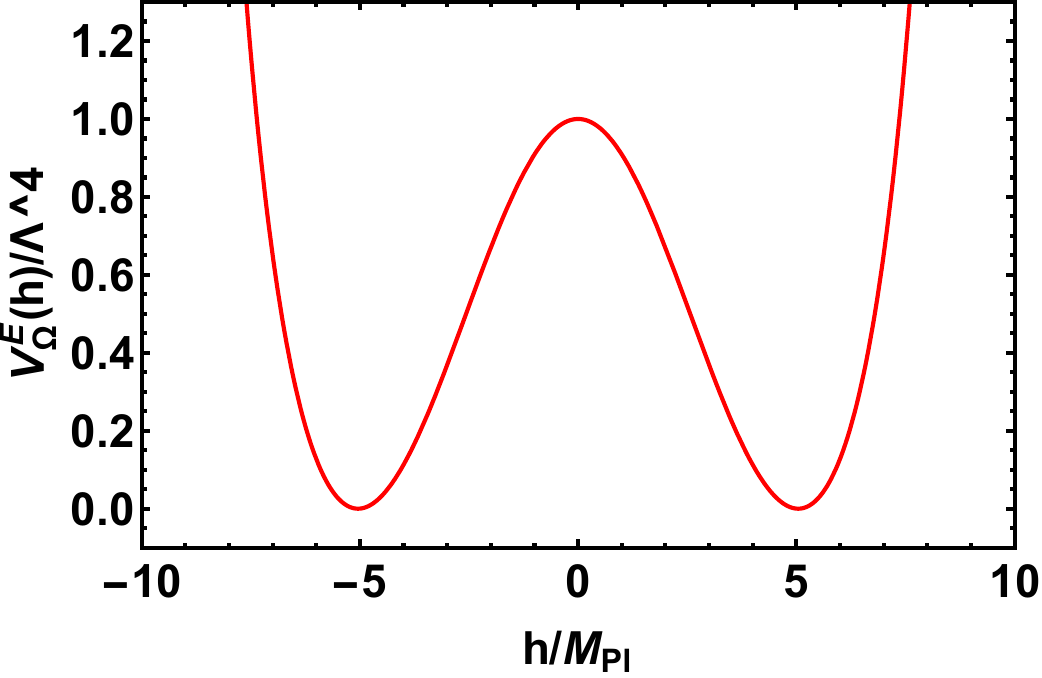}\\
    \vspace{2mm}
  \textbf{e)}
    \includegraphics[width=.4\linewidth]{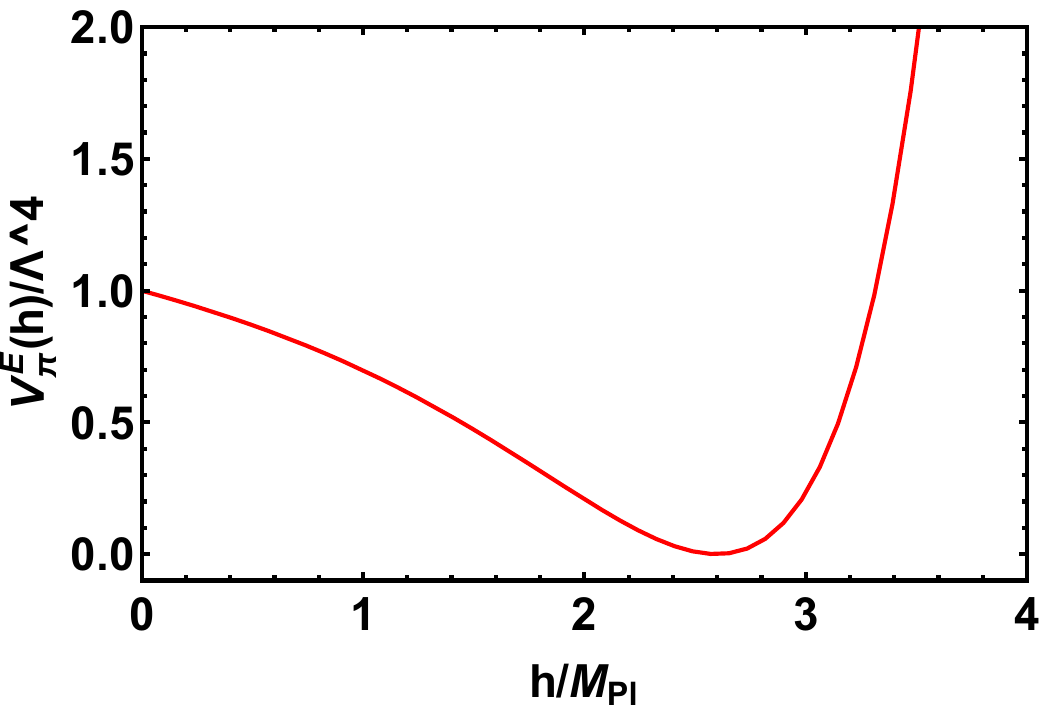}
  \hspace{0mm}
    \textbf{f)}
    \includegraphics[width=.4\linewidth]{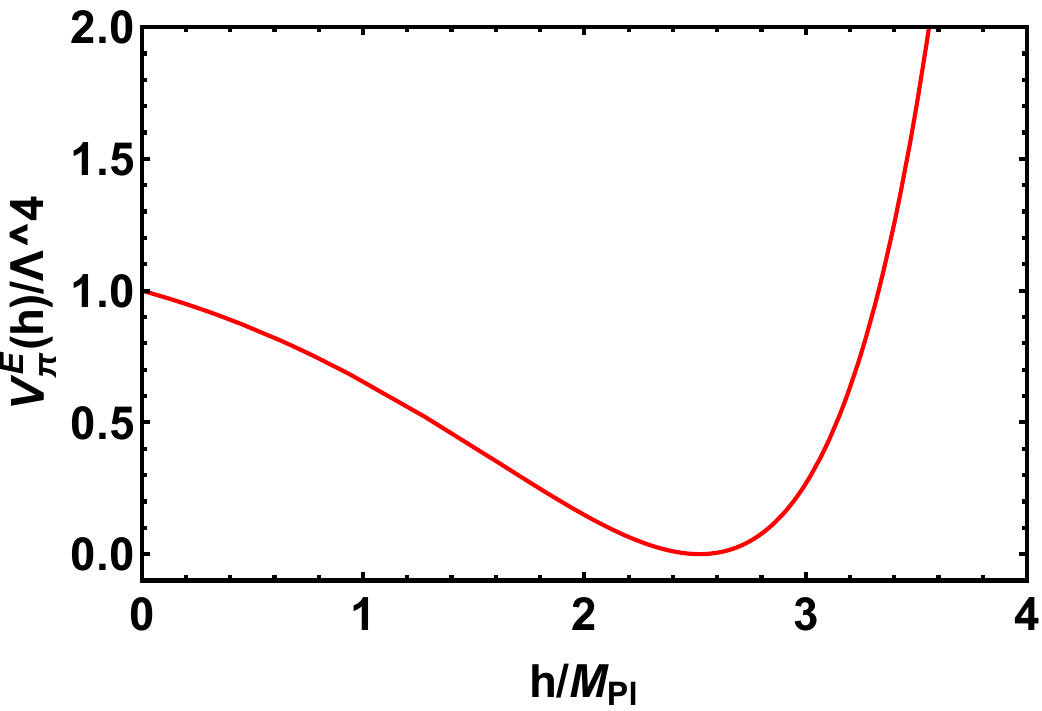}\\
    \vspace{2mm}
  \textbf{g)}
    \includegraphics[width=.42\linewidth]{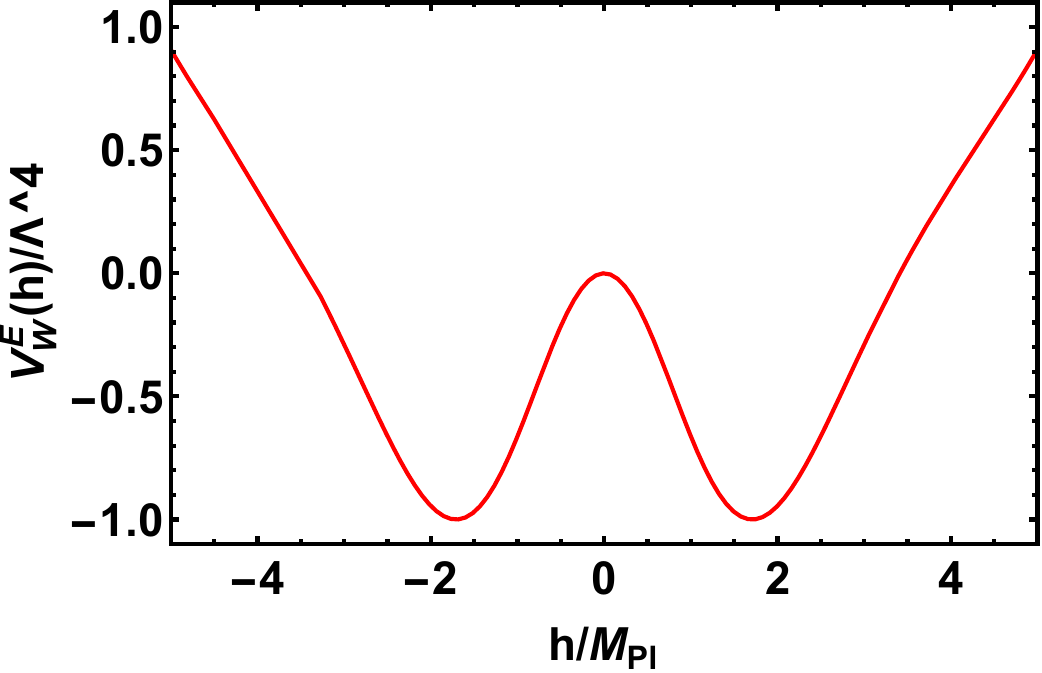}
  \caption{Inflationary potentials introduced in Sect. \ref{sec 4} in a non-minimally coupled scenario: \textbf{a), g)} $V_W(\phi)$, Eq. \eqref{v2}, respectively in a small field and large field regime; \textbf{b)} $V_S(\phi)$, Eq. \eqref{star_like}; \textbf{c), d)} $V_\Omega(\phi)$, Eq. \eqref{v3}, respectively with a positive and negative coupling strength; \textbf{f), g)} $V_\pi(\phi)$, Eq. \eqref{pi model}, respectively with a positive and negative coupling strength. The choice of parameters is shown in Tab. \ref{tab nonmin}. The plots refer to as the potentials $V$ without considering the kinetic term, namely $\mathcal V=V-K_0$.}
  \label{fig nonminimal pot}
\end{figure*}

\begin{figure*}[p]
  \centering
  \textbf{a)}\hspace{0mm}
    \includegraphics[width=.4\linewidth]{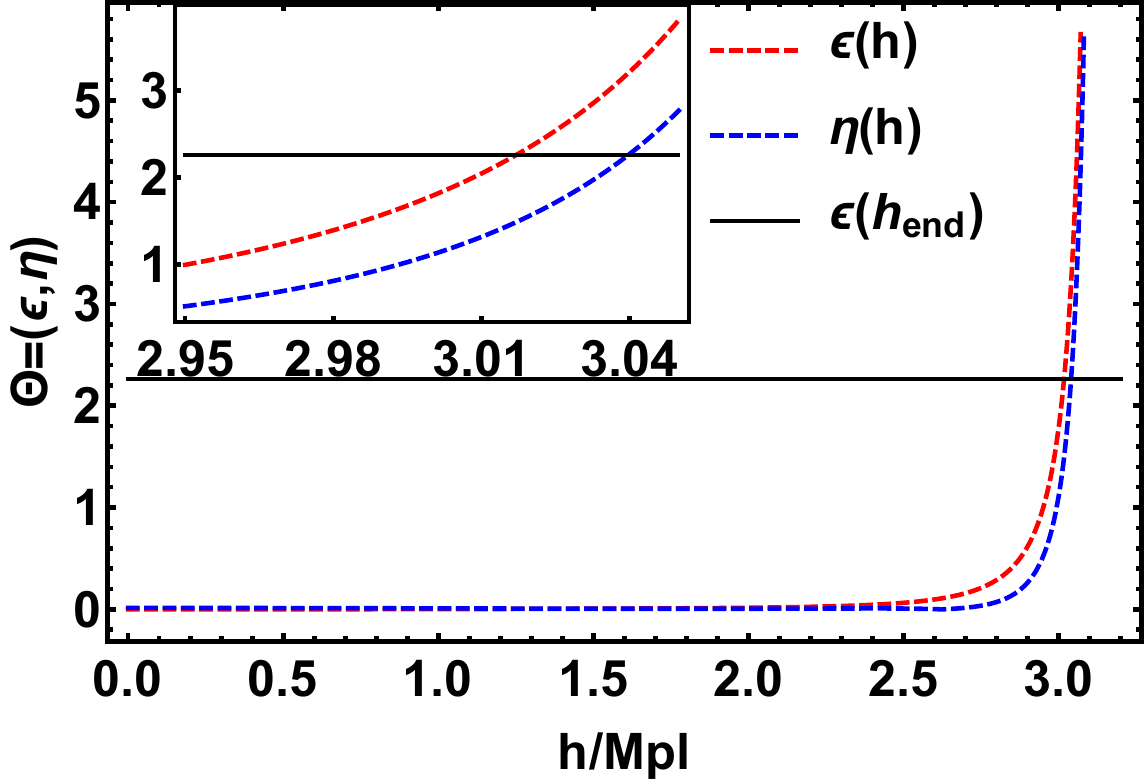}
  \hspace{0mm}
  \textbf{b)}\hspace{0mm}
    \includegraphics[width=.4\linewidth]{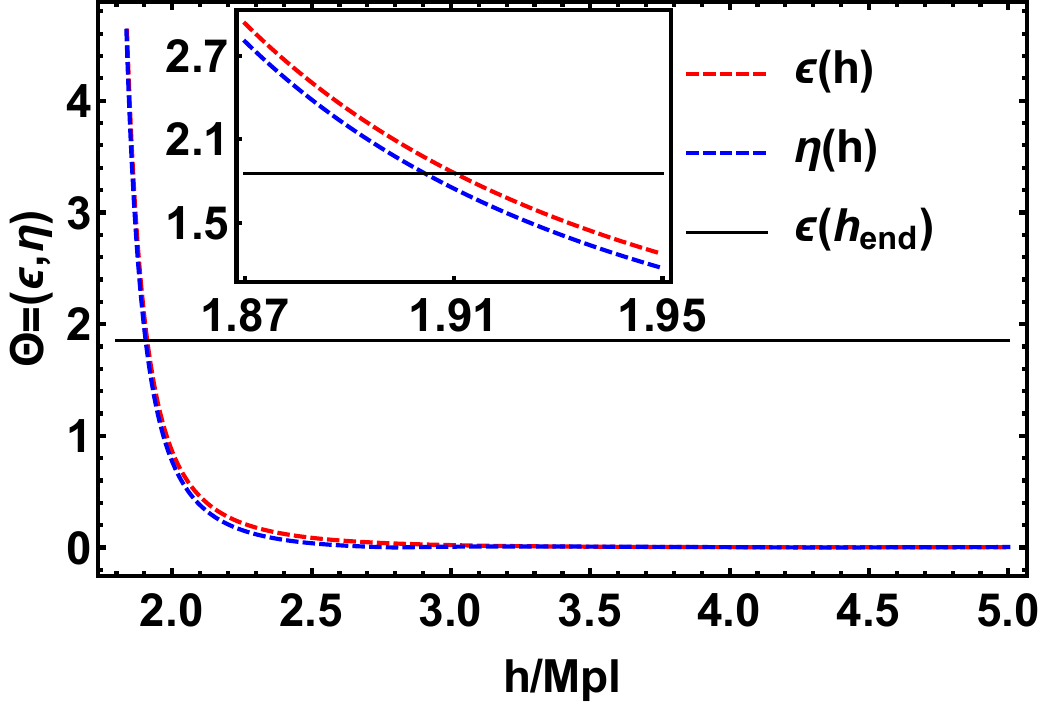}\\
    \vspace{2mm}
  \textbf{c)}
    \includegraphics[width=.4\linewidth]{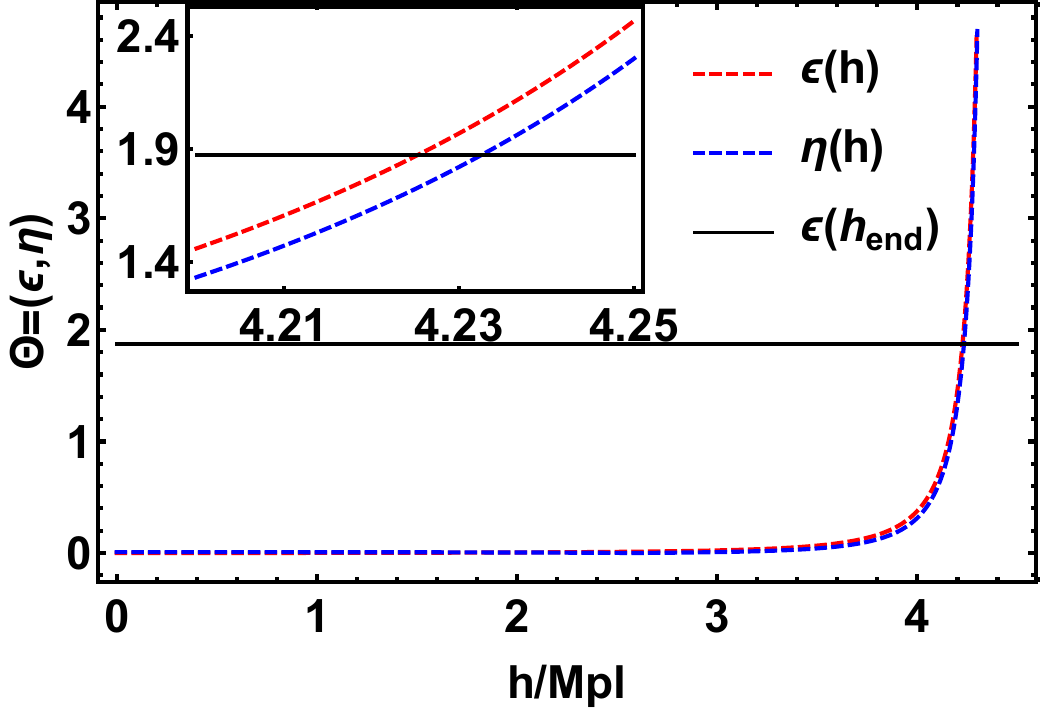}
  \hspace{0mm}
  \textbf{d)}
    \includegraphics[width=.4\linewidth]{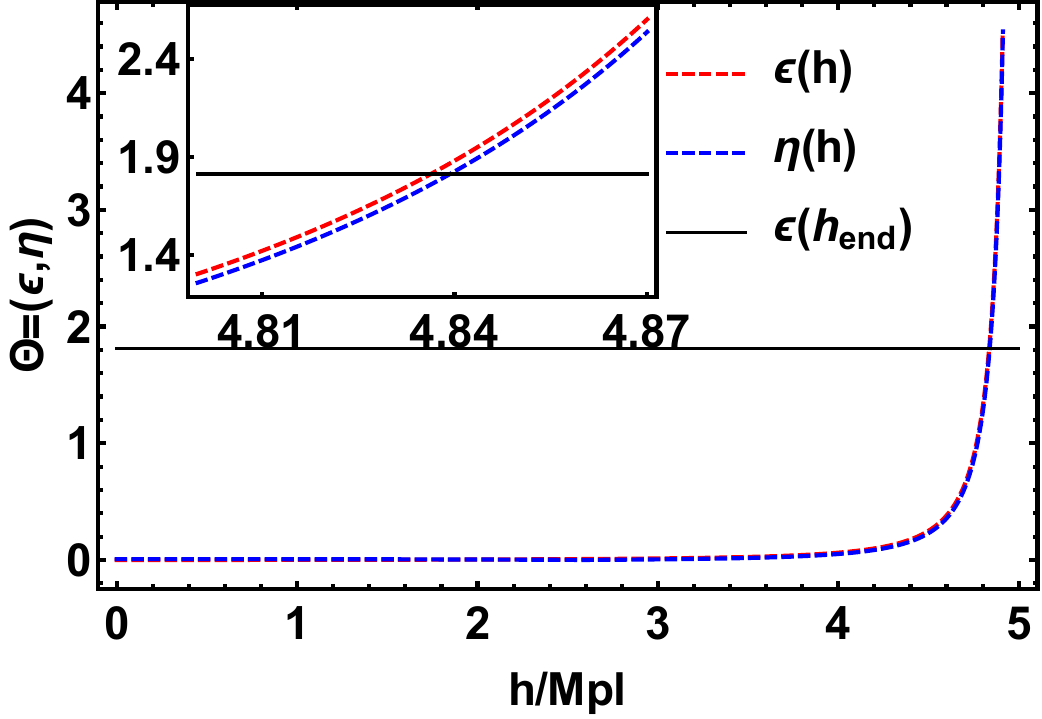}\\
    \vspace{2mm}
  \textbf{e)}
    \includegraphics[width=.4\linewidth]{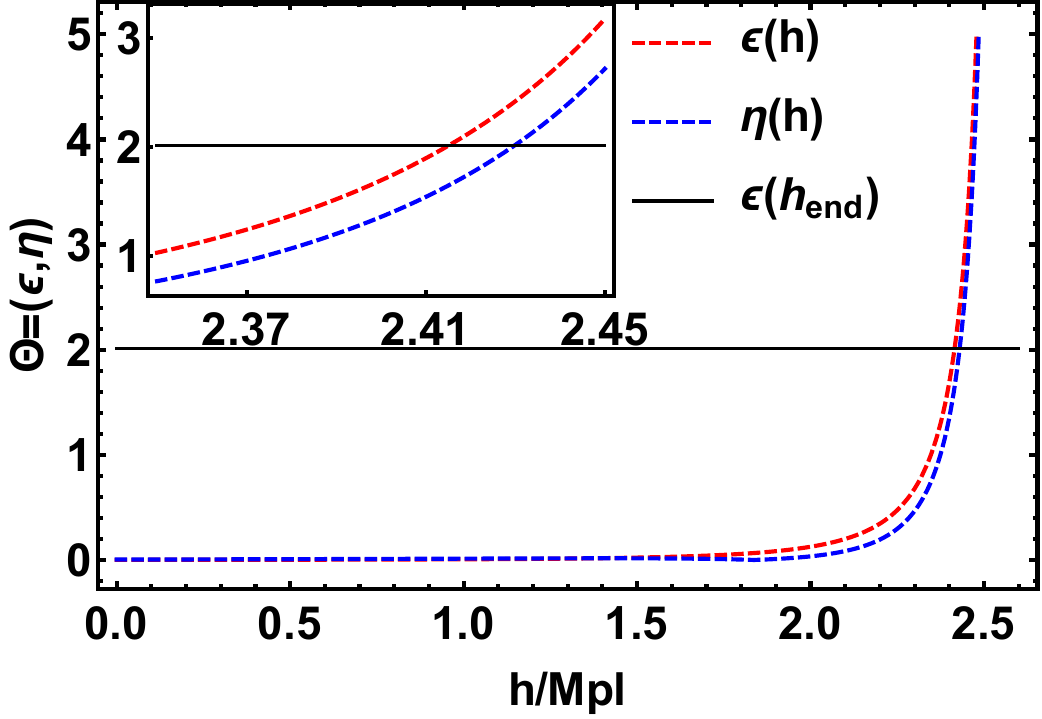}
  \hspace{0mm}
    \textbf{f)}
    \includegraphics[width=.4\linewidth]{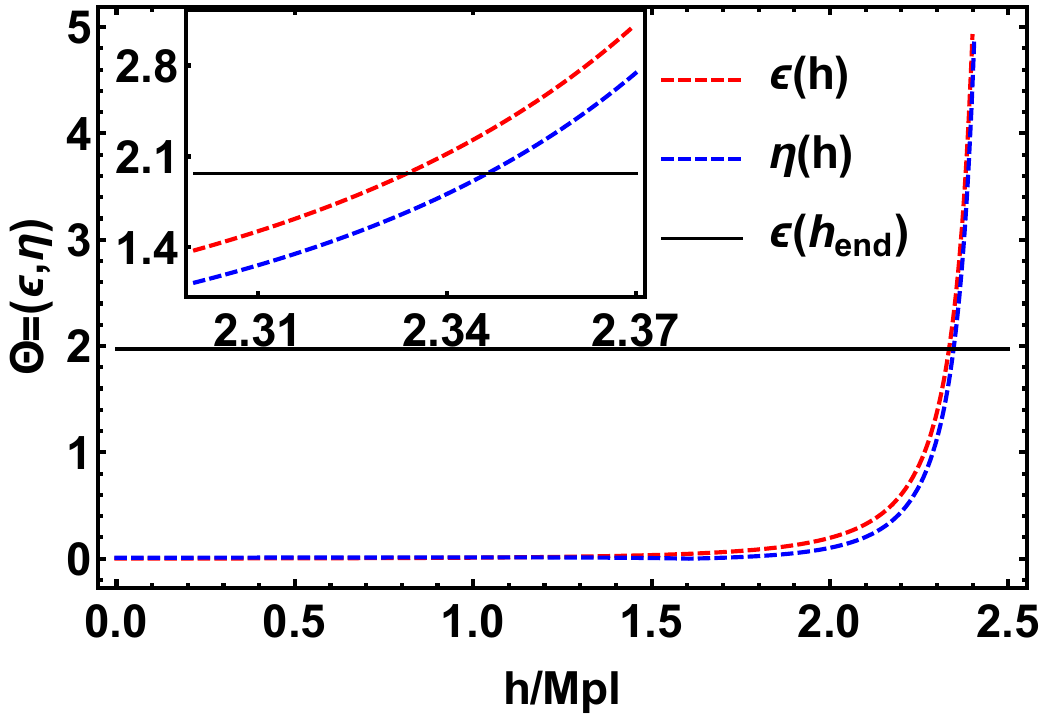}\\
    \vspace{2mm}
  \textbf{g)}
    \includegraphics[width=.4\linewidth]{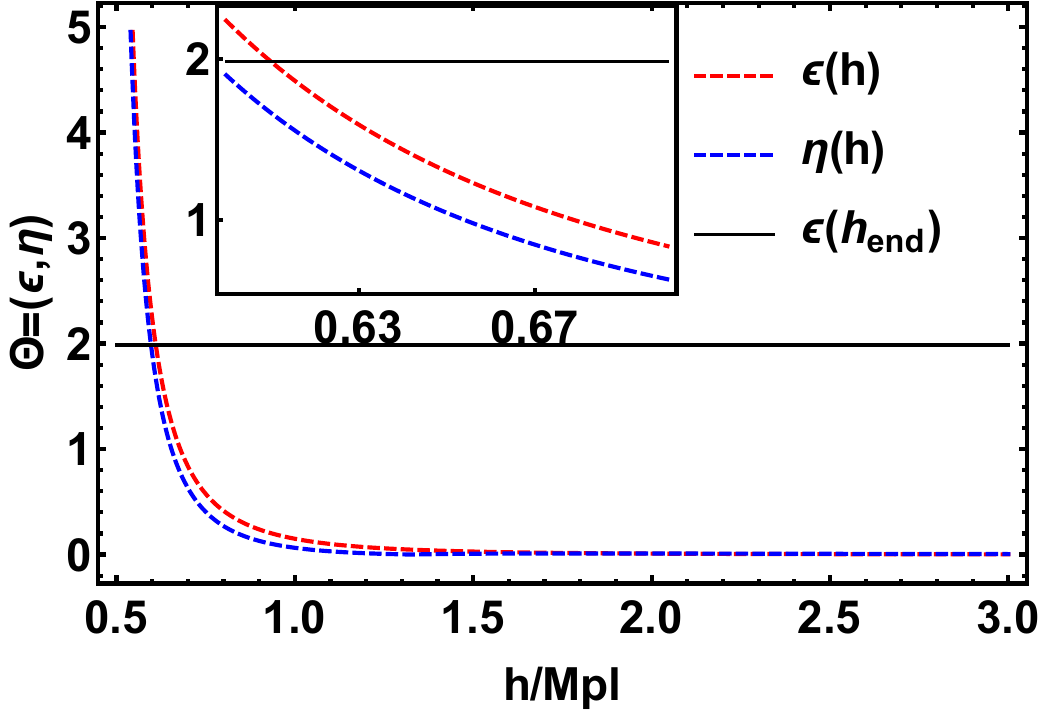}
  \caption{Slow roll parameters for the potentials introduced in Sect. \ref{sec 4} in a non-minimally coupled scenario: \textbf{a), b)} $V_W(\phi)$, Eq. \eqref{v2}, respectively in a small field and large field regime; \textbf{c), d)} $V_\Omega(\phi)$, Eq. \eqref{v3}, respectively with a positive and negative coupling strength; \textbf{e), f)} $V_\pi(\phi)$, Eq. \eqref{pi model}, respectively with a positive and negative coupling strength; \textbf{g)} $V_S(\phi)$, Eq. \eqref{star_like}; The choice of parameters is shown in Tab. \ref{tab nonmin}.}
  \label{fig nonminimal etaeps}
\end{figure*}

\begin{figure*}[p]
  \centering
  \textbf{a)}\hspace{0mm}
    \includegraphics[width=.4\linewidth]{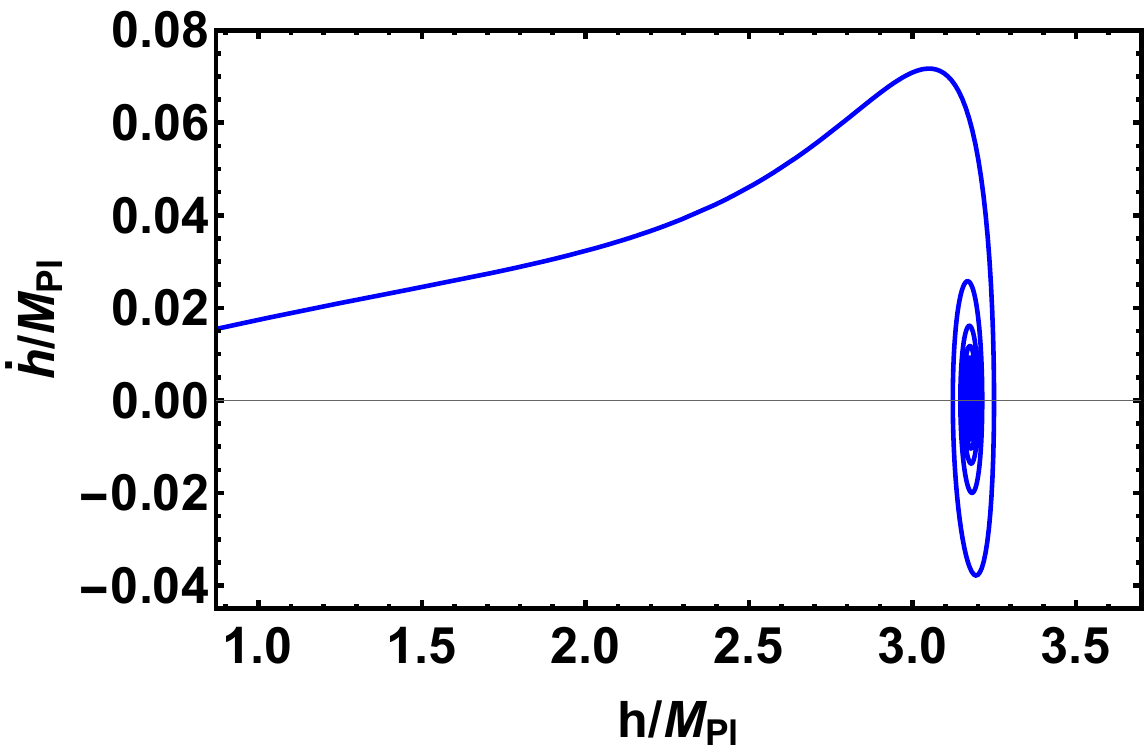}
  \hspace{0mm}
  \textbf{b)}\hspace{0mm}
    \includegraphics[width=.4\linewidth]{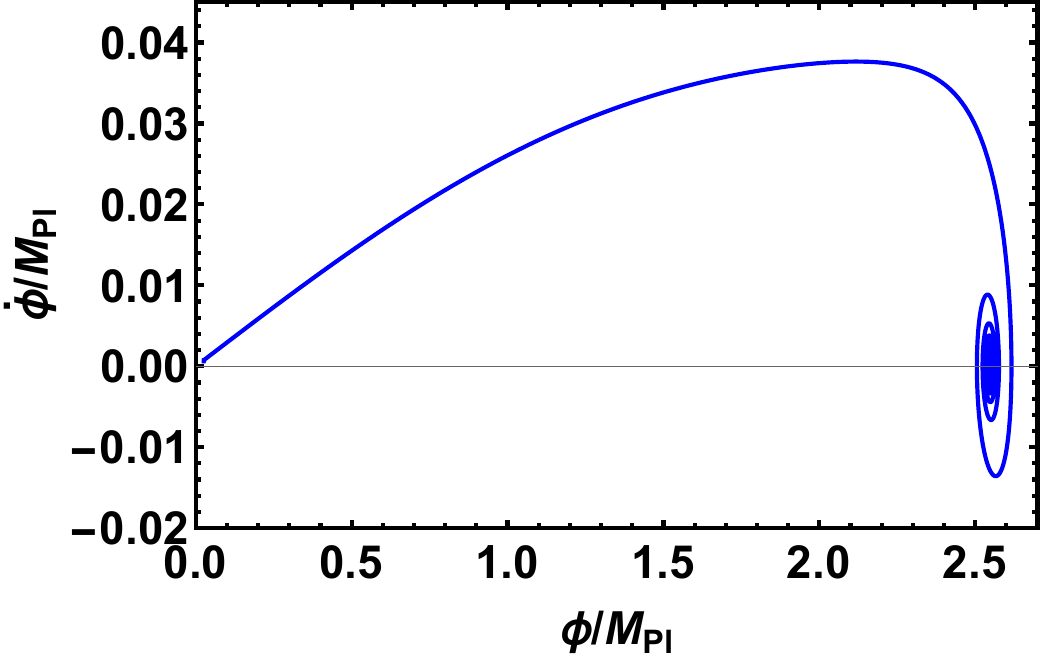}\\
    \vspace{2mm}
  \textbf{c)}
    \includegraphics[width=.4\linewidth]{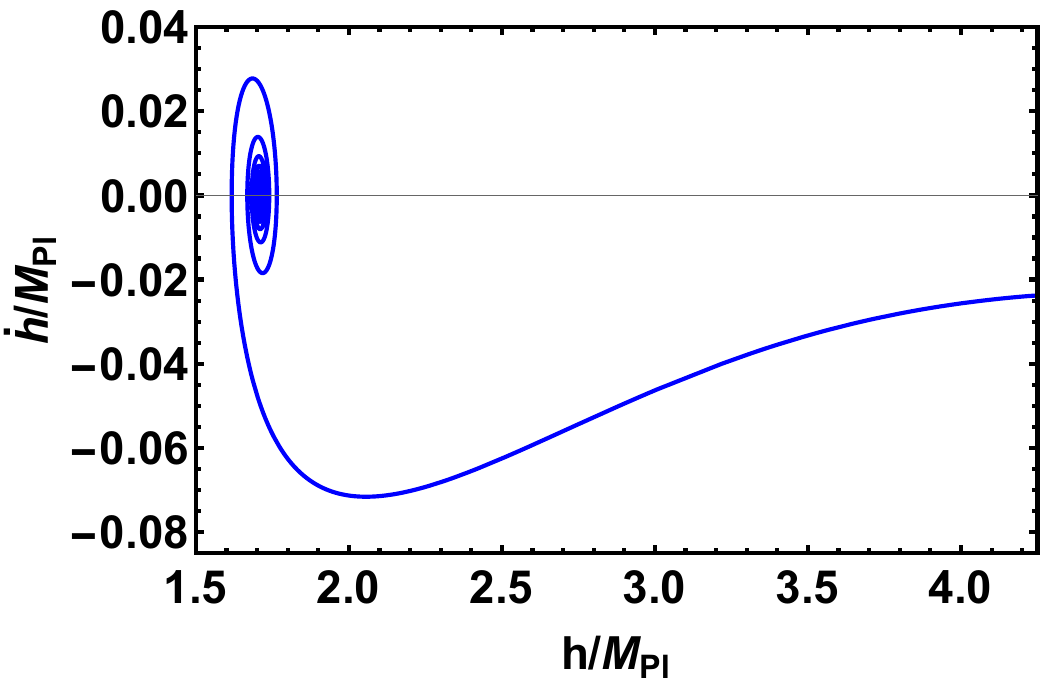}
  \hspace{0mm}
  \textbf{d)}
    \includegraphics[width=.4\linewidth]{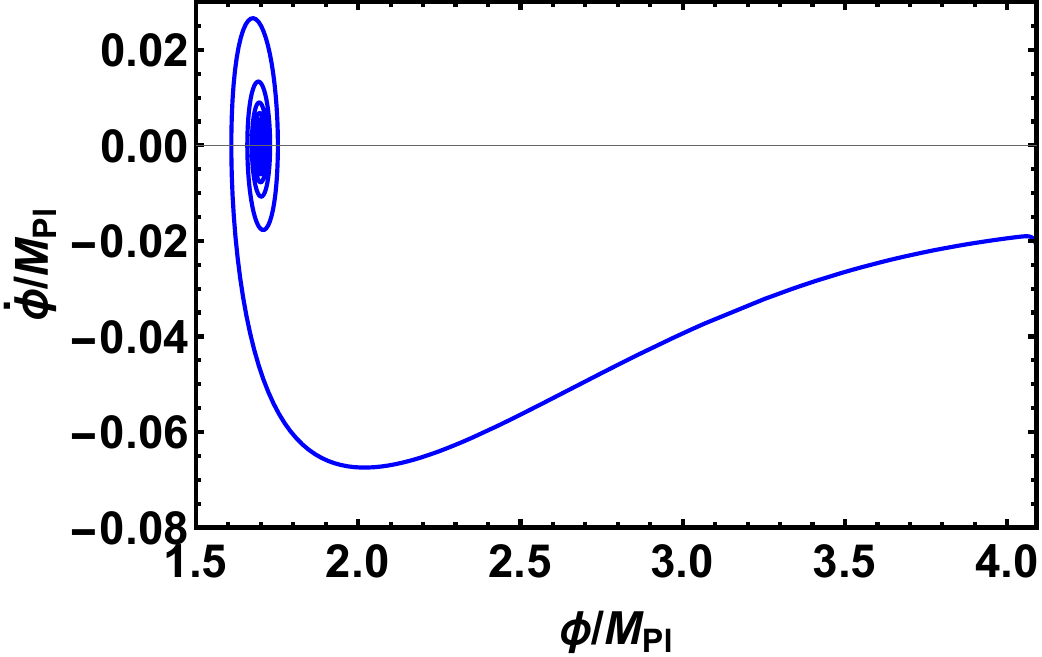}\\
    \vspace{2mm}
  \textbf{e)}\hspace{1mm}
    \includegraphics[width=.4\linewidth]{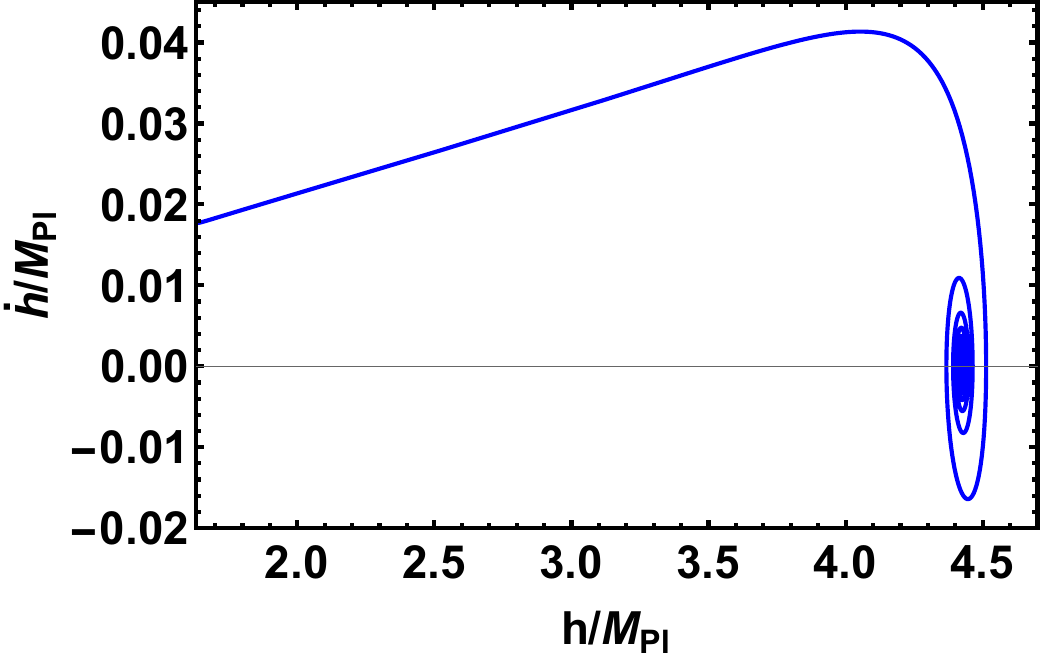}
  \hspace{0mm}
    \textbf{f)}\hspace{1mm}
    \includegraphics[width=.41\linewidth]{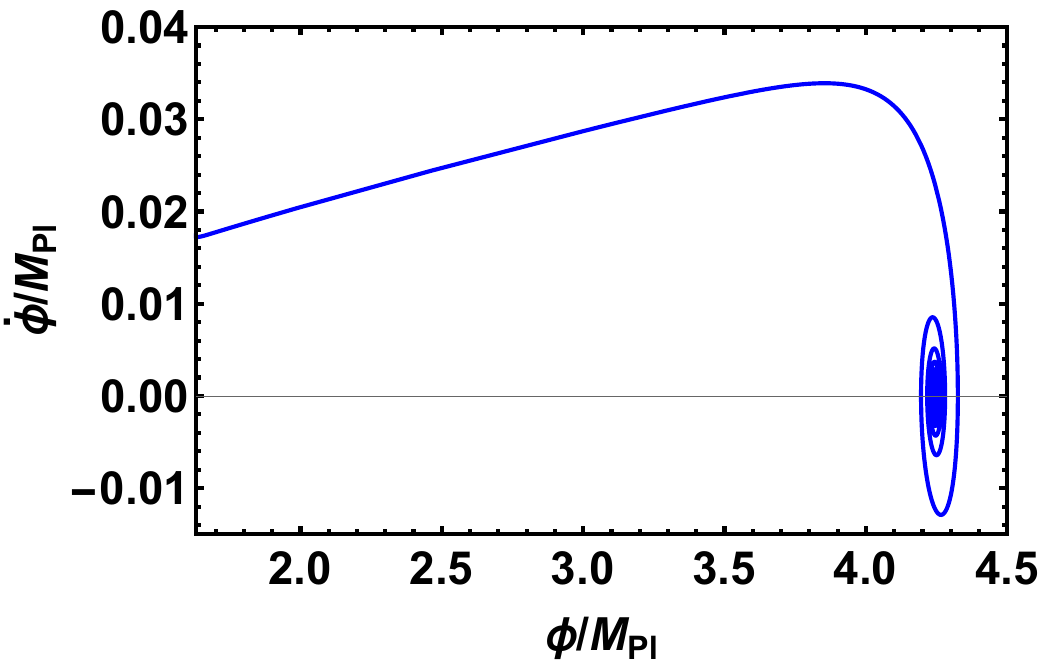}\\
    \vspace{2mm}
  \textbf{g)}\hspace{.5mm}
    \includegraphics[width=.41\linewidth]{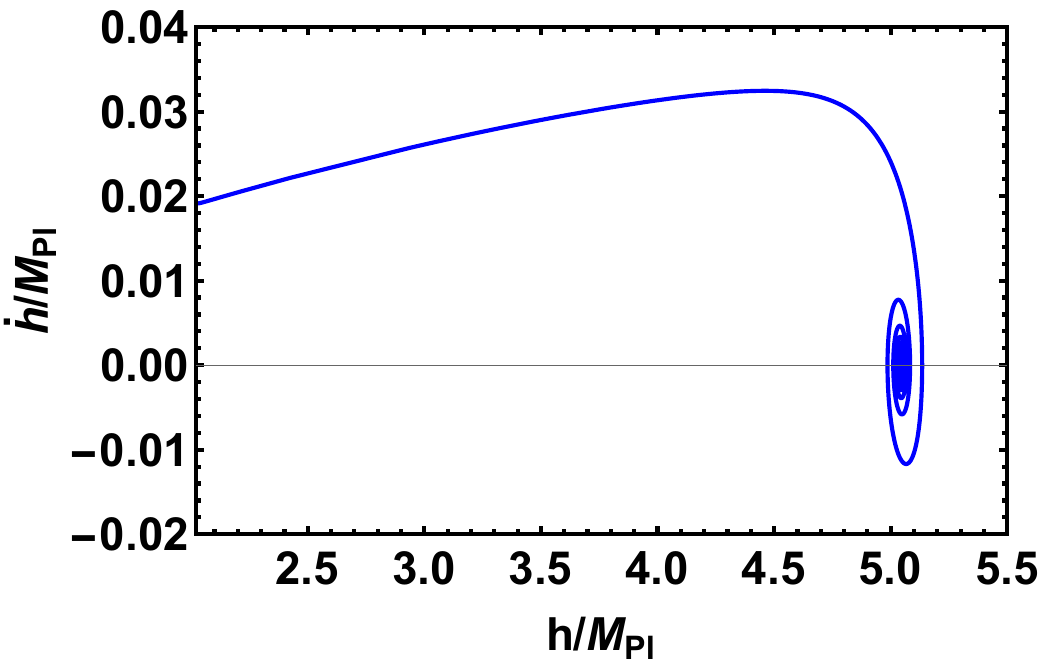}  \hspace{0mm}
    \textbf{h)}
    \includegraphics[width=.405\linewidth]{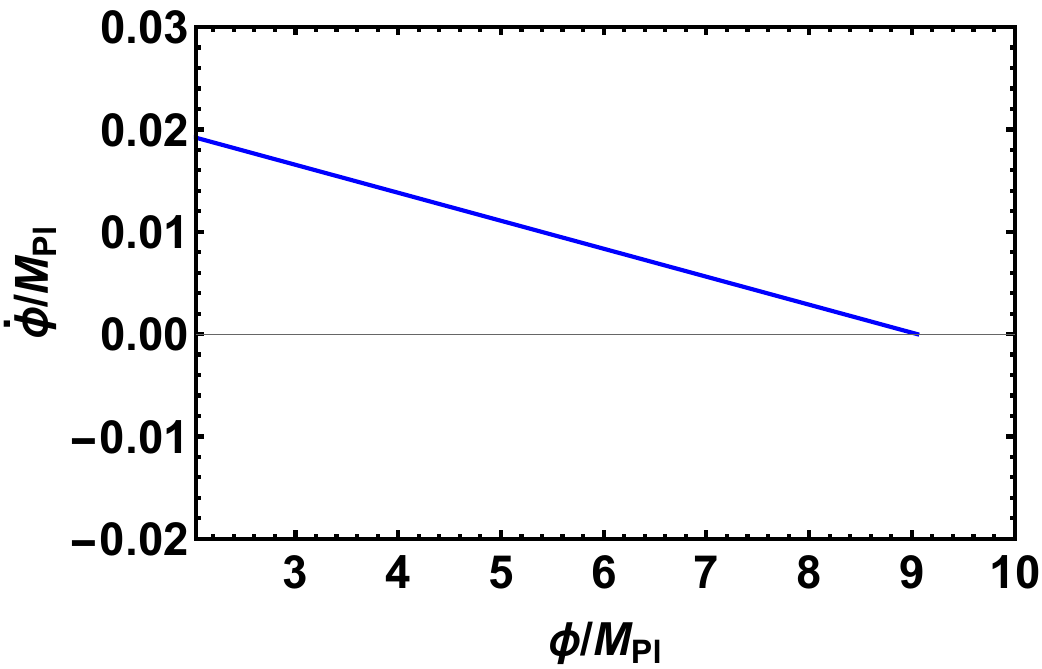}
  \caption{Inflation phase space of the potentials introduced in Sect. \ref{sec 4} in a non-minimally coupled scenario: on the left in the Einstein frame, while on the right in the Jordan frame. \textbf{a), b)} $V_W(\phi)$, Eq. \eqref{v2}, in a small field regime; \textbf{c), d)} $V_W(\phi)$, Eq. \eqref{v2}, in a large field regime; \textbf{e), f)} $V_\Omega(\phi)$, Eq. \eqref{v3}, with a positive coupling strength; \textbf{g), h)} $V_\Omega(\phi)$, Eq. \eqref{v3}, with a negative coupling strength. The choice of parameters is shown in Tab. \ref{tab nonmin}.}
  \label{fig nonminimal sym dyn}
\end{figure*}

\begin{figure*}[p]
  \centering
  \textbf{a)}\hspace{0mm}
    \includegraphics[width=.445\linewidth]{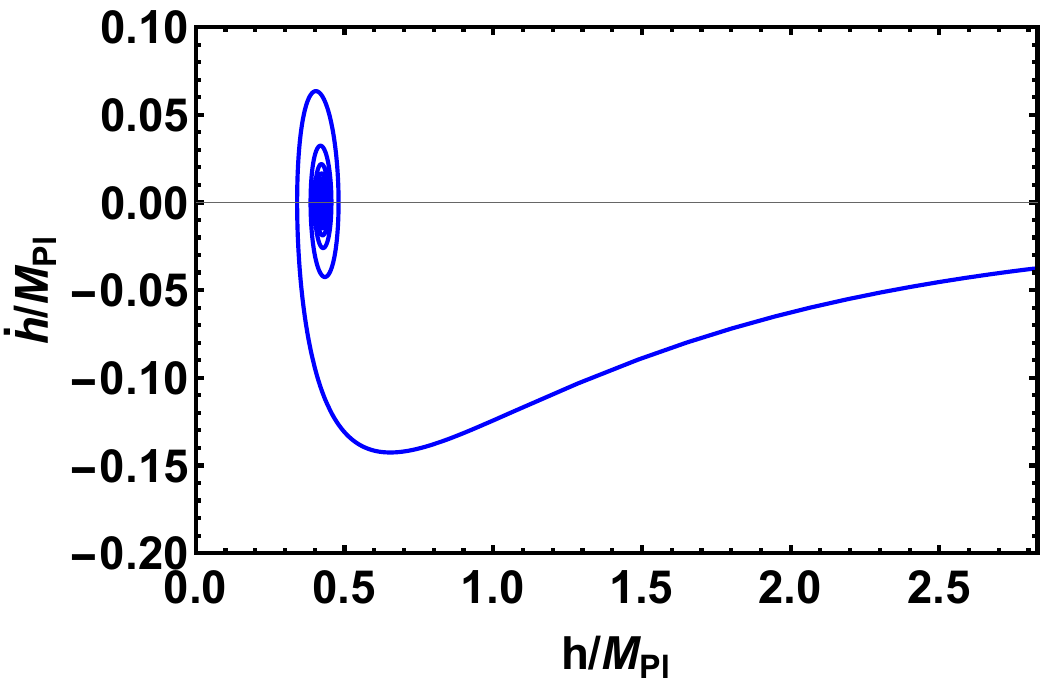}
  \hspace{0mm}
  \textbf{b)}\hspace{1mm}
    \includegraphics[width=.45\linewidth]{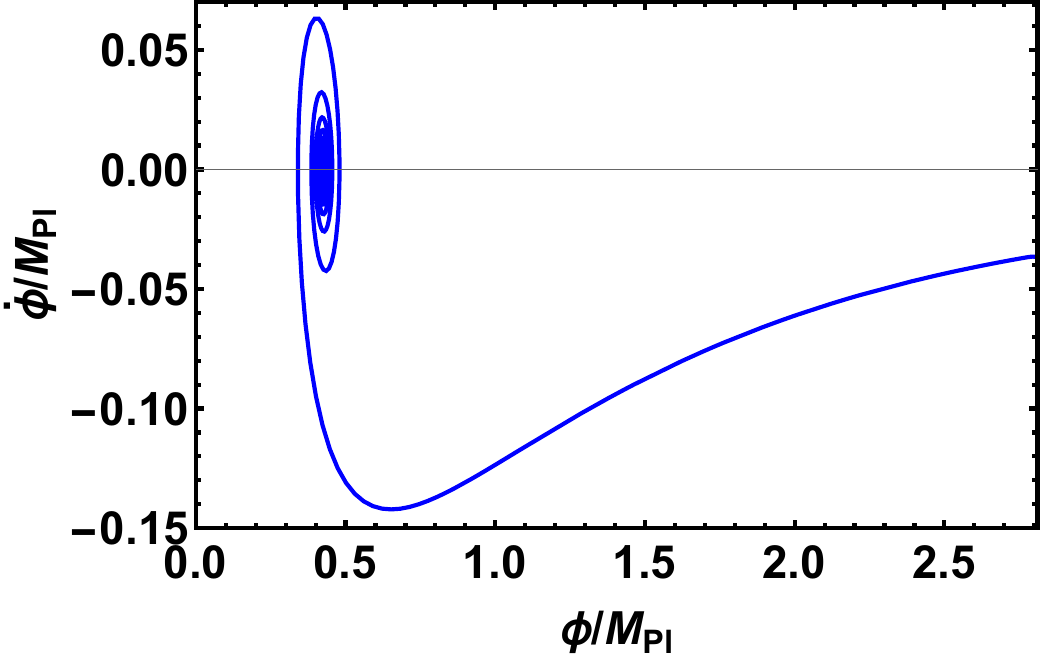}\\
    \vspace{7mm}
  \textbf{c)}\hspace{2mm}
    \includegraphics[width=.465\linewidth]{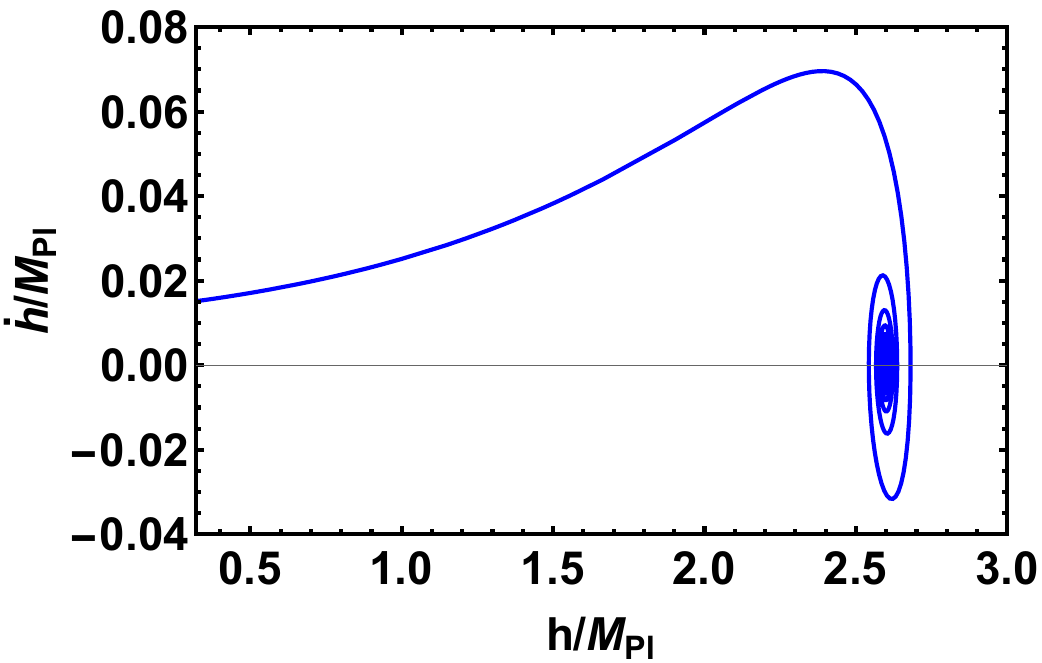}
  \hspace{0mm}
  \textbf{d)}\hspace{0.mm}
    \includegraphics[width=.465\linewidth]{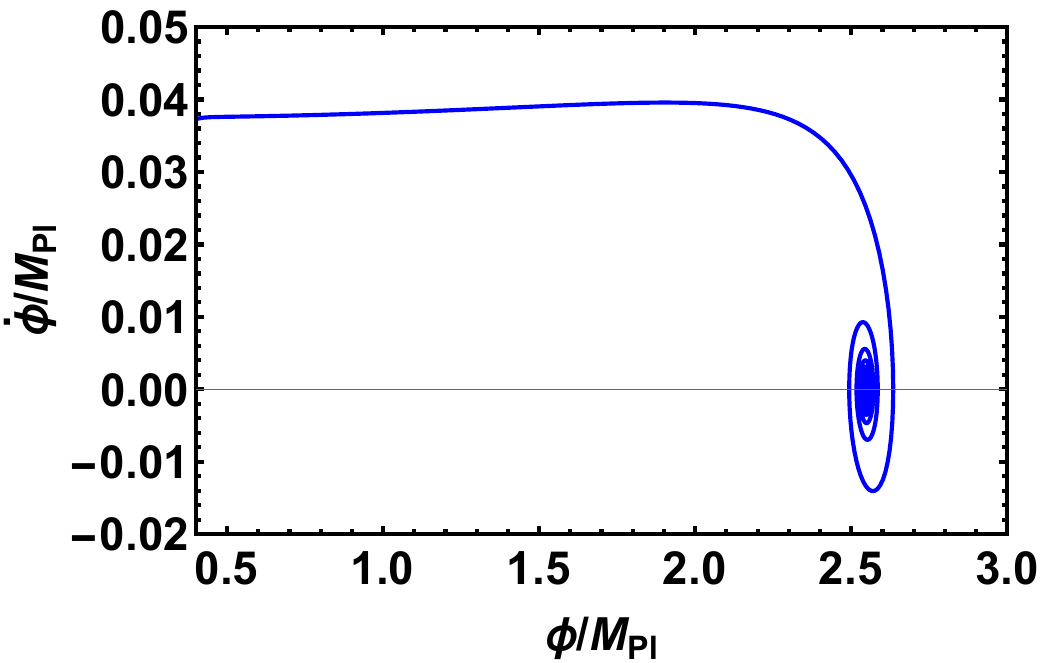}\\
    \vspace{7mm}
  \textbf{e)}\hspace{1mm}
    \includegraphics[width=.465\linewidth]{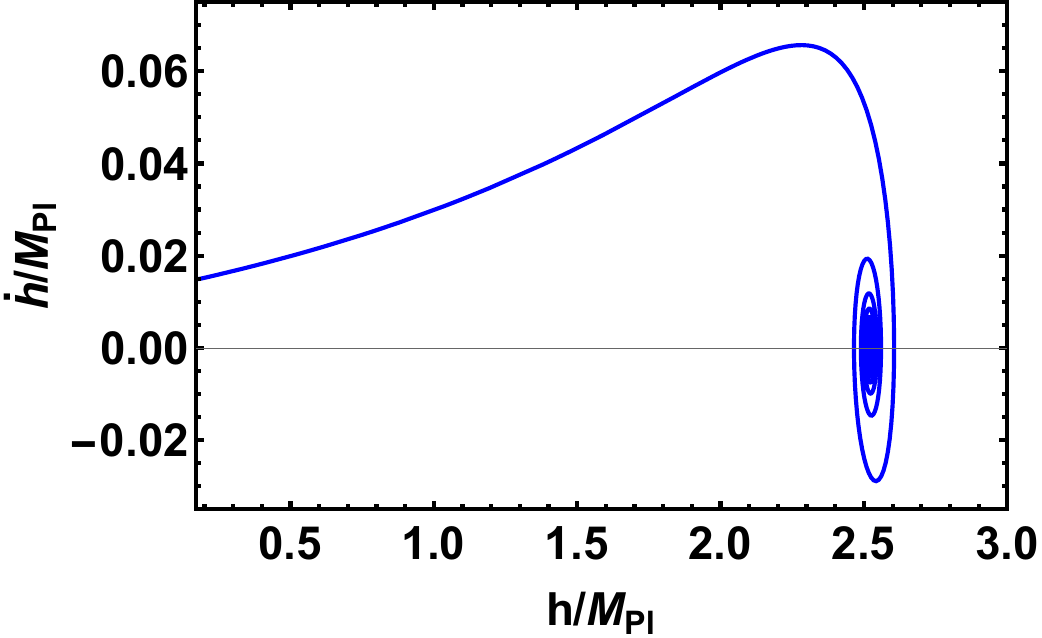}
  \hspace{0mm}
    \textbf{f)}\hspace{0.mm}
    \includegraphics[width=.45\linewidth]{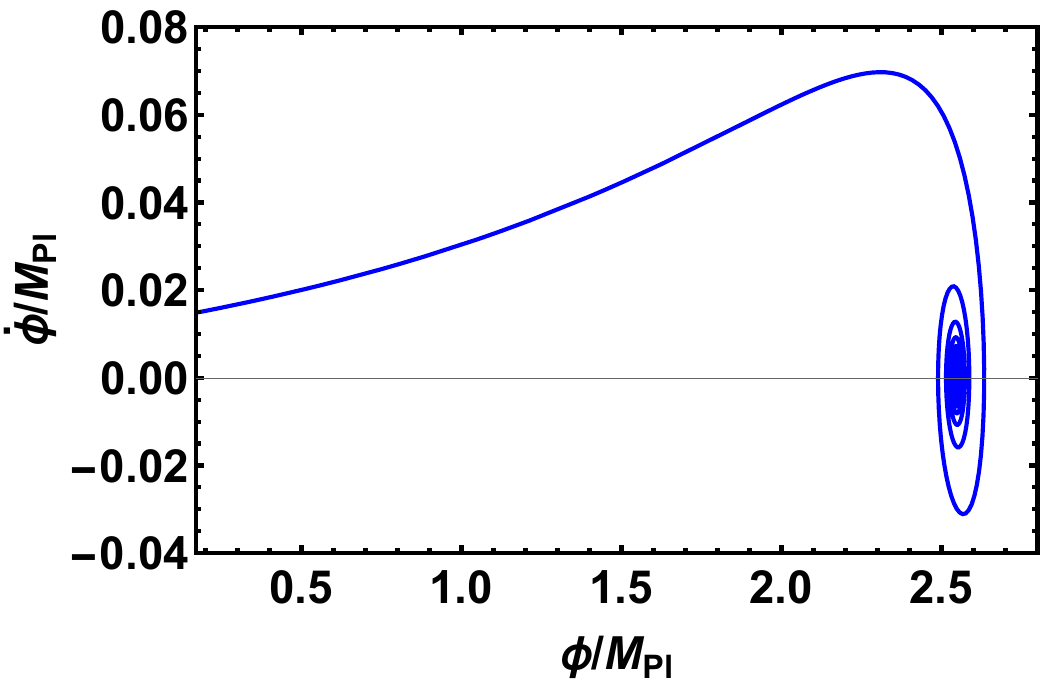}
  \caption{Inflation phase space of the potentials introduced in Sect. \ref{sec 4} in a non-minimally coupled scenario: on the left in the Einstein frame, while on the right in the Jordan frame. \textbf{a), b)} $V_S(\phi)$, Eq. \eqref{star_like}; \textbf{c), d)} $V_\pi(\phi)$, Eq. \eqref{pi model}, with a positive coupling strength; \textbf{e), f)} $V_\pi(\phi)$, Eq. \eqref{pi model}, with a negative coupling strength. The choice of parameters is shown in Tab. \ref{tab nonmin}.}
  \label{fig nonminimal dyn}
\end{figure*}


\subsection{Non-minimally coupled quasi-quintessence inflation}\label{subsec 52}

At this point we study whether the non-minimal coupling, $\frac{1}{2}\xi R\phi^2$, can lead to a consistent inflationary scenario or not, in the limit of weak interactions, $|\xi|\ll1$ and we focus on both the Einstein and the Jordan frame.

We start performing the transformation between the field $\phi$ and $h$, Eqs. \eqref{small xi}-\eqref{small neg xi}, to write the potentials in the Einstein frame, as in Eq. \eqref{pot}:
\begin{align}
     \mathcal V_E(h)&=\frac{\mathcal V(\phi(h))}{\left(1-\xi\chi\phi^2(h)\right)^2},\\
     \phi^{(+)}&\simeq\frac{1}{\sqrt{\chi\xi}}\sin(\sqrt{\chi\xi}h),\\
    \phi^{(-)}&\simeq-\frac{1}{\sqrt{\chi\xi(6\xi-1)}}\sinh\left(\frac{\sqrt{\chi\xi}h}{\sqrt{6\xi-1}}\right).
\end{align}
In order to satisfy the aforementioned conditions for the potentials we fix the free parameters as reported in Tab. \ref{tab nonmin}. The transformed potentials are plotted in Figs. \ref{fig nonminimal pot}.
As evident from these figures, the coupling makes the potential walls infinite for both the models. Thus, the two potentials manifest a similar behavior, \emph{i.e.},  showing a quite analogous dynamics. Following the previous prescription, we compute the slow roll parameters and the initial and final conditions, as prompted in Figs. \ref{fig nonminimal etaeps}.

In the Jordan frame, because of the scalar curvature, the slow roll parameters are no longer defined in the usual way, via Eqs. \eqref{eps}-\eqref{eta}. Thus, taking the previous initial conditions on the field $h$, from Eqs. \eqref{small xi}-\eqref{small neg xi}, we have
\begin{align}
    \dot\phi&\simeq\cos{(\sqrt{\chi\xi}h)}\dot h,\quad&\xi>0,\\
    \dot\phi&\simeq(1-6\xi)\cosh\left(-\frac{\sqrt{\chi\xi}h}{\sqrt{6\xi-1}}\right)\dot{h},\quad&\xi<0.
\end{align}
All the initial and final conditions for this case are written in Tab. \ref{tab nonmin}.

\begin{table}[ht]
  \centering
  \begin{tabular}{c|c|c|c|c|c}
    \hline\hline
    &$\xi$&$\phi_0/M_{Pl}$&$\alpha$&$\chi$ &$V_0$  \\
    \hline
    \hline
    $V_S$ & $2\cdot10^{-4}$ & $\sqrt{\frac{3}{2}}\frac{\ln 2}{2}$ & $\sqrt{\frac{2}{3}}\frac{1}{M_{Pl}}$ &26.56 & $-{\chi\phi_0^4}/{4}$ \\

    $V^{(+)}_\pi$ & $7\cdot10^{-4}$ & $3\sqrt{\frac{3}{2}}{\ln 2}$ & $\sqrt{\frac{2}{3}}\frac{1}{M_{Pl}}$ &0.02  & 0 \\

    $V^{(-)}_\pi$ & $-5\cdot10^{-4}$ & $3\sqrt{\frac{3}{2}}{\ln 2}$ & $\sqrt{\frac{2}{3}}\frac{1}{M_{Pl}}$ &0.02  & 0 \\

    $V_W^{small}$ & $5\cdot10^{-3}$ & $3\sqrt{\frac{3}{2}}{\ln 2}$ & $\frac{\ln2}{\phi_0^2}$ &0.02  &  0\\

    $V_W^{large}$ & $5\cdot10^{-4}$ & $2\sqrt{\frac{3}{2}}{\ln 2}$ & $\frac{\ln2}{\phi_0^2}$ & 0.10 & $-{\chi\phi_0^4}/{4}$ \\

    $V_\Omega^{(+)}$  & $5\cdot10^{-4}$ & $5\sqrt{\frac{3}{2}}{\ln 2}$ & $\frac{\ln2}{\phi_0^2}$ &$3\cdot10^{-3}$  & 0 \\

    $V_\Omega^{(-)}$ & $-1\cdot10^{-4}$ & $6\sqrt{\frac{3}{2}}{\ln 2}$ & $\frac{\ln2}{\phi_0^2}$ & $1\cdot10^{-3}$ &  0\\

    \hline\hline
     &$h_{in}/M_{Pl}$ & $h_{end}/M_{Pl}$ & $\dot h_{in}/M_{Pl}$& $\phi_{in}/M_{Pl}$& $\dot\phi_{in}/M_{Pl}$\\
    \hline
    \hline
    $V_S$ &  $2.83$ & $0.61$ & $-0.04$& $2.81$ & $-0.04$ \\

    $V^{(+)}_\pi$ & $0.32$ & $2.42$ & $0.02$& $0.32$ & $0.02$ \\

    $V^{(-)}_\pi$ & $0.17$ & $2.33$ & $0.01$& $0.17$ & $0.01$ \\

    $V_W^{small}$ & $0.87$ & $3.02$ & $0.02$& $0.86$& $0.01$ \\

    $V_W^{large}$ & $4.25$ & $1.91$ & $-0.02$& $4.09$ & $-0.02$ \\

    $V_\Omega^{(+)}$ & $1.63$ & $4.23$ & $0.02$& $1.62$ & $0.02$ \\

    $V_\Omega^{(-)}$ & $2.02$ & $4.84$ & $0.02$& $2.02$ & $0.02$ \\
    \hline\hline
  \end{tabular}
  \caption{Choice of the parameters $\xi,\>\alpha,\>\phi_0,\>\chi,\>V_0$ for the inflationary potentials introduced in Sect. \ref{sec 4}. Initial and final conditions of the minimally coupled inflationary stage. The label $\pm$ indicates where the coupling strength is positive or negative. The choice of the parameters $\xi,\>\phi_0$, and consequently of $\chi$, is made to provide a good agreement with the slow roll conditions first and with observational constraints. Specifically, $V_0$ and $\alpha$ are fixed as explained in Sects. \ref{sec 3} and \ref{sec 4}, whereas the initial and final conditions are recovered as developed in Sects. \ref{sec 5} and \ref{sec 6}.}
  \label{tab nonmin}
\end{table}

Finally we numerically solve the equations of motion, Eqs. \eqref{infl dyn}-\eqref{Jordan dyn}, respectively within the Einstein and the Jordan frame, as drawn in Figs. \ref{fig nonminimal sym dyn} and \ref{fig nonminimal dyn}.

The non-minimally coupled quasi-quintessence model proves to be successful in generating a suitable inflationary stage. Some interesting features are that the symmetry breaking potentials exhibit similar dynamics due to the coupling between the scalar field and curvature and that the energy scales where inflation occurs are much higher with respect the others potentials.

The evolution of the inflaton field in the phase space is characterized by a slow roll phase, followed by a well-defined attractor behavior around the minimum of the potential for all the potentials addressed, but $V_\Omega$ that, despite an inflationary stage in both frames, does not present a chaotic graceful exit in the Jordan frame.

Further, the analyses conducted in both the Einstein frame and  Jordan frame consistently confirm the equivalence between the two frames. Phrasing it differently, the dynamics of the inflationary stage and the resulting predictions for observables remain consistent and unaffected by the choice of frame.


\begin{figure*}[p]
  \centering
  \textbf{a)}\hspace{0mm}
    \includegraphics[width=.4\linewidth]{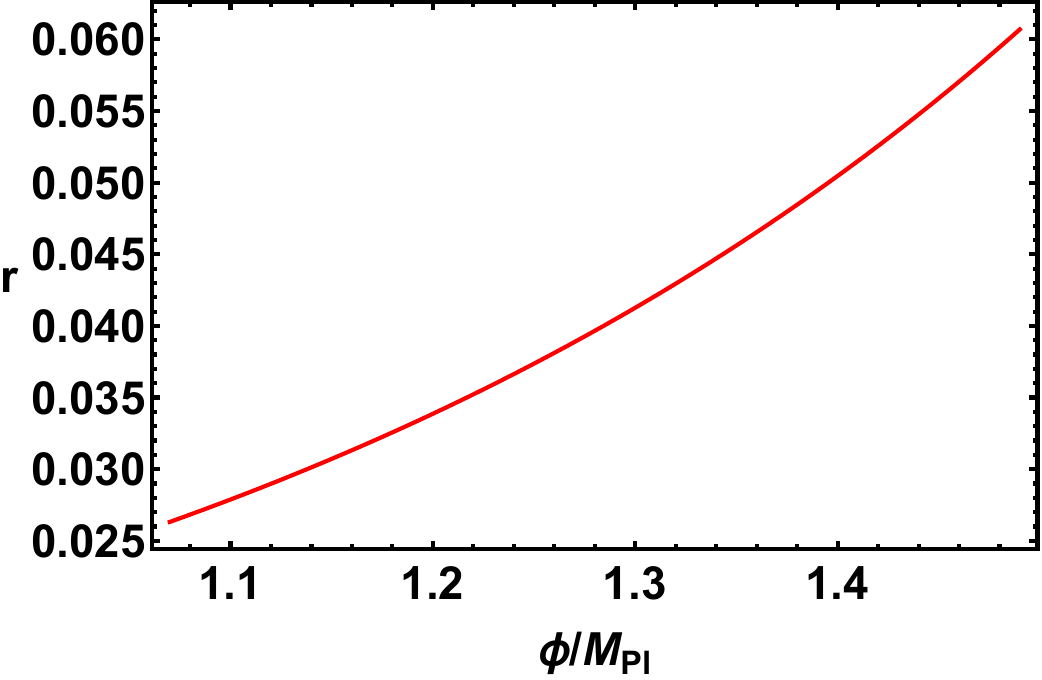}
  \hspace{0mm}
  \textbf{b)}\hspace{0mm}
    \includegraphics[width=.415\linewidth]{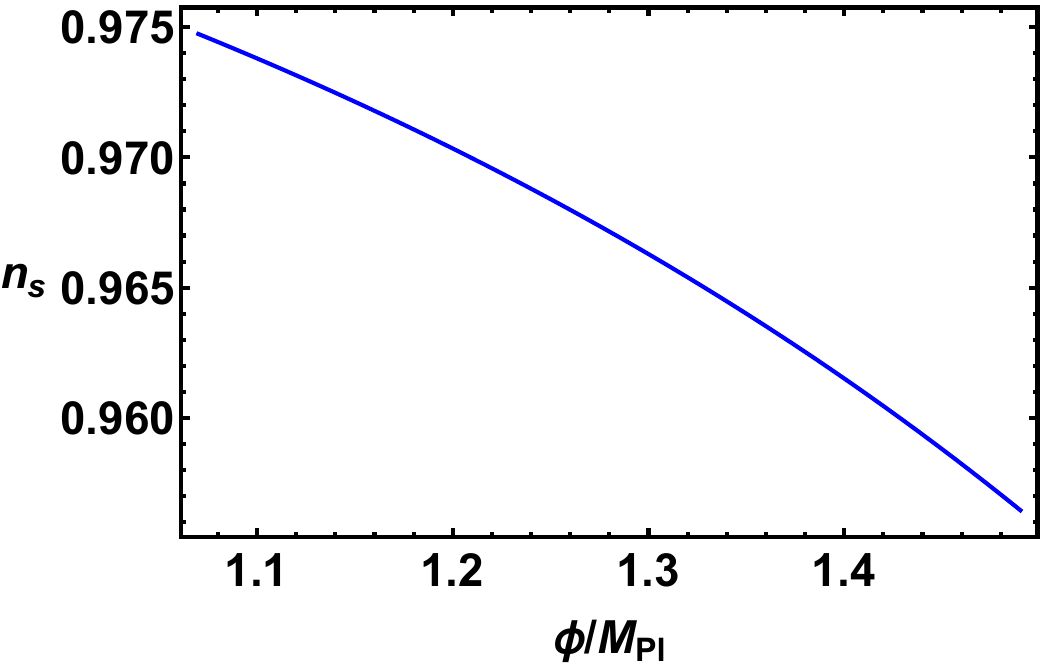}\\
    \vspace{7mm}
  \textbf{c)}
    \includegraphics[width=.4\linewidth]{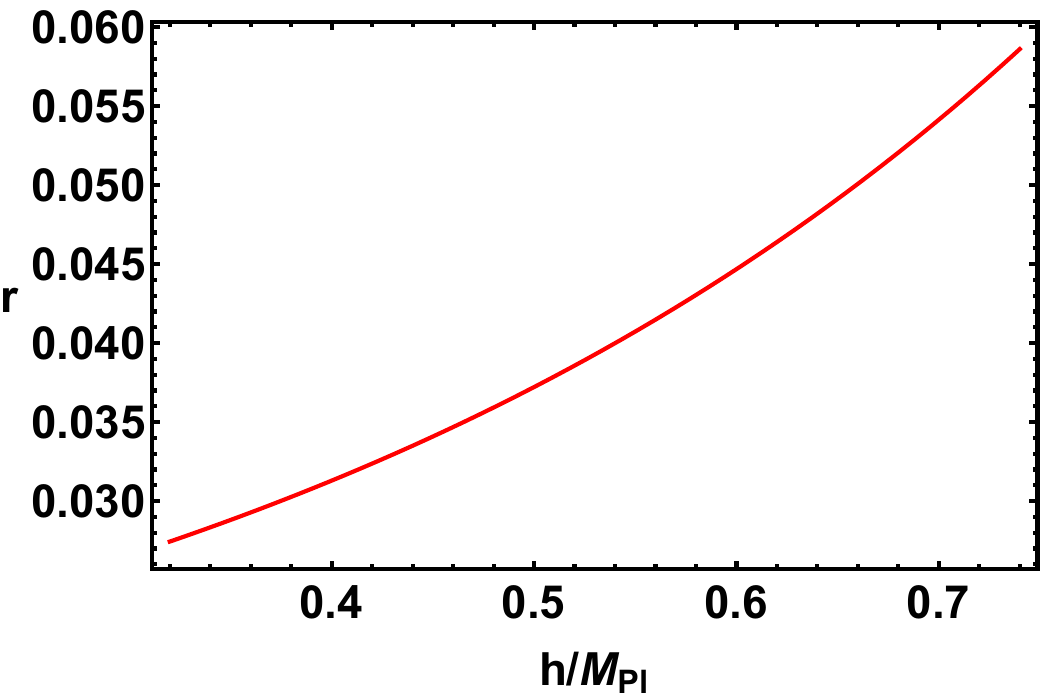}
  \hspace{0mm}
  \textbf{d)}
    \includegraphics[width=.415\linewidth]{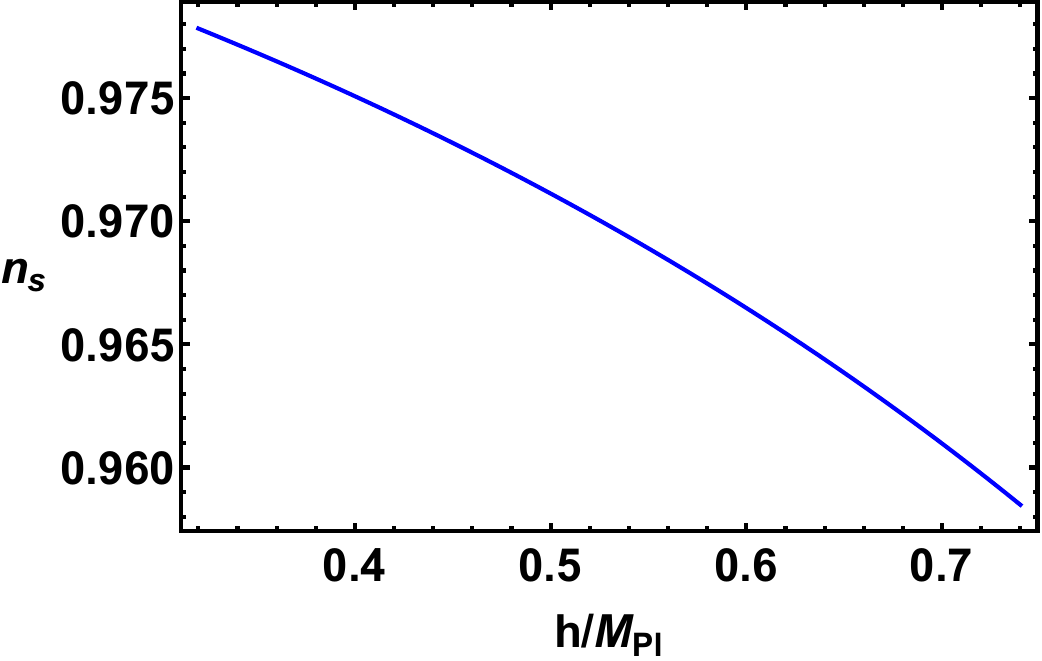}\\
    \vspace{7mm}
  \textbf{e)}\hspace{1mm}
    \includegraphics[width=.4\linewidth]{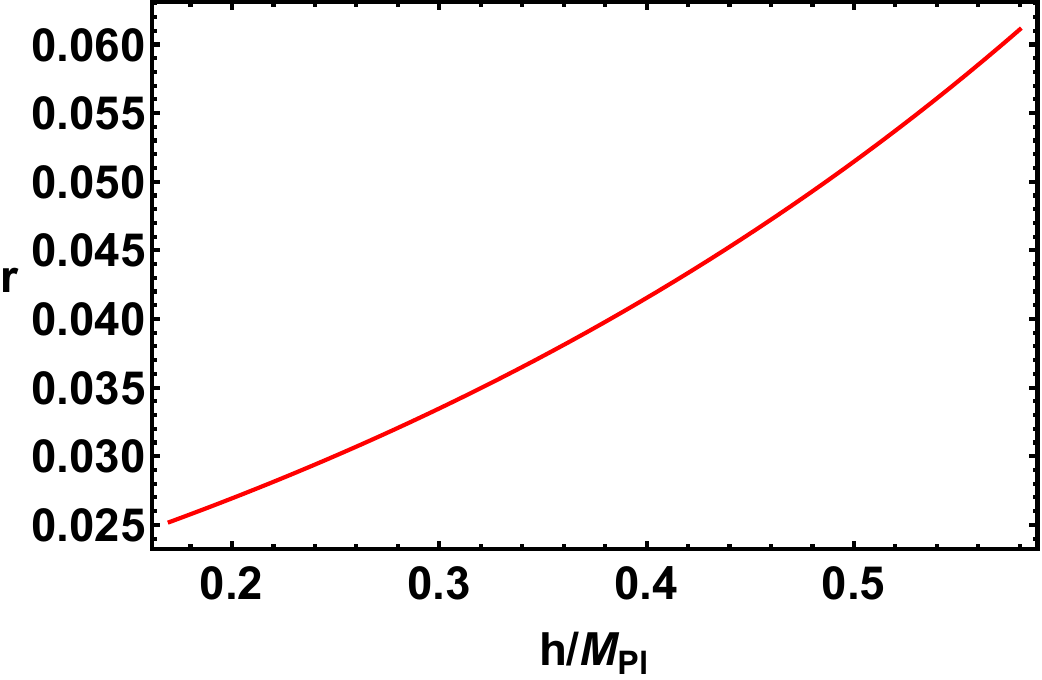}
  \hspace{0mm}
    \textbf{f)}\hspace{1mm}
    \includegraphics[width=.415\linewidth]{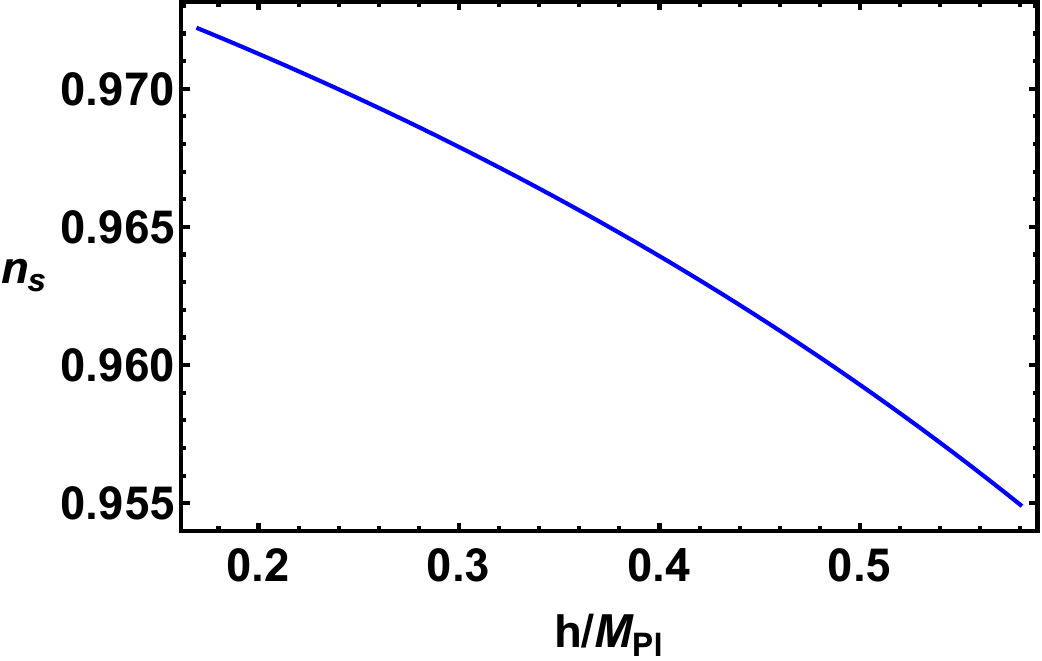}\\
  \caption{On the left the tensor-to-scalar ratio, $r$ defined in Eq. \eqref{r}, and on the right the spectral index, $n_s$ defined in Eq. \eqref{ns}, of the potential $V_\pi(\phi)$ introduced in Eq. \eqref{pi model}: \textbf{a), b)} minimally coupled scenario; \textbf{c), d)} non-minimally coupled scenario with a positive coupling strength; \textbf{e), f)} non-minimally coupled scenario with a negative coupling strength. The choice of parameters is shown in Tab. \ref{tab nonmin}.}
  \label{fig rns pi}
\end{figure*}

\begin{figure*}[p]
  \centering
  \textbf{a)}\hspace{0mm}
    \includegraphics[width=.39\linewidth]{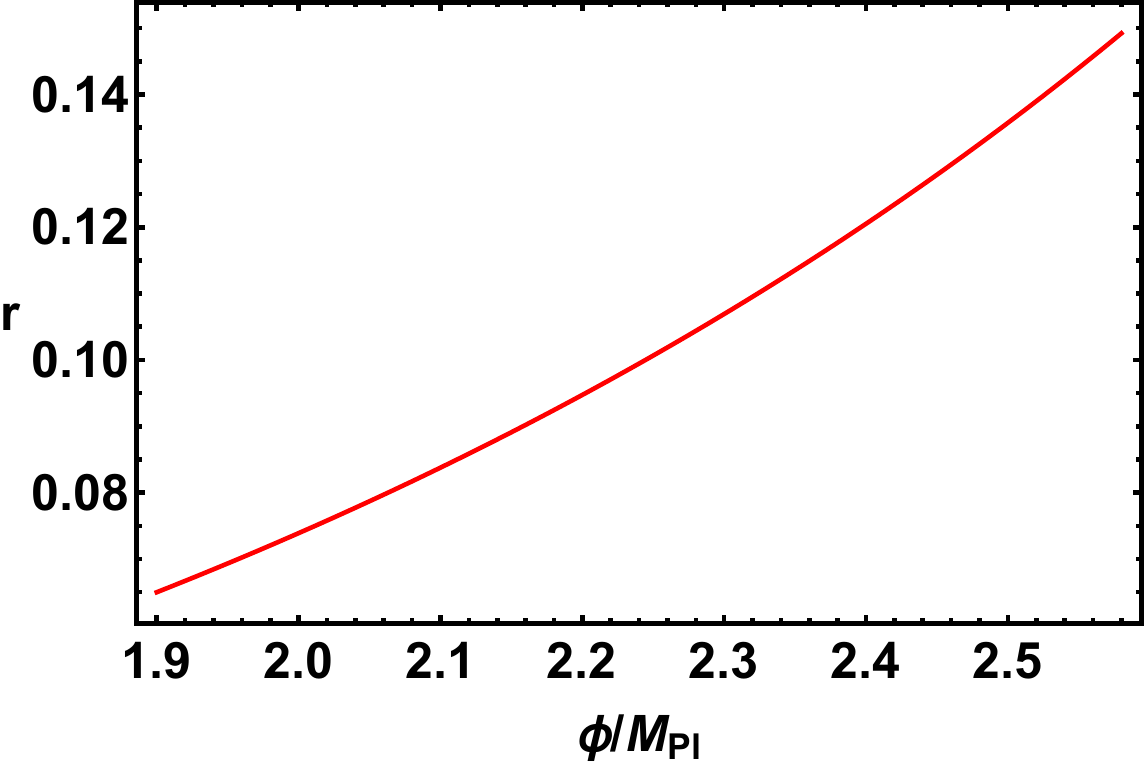}
  \hspace{0mm}
  \textbf{b)}\hspace{1mm}
    \includegraphics[width=.405\linewidth]{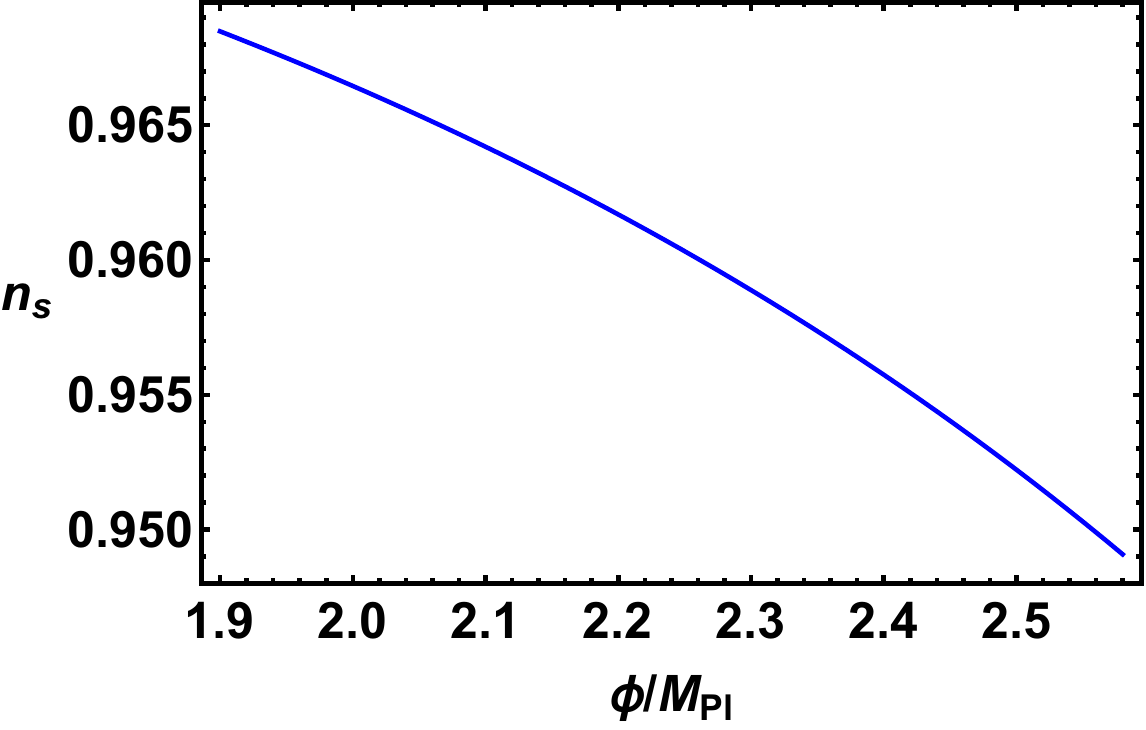}\\
    \vspace{5mm}
  \textbf{c)}\hspace{1mm}
    \includegraphics[width=.415\linewidth]{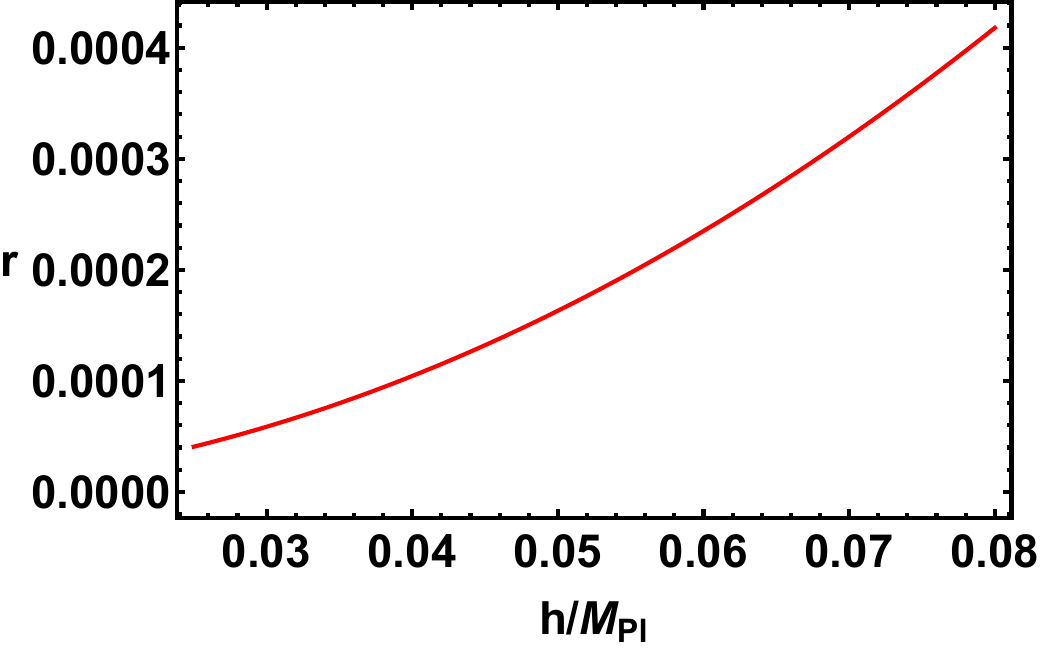}
  \hspace{0mm}
  \textbf{d)}\hspace{0mm}
    \includegraphics[width=.43\linewidth]{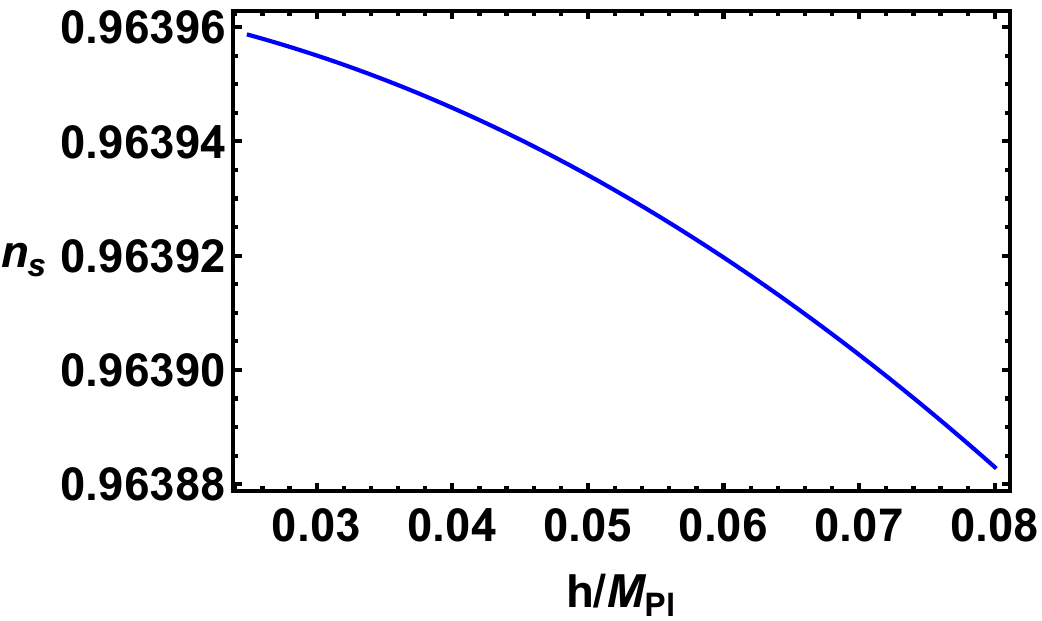}\\
    \vspace{5mm}
  \textbf{e)}\hspace{0mm}
    \includegraphics[width=.39\linewidth]{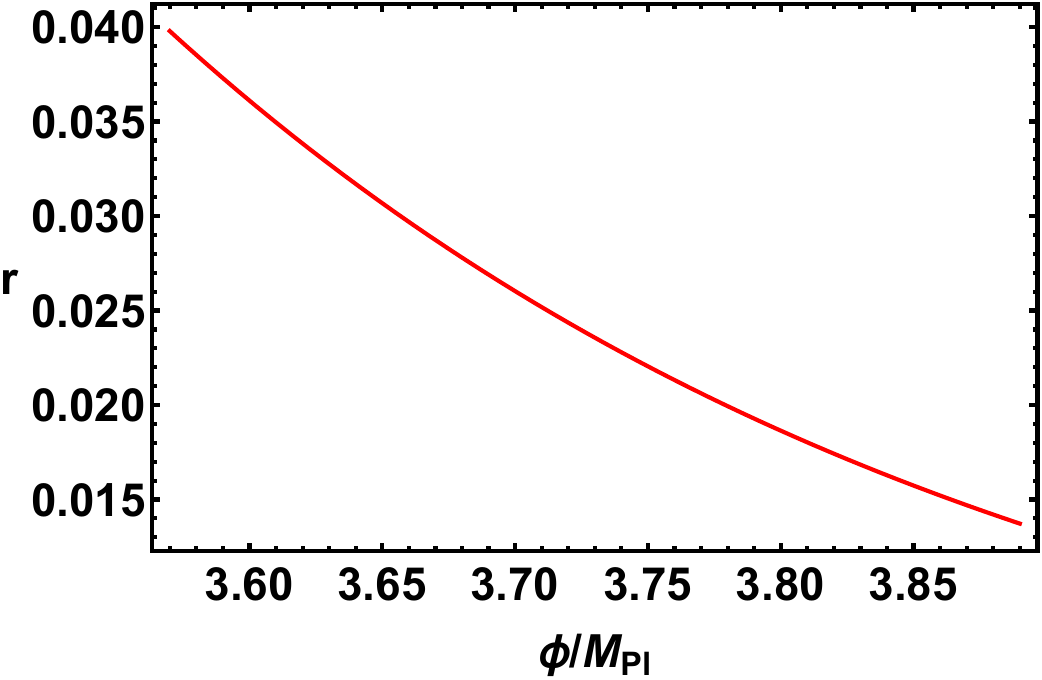}
  \hspace{0mm}
    \textbf{f)}\hspace{1mm}
    \includegraphics[width=.405\linewidth]{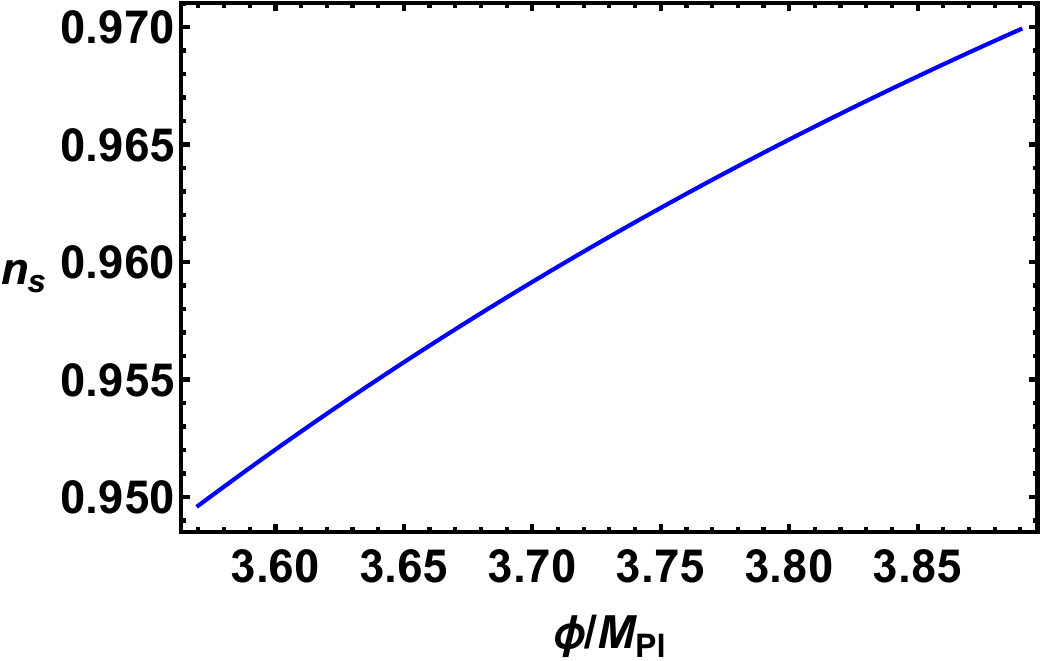}\\
    \vspace{5mm}
  \textbf{g)}\hspace{0mm}
    \includegraphics[width=.39\linewidth]{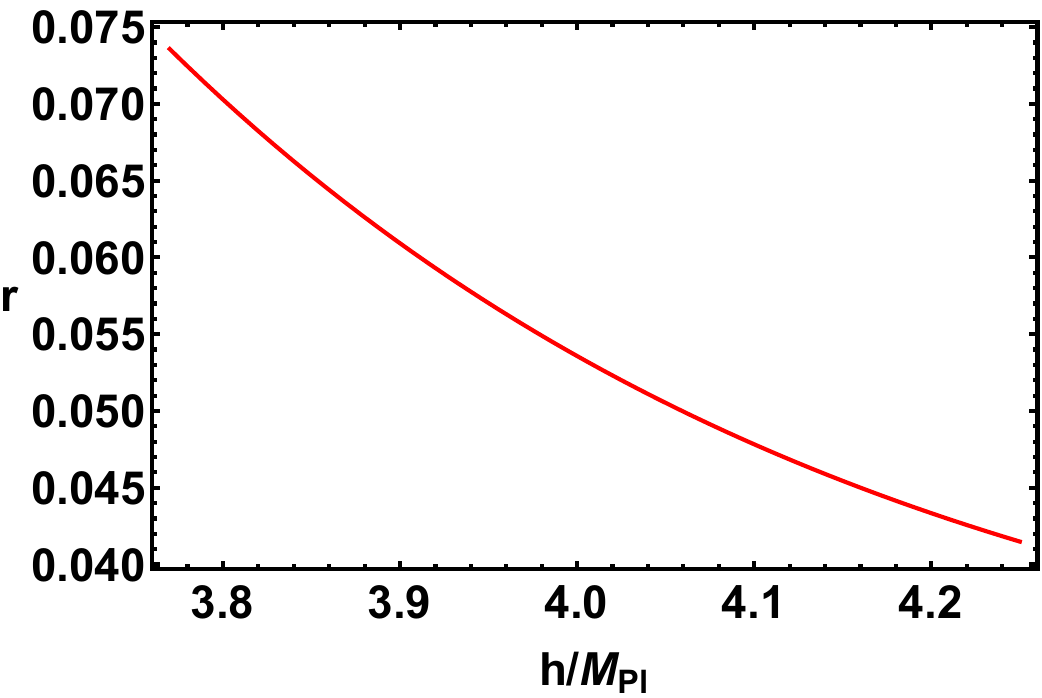}  \hspace{0mm}
    \textbf{h)}\hspace{1mm}
    \includegraphics[width=.405\linewidth]{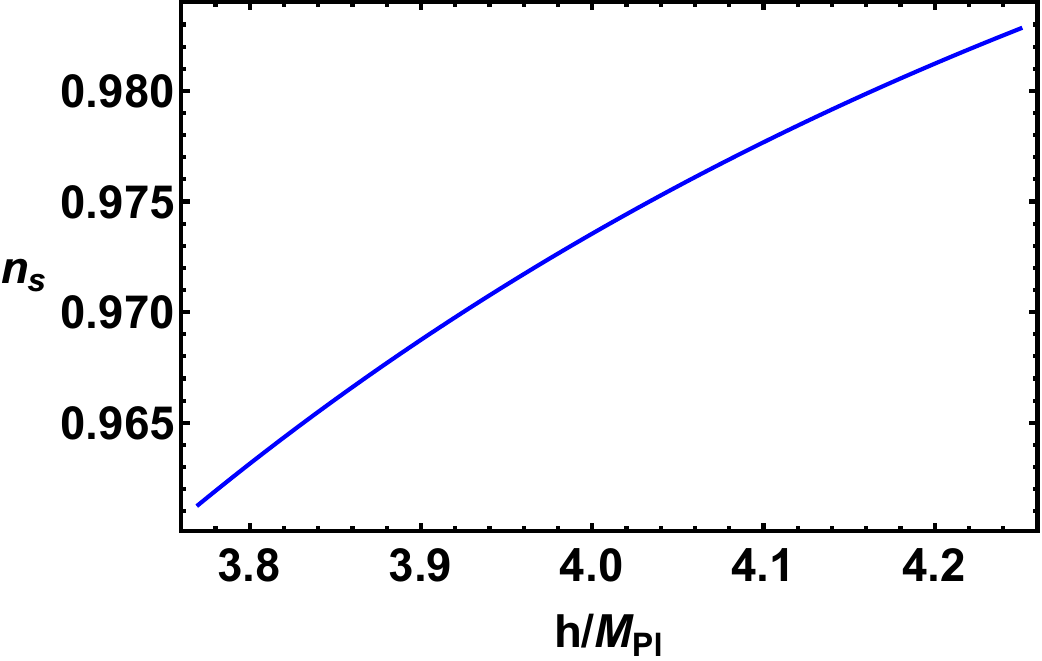}
  \caption{On the left the tensor-to-scalar ratio, $r$ defined in Eq. \eqref{r}, and on the right the spectral index, $n_s$ defined in Eq. \eqref{ns}, of the potential $V_W(\phi)$ introduced in Eq. \eqref{v2}: \textbf{a), b)} small field regime in a minimally coupled scenario; \textbf{c), d)} small field regime in a non-minimally coupled scenario; \textbf{e), f)} large field regime in a minimally coupled scenario; \textbf{g), h)} large field regime in a non-minimally coupled scenario with a positive coupling strength. The choice of parameters is shown in Tab. \ref{tab nonmin}.}
  \label{fig rns w}
\end{figure*}

\begin{figure*}[p]
  \centering
  \textbf{a)}\hspace{1.mm}
    \includegraphics[width=.4\linewidth]{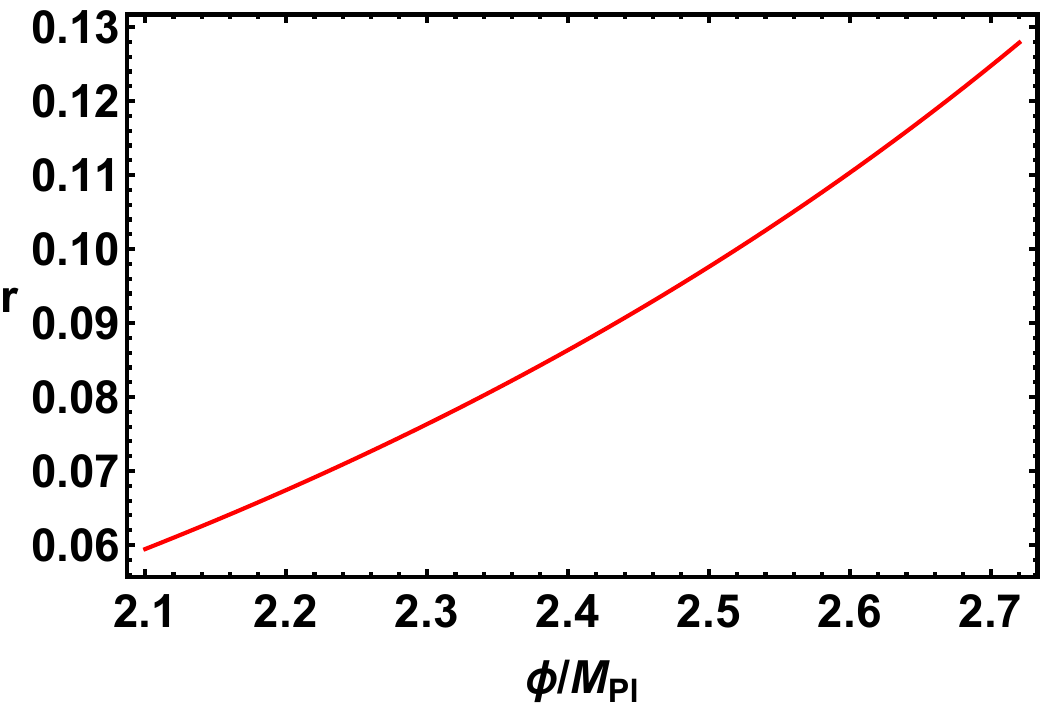}
  \hspace{0mm}
  \textbf{b)}\hspace{0.8mm}
    \includegraphics[width=.415\linewidth]{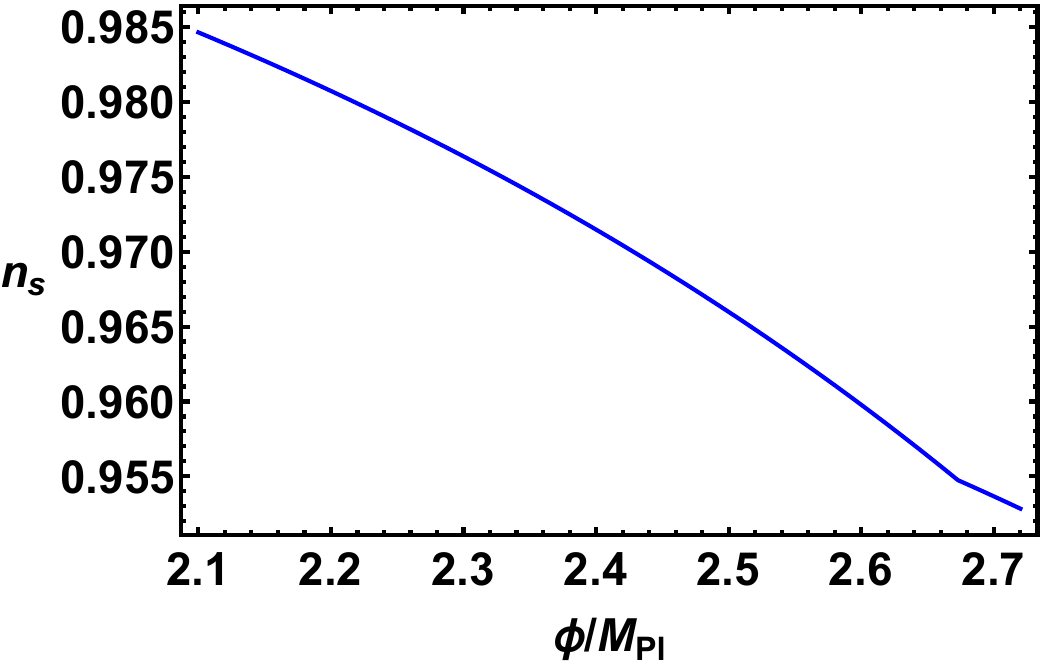}\\
    \vspace{7mm}
  \textbf{c)}
    \includegraphics[width=.4\linewidth]{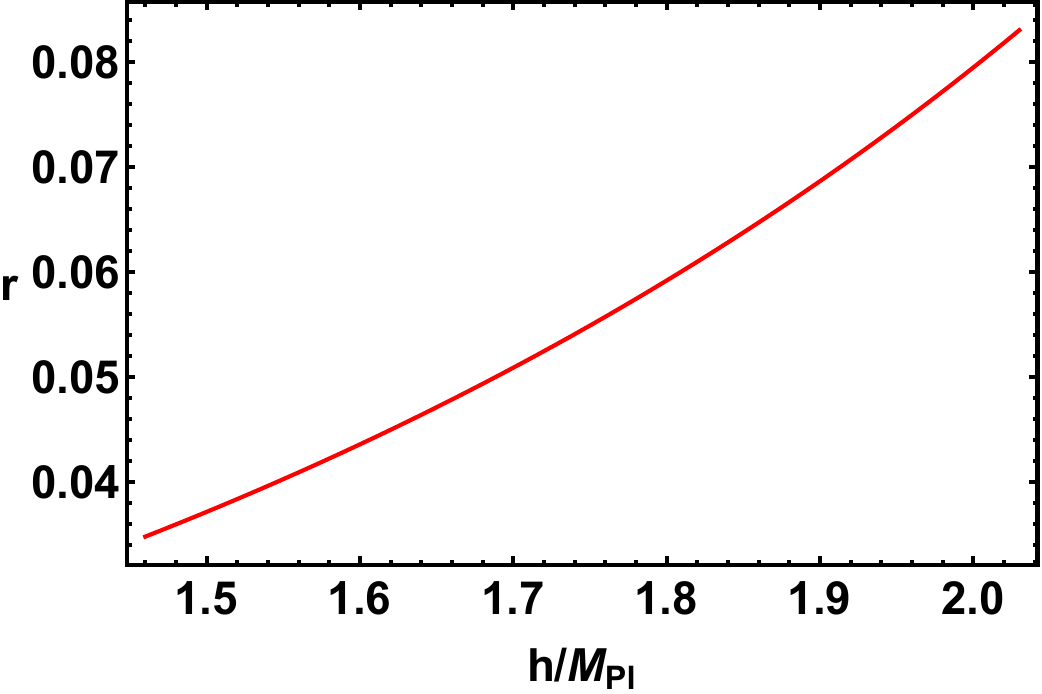}
  \hspace{0mm}
  \textbf{d)}\hspace{1.3mm}
    \includegraphics[width=.415\linewidth]{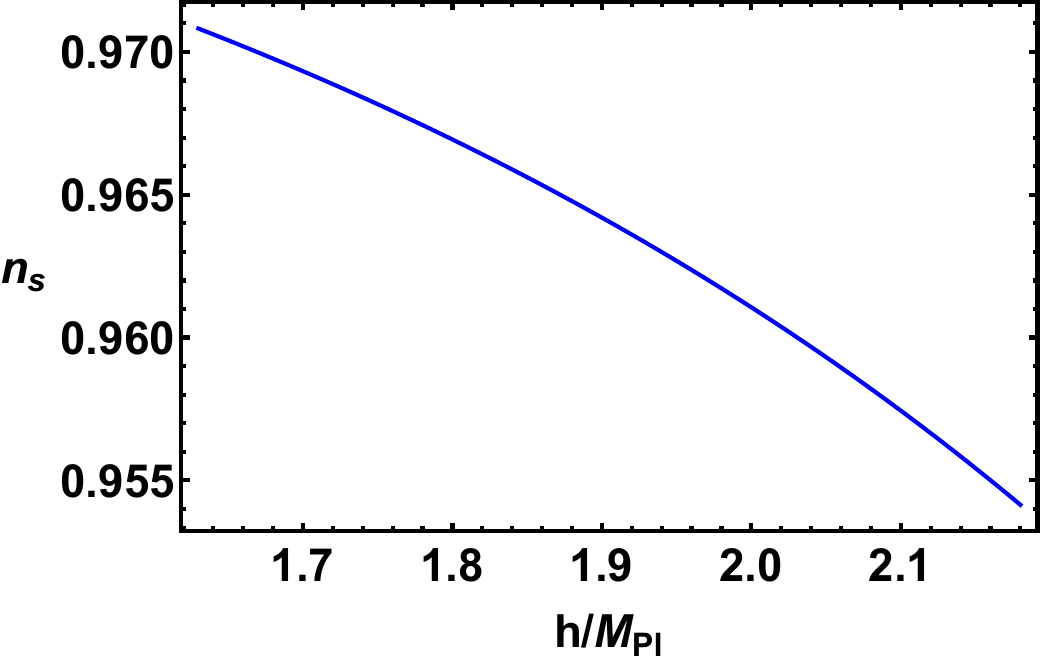}\\
    \vspace{7mm}
  \textbf{e)}\hspace{1.mm}
    \includegraphics[width=.4\linewidth]{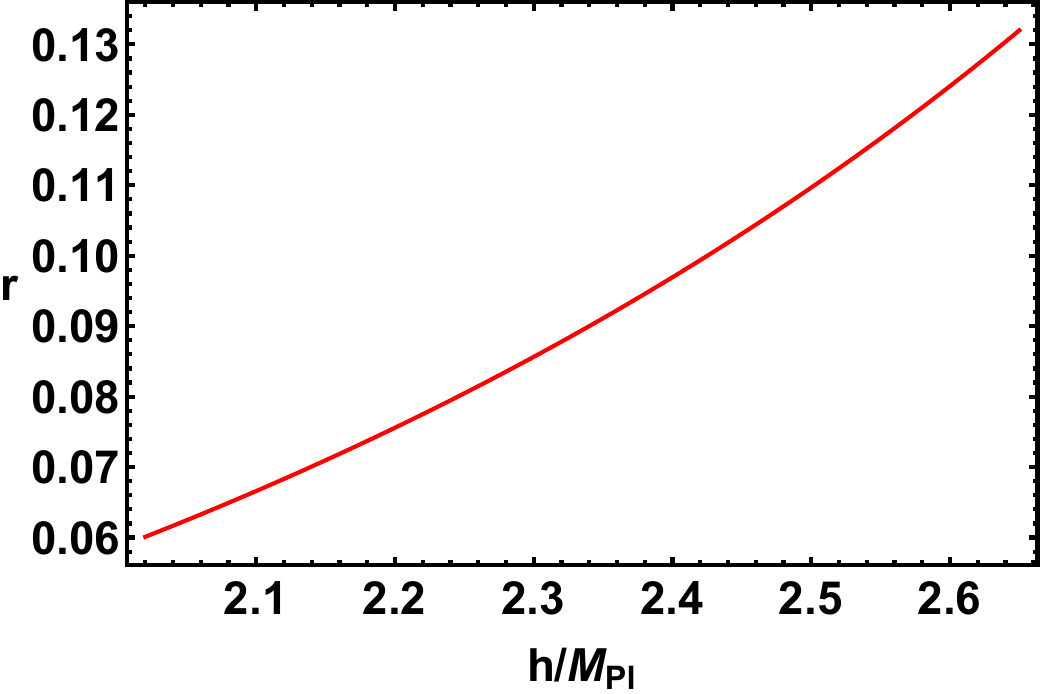}
  \hspace{0mm}
    \textbf{f)}\hspace{1.3mm}
    \includegraphics[width=.415\linewidth]{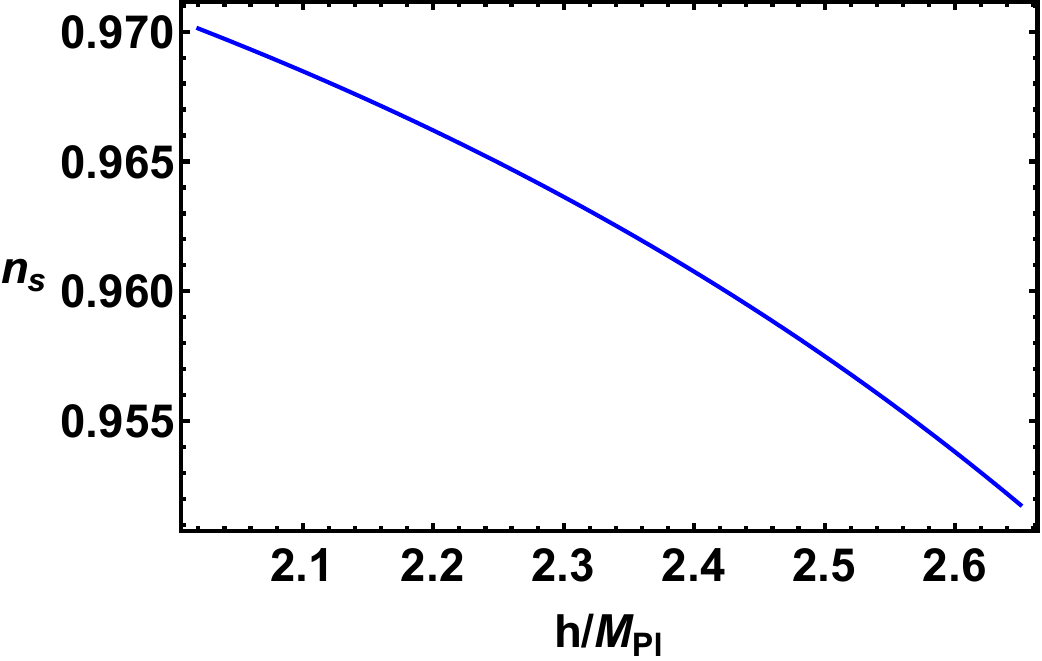}\\
  \caption{On the left the tensor-to-scalar ratio, $r$ defined in Eq. \eqref{r}, and on the right the spectral index, $n_s$ defined in Eq. \eqref{ns}, of the potential $V_\Omega(\phi)$ introduced in Eq. \eqref{v3}: \textbf{a), b)} minimally coupled scenario; \textbf{c), d)} non-minimally coupled scenario with a positive coupling strength; \textbf{e), f)} non-minimally coupled scenario with a negative coupling strength. The choice of parameters is shown in Tab. \ref{tab nonmin}.}
  \label{fig rns omega}
\end{figure*}

\begin{figure*}[ht]
  \centering
  \textbf{a)}\hspace{0mm}
    \includegraphics[width=.43\linewidth]{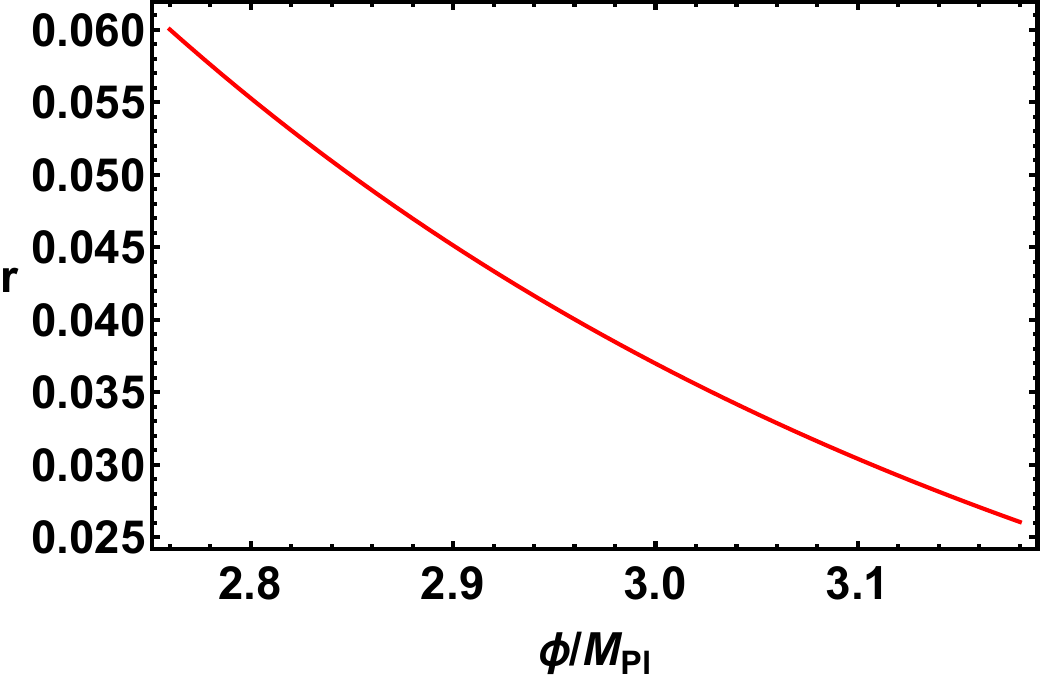}
  \hspace{0mm}
  \textbf{b)}\hspace{0mm}
    \includegraphics[width=.445\linewidth]{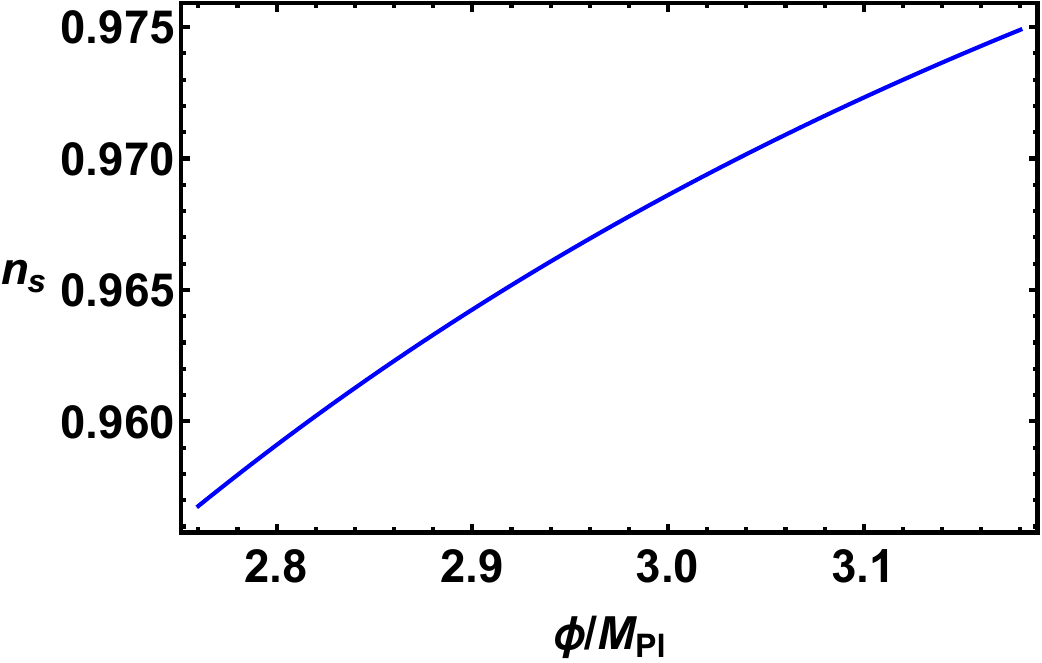}\\
    \vspace{5mm}
  \textbf{c)}
    \includegraphics[width=.43\linewidth]{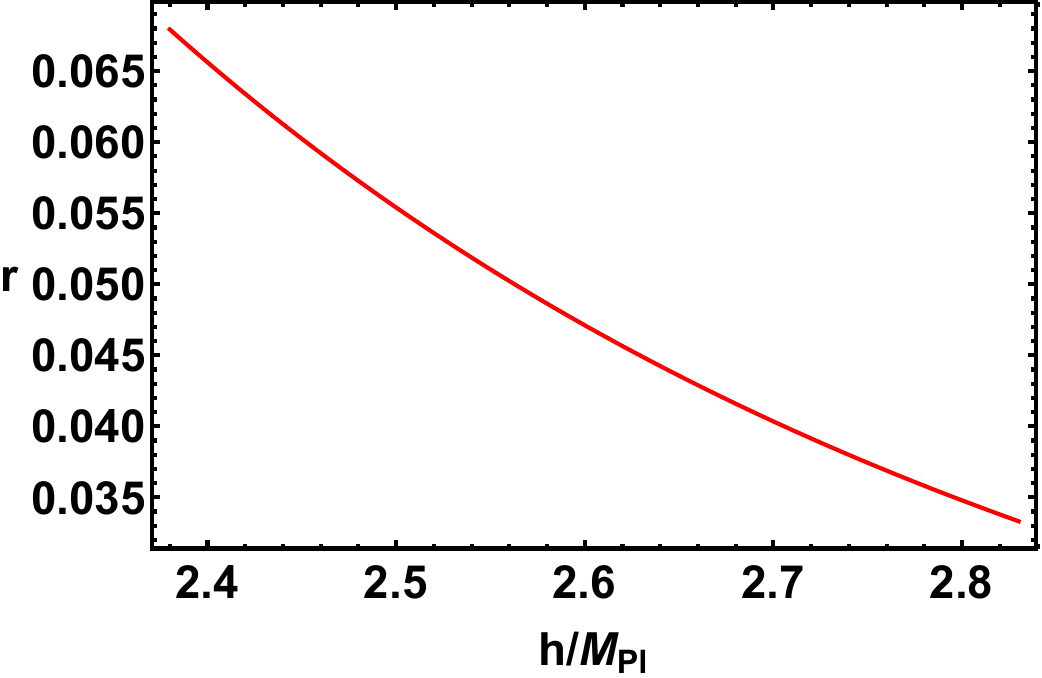}
  \hspace{0mm}
  \textbf{d)}
    \includegraphics[width=.445\linewidth]{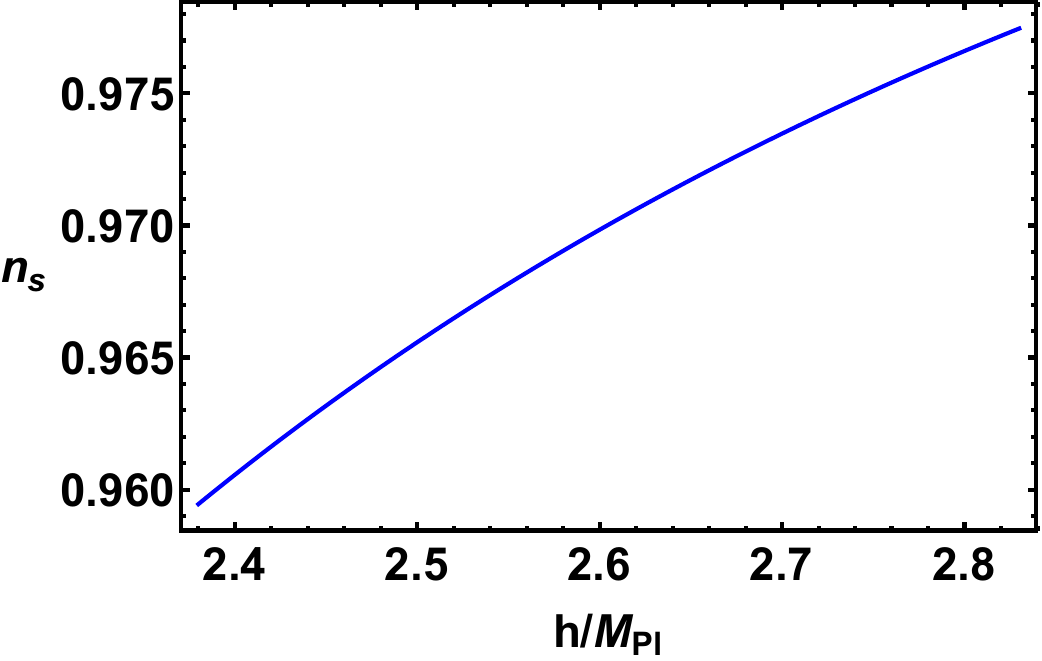}\\
  \caption{On the left the tensor-to-scalar ratio, $r$ defined in Eq. \eqref{r}, and on the right the spectral index, $n_s$ defined in Eq. \eqref{ns}, of the potential $V_S(\phi)$ introduced in Eq. \eqref{star_like}: \textbf{a), b)} minimally coupled scenario; \textbf{c), d)} non-minimally coupled scenario; \textbf{e), f)} non-minimally coupled scenario with a negative coupling strength. The choice of parameters is shown in Tab. \ref{tab nonmin}.}
  \label{fig rns star}
\end{figure*}

\section{Theoretical discussion and predictions}\label{Sec 8}

From the Planck mission data \cite{planck}, we argue an experimental feedback regarding inflationary models and from the cosmic microwave background anisotropies and their power spectrum, we can extract  information that can be encoded in two key parameters: the tensor-to-scalar ratio, denoted as $r$, and the spectral index, denoted as $n_s$ \cite{Lyth_1999, Cook_2015}.

These parameters play a crucial role in characterizing the properties of inflationary models. In the slow roll regime, the tensor-to-scalar ratio and spectral index can be expressed as follows:
\begin{subequations}
    \begin{align}
    r&=16\epsilon^*,\label{r}\\
    n_s&=1+2\eta^*-6\epsilon^*,\label{ns}
    \end{align}
\end{subequations}

where the slow roll parameters, $\epsilon^*$ and $\eta^*$, are evaluated at the horizon crossing, \emph{i.e.}, at the moment when the physical wavelength of fluctuations reaches the Hubble horizon during the slow roll phase. The fluctuations at this stage are directly related to cosmic microwave background anisotropies. Typically, the horizon crossing occurs between the onset of the slow roll phase and an e-folding number up to approximately $40$.

The Planck mission has provided several bounds on these parameters, which vary depending on the chosen background model and the specific data set used. The corresponding constraints and results have been reported in Tab. \ref{tab:} for completeness.

By comparing the predicted values of $r$ and $n_s$ from inflationary models with the measured values derived from early time data, we can assess the viability and consistency of our inflationary scenarios.
Thus, we compute these parameters for all the potentials from the beginning of inflation to $\phi^{**}$ at $N\simeq40$.

\begin{table}[h]
  \centering
  \begin{tabular}{c|c|c}
    \hline\hline
     & $r$  & $n_s$ \\
    \hline
    Planck TT,TE,EE&&\\+lowEB+lensing & $<0.11$ & $0.9659\pm0.0041$  \\\hline
    Planck TT,TE,EE&&\\+lowE+lensing+BK15 & $<0.061$ & $0.9651\pm0.0041$ \\\hline
    Planck TT,TE,EE&&\\+lowE+lensing+BK15+BAO& $<0.063$& $0.9668\pm0.0037$ \\
    \hline\hline
  \end{tabular}
  \caption{Three different bounds obtained with three different data sets, considering as background model the $\Lambda CDM+r$ benchmark. Bounds provided by Planck mission, as reported in Ref. \cite{planck}.}
  \label{tab:}
\end{table}

Our computed plots, in Figs. \ref{fig rns pi}-\ref{fig rns star}, clearly demonstrate that inflationary models satisfy the observational data, as they provide a good fit with the measured values of the tensor-to-scalar ratio $r$ and the spectral index $n_s$.

However, in some cases, we obtain values that conform to only certain data set bounds and not to all of them. Nevertheless, it is important to note that this prescription for the models introduces a fine-tuning issue. The values of $r$ and $n_s$ are strongly dependent on the parameters $\phi_0$ and $\xi$. In particular, achieving a good fit to the observational data requires careful tuning of these parameters. This implies that a specific combination of values for $\phi_0$ and $\xi$ is needed to obtain the desired inflationary predictions. Such fine-tuning may be seen as a limitation for the models, as it suggests a sensitivity to the specific choices of these parameters.

All these results, specifying which model behaves better than the others, are summarized in Tab. \ref{Comparison2}.


    \begin{table*}[p]
    \centering
\footnotesize
\setlength{\tabcolsep}{0.5em}
\renewcommand{\arraystretch}{2}
\begin{tabular}{lcccccccccc}
\hline
\hline
$\xi=0$ &  $V_S$    &  $V_\pi$    &  $V_W^{small}$  &  $V_W^{large}$  &  $V_\Omega$  \\

\hline
$V_0$       & $-\chi\phi_0^4/4$
            & 0
            & 0
            & $-\chi\phi_0^4/4$
            & 0
                \\
Inflation   & large field
            & small field
            & small field
            & large field
            & small field
                \\
$n_s$       & $\checkmark$
            & $\checkmark$
            & $\checkmark$
            & $\checkmark$
            & $\checkmark$
                \\
$r$         & $\checkmark$
            & $\checkmark$
            & $\checkmark$
            & $\checkmark$
            & $\checkmark$
                \\
graceful exit
            & $\checkmark$
            & $\checkmark$
            & $\checkmark$
            & $\checkmark$
            & $\checkmark$
                \\
energy scales   & $\phi_{in}\simeq3.18M_{Pl}$
                & $\phi_{in}\simeq1.07M_{Pl} $
                & $\phi_{in}\simeq1.91M_{Pl} $
                & $\phi_{in}\simeq3.89M_{Pl} $
                & $\phi_{in}\simeq2.10M_{Pl} $
                \\
               & $\phi_{end}\simeq1.03M_{Pl}$
                & $\phi_{end}\simeq3.21M_{Pl}$
                & $\phi_{end}\simeq4.88M_{Pl}$
                & $\phi_{end}\simeq1.90M_{Pl}$
                & $\phi_{end}\simeq4.89M_{Pl}$
                \\
fine-tuning
            & $\checkmark$
            & $\checkmark$
            & $\checkmark$
            & $\checkmark$
            & $\checkmark$
                \\
coincidence
            & $\checkmark$
            & \ding{55}
            & \ding{55}
            & $\checkmark$
            & \ding{55}
                \\
\hline
\hline
$\xi>0$ &  $V_S$    &  $V_\pi$    &  $V_W^{small}$  &  $V_W^{large}$  &  $V_\Omega$  \\

\hline
$V_0$       & $-\chi\phi_0^4/4$
            & 0
            & 0
            & $-\chi\phi_0^4/4$
            & 0
                \\
Inflation   & large field
            & small field
            & small field
            & large field
            & small field
                \\
$n_s$       & $\checkmark$
            & $\checkmark$
            & $\checkmark$
            & $\checkmark$
            & $\checkmark$
                \\
$r$         & $\checkmark$
            & $\checkmark$
            & $\checkmark$
            & $\checkmark$
            & $\checkmark$
                \\
graceful exit
            & $\checkmark$
            & $\checkmark$
            & $\checkmark$
            & $\checkmark$
            & $\checkmark$
                \\
energy scales   & $h_{in}\simeq2.83M_{Pl}$
                & $h_{in}\simeq0.32M_{Pl}$
                & $h_{in}\simeq0.87M_{Pl} $
                & $h_{in}\simeq4.25M_{Pl} $
                & $h_{in}\simeq1.63M_{Pl} $
                \\
                & $h_{end}\simeq0.61M_{Pl}$
                & $h_{end}\simeq2.42M_{Pl}$
                & $h_{end}\simeq3.02M_{Pl}$
                & $h_{end}\simeq1.91M_{Pl}$
                & $h_{end}\simeq4.23M_{Pl}$
                \\
coupling strength
                & $0.0002$
                & $0.0007$
                & $0.0050$
                & $0.0005$
                & $0.0005$
                \\
fine-tuning
            & $\checkmark$
            & $\checkmark$
            & $\checkmark$
            & $\checkmark$
            & $\checkmark$
                \\
coincidence
            & $\checkmark$
            & \ding{55}
            & \ding{55}
            & $\checkmark$
            & \ding{55}
                \\
\hline
\hline
$\xi<0$ &  $V_S$    &  $V_\pi$    &  $V_W^{small}$  &  $V_W^{large}$  &  $V_\Omega$  \\

\hline
$V_0$       & $-\chi\phi_0^4/4$
            & 0
            & 0
            & $-\chi\phi_0^4/4$
            & 0
                \\
Inflation   & \ding{55}
            & small field
            & \ding{55}
            & \ding{55}
            & small field
                \\
$n_s$       & \ding{55}
            & $\checkmark$
            & \ding{55}
            & \ding{55}
            & $\checkmark$
                \\
$r$         & \ding{55}
            & $\checkmark$
            & \ding{55}
            & \ding{55}
            & $\checkmark$
                \\
graceful exit
            & \ding{55}
            & $\checkmark$
            & \ding{55}
            & \ding{55}
            & \ding{55}
                \\
energy scales   & \ding{55}
                & $h_{in}\simeq0.17M_{Pl}$
                & \ding{55}
                & \ding{55}
                & $h_{in}\simeq2.02M_{Pl}$
                \\
   &
                & $h_{end}\simeq2.42M_{Pl}$
                &
                &
                & $h_{end}\simeq4.84M_{Pl}$
                \\
coupling strength
                & \ding{55}
                & $-0.0005$
                & \ding{55}
                & \ding{55}
                & $-0.0001$
                \\
fine-tuning
            & \ding{55}
            & $\checkmark$
            & \ding{55}
            & \ding{55}
            & $\checkmark$
                \\
coincidence
            & \ding{55}
            & \ding{55}
            & \ding{55}
            & \ding{55}
            & \ding{55}
                \\
\hline
\hline
\end{tabular}
\caption{Resuming table of the inflationary features addressed in this work. We compare the results obtained for all the potentials introduced in Sect. \ref{sec 4} in the minimally and non-minimally coupled scenario with both positive and negative coupling strengths. Specifically, in this table are shown the values of the offsets, the coupling strengths and the energy scales. Furthermore, it is reported whether the potentials lead to an inflationary stage, eventually if they provide a graceful exit, and if they satisfy the observational constraints. Finally, we highlight when the potentials solve the cosmological constant problem within the quasi-quintessence picture. The choices over the free parameters, $\xi$ and $\phi_0$, are made  to provide viable agreement with the slow-toll conditions and with observational constraints. Moreover, $V_0$ is fixed as explained in Sects. \ref{sec 3} and \ref{sec 4},  whereas the initial and final conditions are recovered as shown in Sects. \ref{sec 5} and \ref{sec 6}.}
\label{Comparison2}
\end{table*}

From our analyses of the inflationary dynamics in both minimal and non-minimal coupling scenarios, we develop the following outcomes:
\begin{itemize}
    \item[-] all the considered potentials exhibit a well-defined inflationary scenario in both the minimally  and non-minimally coupled cases, where the coupling strength is positive, with magnitude of $|\xi|\sim10^{-3}\div10^{-4}$;
    \item[-] the non-minimally coupled case with a negative coupling strength is not suitable with most of the potentials. Only $V_\pi$ and $V_\Omega$ manage to provide an inflationary stage, with slight limitations;
    \item[-] for large field potentials, the inflationary evolution perfectly aligns with the quasi-quintessence scenario, where inflation occurs during a phase transition. This approach looks similar to the old inflation model, although after the slow roll stage, a graceful exit from inflation takes place naturally. Consequently, the potentials unify the pictures of old and chaotic inflation into the same standards;
    \item[-] even though small field potentials show viable inflationary stages, the small field is disfavored within the quasi-quintessence paradigm and the overall byproduct is that inflation would not coincide with the transition;
    \item[-] among all the explored models, symmetry breaking potentials emerge as the least compatible with observational constraints;
    \item[-] the inflationary dynamics are independent of the choice of frame, suggesting an equivalence between the Jordan and Einstein frames. This may be seen as an indirect confirmation that during the transition particles, even different from baryons, can emerge.
\end{itemize}

Considering these findings, we select the Starobinsky-like potential as the most promising, exactly as recent observations state.

Notice that our scenario slightly departs from the pure Starobinsky model, since the free constants appear slightly different.

More importantly, the way we inferred the model itself is \emph{completely different} from a second-order generalization of Einstein's gravity, \emph{i.e.}, from the original technique developed to obtain it. In this respect, we notice that our potential can adapt to couple with the Ricci curvature, while the original Starobinsky potential cannot be coupled since it is derived from geometrical perspective via a conformal transformation \cite{DiValentino:2016nni}.

More broadly, we introduced a class of potentials analogous to $\alpha-$attractor models, but starting with totally different physical hypothesis.

In such a way, our results imply that the quasi-quintessence scenario seems to prefer \emph{large fields}, originating from the generic double exponential ansatz, in Eq. \eqref{proposed} with $c=d=1$.

Next, the analyses of the tensor-to-scalar ratio and the spectral index reveal that the choice of coupling has a significant impact on the compatibility with the cosmic microwave background  anisotropies. Although the potentials yield consistent values for the tensor-to-scalar ratio and spectral index, we observe a strong dependence on the parameter choice, except for the Starobinsky-like potential.

Overall, these findings support the viability of the quasi-quintessence framework for inflationary scenarios and highlight the role of non-minimal coupling in providing consistent inflationary dynamics with desirable properties.

Our study also enriches the understanding toward obtaining the Starobinsky-like picture without passing from extensions of Einstein's gravity and/or phenomenological extensions, see e.g. \cite{Ivanov:2021chn,Blumenhagen:2015qda,Brinkmann:2023eph,Rodrigues-da-Silva:2021jab}.


\section{Outlooks and perspectives}\label{Sec 9}

In this work, we investigated a scalar field model that predicts the existence of a quasi-quintessence fluid. Notably, we showed that this fluid exhibits a vanishing speed of perturbations, implying that its properties resemble those of a matter-like fluid. This characteristic holds significant implications, particularly within the realm of small perturbation theory.

We put forth that, from the quasi-quintessence puzzle, potentials addressing the cosmological constant problem can be formulated. We achieved this goal by applying the standard thermodynamics and the shift-symmetry over the field itself. Additionally, by introducing a fourth-order symmetry breaking potential, we demonstrated a method to cancel out  cosmological constant surplus through a phase transition induced by this symmetry breaking mechanism.

We remarked that, during the transition, the quasi-quintessence fluid can exhibit a phase of strong acceleration, reinterpreted in terms of inflation. Hence, we discussed the kinds of potentials associated with this phase, constructing them from very general assumptions. Particularly, we modeled fourth possible potentials, split into two main classes.

On the one hand, the first has been categorized as Starobinsky-like paradigms, where we moved from a general double exponential potential, yielding a large field potential, $V_S$, closely resembling, in its functional form, the pure Starobinsky one. Moreover, within this scenario we incorporated symmetries. The first, delving into a field shift of the form $\phi\rightarrow1/\phi$ that has been immediately dismissed, being non-predictive, whereas another (discrete) symmetry, say $\phi\rightarrow-\phi$, returning into a novel small field approach, baptized $\pi$ potential.

The second class of potentials is rooted in symmetry breaking frameworks, constructed through careful selections of the free parameters within the double exponential hypothesis. In this context, we introduced the $V_W$ and $V_\Omega$ models. The latter exhibits infinite potential walls and is characterized by its low-field nature, whereas the former, featuring finite potential walls,  displays both low and large field behaviors.

Accordingly, we conducted an investigation into the dynamics of inflation in both of these scenarios for small and large fields.

Further, we delved into minimal and non-minimal couplings with scalar curvature. Then, to accomplish this, we incorporated a Yukawa-like interacting term, giving rise to effective coupled potentials.

To this end, in the Einstein frame, we computed the action, deriving the analytical form of the transformed field $h$. Afterwards, we provided approximated expressions for specific limiting case, with a focus on the weak interaction limit, \emph{i.e.}, $|\xi|\ll1$. Conversely, in the Jordan frame, we explicitly found the  modified equation of motion, and assumed a quasi-de Sitter phase characterized by $\dot R\simeq 0$.

In this regard, all the here-investigated potentials
showed a well-defined inflationary epoch in both the minimally and non-minimally coupled cases. However, limits on the coupling constants have also been found, suggesting $|\xi|\sim10^{-3}\div10^{-4}$. In addition, the non-minimally coupled case with a negative coupling strength appeared disfavored, showing that only $V_\pi$ and $V_\Omega$ manage to provide an inflationary stage, albeit with limitations on the fine-tuning of free coefficients.

As a byproduct of our findings, the large field potentials have been found to better adapt to the inflationary evolution in the quasi-quintessence scenario than small fields, that instead appear quite disfavored. For large field, we showed that the involved potentials might exhibit an offset fixed by $V_0=-\frac{\chi\phi_0^4}{4}$, healing \emph{de facto} the classical cosmological constant problem.

In all these cases, we remarked a strategy to unify old with chaotic inflation under the same standards. Indeed, after the slow roll stage, a graceful exit from inflation takes place naturally, for both the coupled and uncoupled  frameworks.  In this respect, we also showed that the inflationary dynamics is independent of the choice of frame, \emph{i.e.}, Jordan or Einstein one. This prerogative may suggest the equivalence between the Jordan and Einstein frames in terms of particle production, as required for cancelling the cosmological constant during the transition.

Concluding, among all the models we investigated, symmetry breaking potentials have emerged as the least compatible with observational constraints.

On the contrary, the most well-aligned model is our Starobinsky-like one. This suggests that our approach provides a conceivable alternative to obtaining the Starobinsky potential, without the need of additional terms in the Hilbert-Einstein action.

In future works, we aim to explore additional potentials, particularly focusing on the nature of the quasi-quintessence fluid. A possible hint that would explain the role played by the fourth-order symmetry breaking potential would be to match $\phi$ with the Higgs field, unifying \emph{de facto} our model with the Higgs inflation \cite{Cheong:2021vdb,Rubio:2018ogq}.

Moreover, we underlined throughout the text that, in order to delete the vacuum energy cosmological constant, the potential might transform into particles, likely different from baryons, that are produced during the inflationary stage. This prerogative is extremely different from the standard model in which particles, under the form of baryons, emerge in the reheating phase. Hence, one aspect that we aim to investigate will be to better characterize particle production during quasi-quintessence inflation.


\section*{Acknowledgements}
OL acknowledges hospitality to the Al-Farabi Kazakh National University during the time in which this paper has been finalized. The authors are grateful to Marco Muccino for the help in the computational part of this work. The work of OL is partially financed by the Ministry of Higher Education and Science of the Republic of Kazakhstan, Grant: IRN AP19680128.

%


\appendix

\section{Recasting the Einstein frame through a conformal transformation }\label{appendix}

In order to get rid of the non-minimally coupled term we perform the following conformal transformation \cite{einsteinframe,einsteinframe1}
\begin{align}
    g_{\mu\nu}^{(E)}&=g_{\mu\nu}\Omega^2(\phi)\>\Rightarrow\> \sqrt{-g^{(E)}}=\sqrt{-g}\Omega^4(\phi),\\
    R&=\Omega^2(\phi)\left[R_E-6\left(\frac{\partial_{\mu}\Omega\partial^\mu\Omega}{\Omega^2(\phi)}+\Box\ln{\Omega(\phi)}\right)\right].
\end{align}
Now the action becomes
\begin{equation}
    S_E=\int{d^4x{\sqrt{-g^{(E)}}}\left[\frac{R^{(E)}}{2\chi}\frac{1-\xi\chi\phi^2}{\Omega^2}-\mathcal{L}^{(E)}\right]},
\end{equation}
with the positions
\begin{align}
    \mathcal{L}^{(E)}&=\frac{\mathcal{L}}{\Omega^4},\\
    X_E&=\frac{1}{2}\partial^\mu\phi\partial_\mu\phi\left(\frac{\Omega^2+6\frac{1-\xi \chi\phi^2}{\chi}\left(\frac{\partial\Omega}{\partial\phi}\right)^2}{\Omega^4}\right),
\end{align}
in which the term $\Box\ln\Omega$ has been omitted since it does not contributes to the equation of motion.

To determine the scale factor we need to recover the Hilbert-Einstein action, $S_{HE}=\frac{1}{2\chi}\int{d^4x{\sqrt{-g}R}}$. Hence we fix
\begin{equation}
    \Omega^2=1-\xi\chi\phi^2. \label{scale factor}
\end{equation}
Moreover we introduce a new field $h$ such that
\begin{align}
    &\frac{\partial h}{\partial \phi}=\frac{\sqrt{\Omega^2+6\frac{1-\xi\chi\phi^2}{\chi}\left(\frac{\partial\Omega}{\partial\phi}\right)^2}}{\Omega^2},\label{jacobian}\\
    &V_{E}(\phi)=\frac{V(\phi)}{\Omega^4(\phi)}.\label{pot}
\end{align}
The last step that we need is explicitly defining the new field $h$. So, from Eqs. \eqref{scale factor}, \eqref{jacobian}, the following relations hold:
\begin{subequations}
\begin{equation}
    \frac{\partial\Omega}{\partial\phi}=-\frac{\xi \chi\phi}{\sqrt{1-\xi \chi\phi^2}},
\end{equation}
\begin{equation}
    \frac{\partial h}{\partial \phi}=\frac{\sqrt{1-\xi(1-6\xi)\chi\phi^2}}{1-\xi \chi\phi^2},
\end{equation}
\end{subequations}
thus the field $h$ becomes
\begin{equation}
\begin{split}
     h(\phi)=&\sqrt{\frac{6}{{\chi}}}\tanh^{-1}{\left(\frac{\sqrt{6\chi}\xi{\phi}}{\sqrt{1+\phi^2\chi\xi(-1+6\xi)}}\right)}+\\
     &-\sqrt{\frac{{-1+6\xi}}{{\chi\xi}}}\sinh^{-1}{\left(\sqrt{\chi\xi(-1+6\xi)}\phi\right)}.
\end{split}
\end{equation}


\vspace{0.5mm}
\section{Einstein's field equations in the Jordan frame}\label{appendix1}

Now, we focus on the Jordan frame. As introduced at the beginning, in this frame, we need to recover new Einstein's field equations through the variation of the total action. Thus, we obtain the usual Hilbert-Einstein Lagrangian variation plus a new term given by the variation of the non-minimal coupling Eq. \eqref{L int}
\begin{equation}
    \delta\left(\xi R\phi^2\sqrt{-g}\right)=\xi\delta R_{\mu\nu}g^{\mu\nu}\phi^2\sqrt{-g}+\xi G_{\mu\nu}\phi^2\sqrt{-g}\delta g^{\mu\nu},
\end{equation}
where the Einstein tensor, $G_{\mu\nu}\equiv R_{\mu\nu}-{1\over2}g_{\mu\nu}R$, has been used. It is easy to prove that
\begin{equation}
   g^{\mu\nu}\delta R_{\mu\nu}=\left(g_{\mu\nu}\partial_\alpha\partial^\alpha-\partial_\mu\partial_\nu\right)\delta g^{\mu\nu}.
\end{equation}
Recalling the energy-momentum tensor definition, we find
\begin{equation}
    T^{int}_{\mu\nu}=\frac{-2}{\sqrt{-g}}\frac{\delta S_{int}}{\delta g^{\mu\nu}}=\xi\left(g_{\mu\nu}\partial_\alpha\partial^\alpha-\partial_\mu\partial_\nu\right)\phi^2+\xi\phi^2G_{\mu\nu}.
\end{equation}
Hence, bearing in mind the energy-momentum tensor for the quasi-quintessence field, Eq. \eqref{eq:no10}, the new equations become
\begin{equation}
\begin{split}
     \frac{1-\chi\xi\phi^2}{\chi}G_{\mu\nu}&=\left(2-2\xi\right)\partial_\mu\phi\partial_\nu\phi+2\xi\phi(g_{\mu\nu}\square-\partial_\mu\partial_\nu)\phi+\\
    &\quad-g_{\mu\nu}\left[\left(\frac{1}{2}-2\xi\right)\partial_\alpha\phi\partial^\alpha\phi+\mathcal{V}(\phi)\right]. \label{efej}
\end{split}
\end{equation}

\end{document}